\documentclass[11pt, a4paper]{article}
\pdfoutput=1
\usepackage[utf8]{inputenc}
\usepackage{jheppub}
\hypersetup{pdfencoding=unicode, bookmarksopen=true, bookmarksnumbered}
\usepackage{microtype}
\usepackage{amsthm,amsbsy,amsfonts,mathrsfs,enumerate,float,wrapfig,amsmath,mathtools,graphicx,framed,tcolorbox}
\usepackage{tikz}
\usepackage{subcaption}
\usepackage{graphbox}

\DeclareMathOperator{\PE}{PE}
\DeclareMathOperator*{\Res}{Res}

\usepackage{tikz}
\usetikzlibrary{positioning}
\usetikzlibrary{arrows}
\title{Probing Quantum Curves and Transitions in 5d SQFTs via Defects and Blowup Equations}

\author[a,b,c]{Hee-Cheol Kim,}
\author[d]{Minsung Kim,}
\author[e]{Sung-Soo Kim,}
\author[f,g,d]{Kimyeong Lee,}
\author[h,i,d]{Xin Wang}
\affiliation[a]{Department of Physics, POSTECH, Pohang 37673, Korea}
\affiliation[b]{Asia Pacific Center for Theoretical Physics, Postech, Pohang 37673, Korea}
\affiliation[c]{Jefferson Physical Laboratory, Harvard University, Cambridge, MA 02138, USA}
\affiliation[d]{Quantum Universe Center, Korea Institute for Advanced Study, Seoul 02455, Korea}
\affiliation[e]{School of Physics, University of Electronic Science and Technology of China,\\
Chengdu, Sichuan 611731, China}
\affiliation[f]{Beijing Institute of Mathematical Sciences and Applications (BIMSA),
Huaibei Town, Huairou District, Beijing 101408, China}
\affiliation[g]{School of Physics, Korea Institute for Advanced Study, Seoul 02455, Korea}
\affiliation[h]{Interdisciplinary Center for Theoretical Study, University of Science and Technology of China, Hefei, Anhui 230026, China}
\affiliation[i]{Peng Huanwu Center for Fundamental Theory, Hefei, Anhui 230026, China}

\emailAdd{heecheol@postech.ac.kr}
\emailAdd{minsung1@kias.re.kr}
\emailAdd{sungsoo.kim@uestc.edu.cn}
\emailAdd{klee@bimsa.cn}
\emailAdd{wxin@ustc.edu.cn}

\abstract{
We investigate codimension-2 defect partition functions and quantum Seiberg-Witten curves in 5d rank-1 supersymmetric QFTs, including non-Lagrangian and Kaluza-Klein theories. Using generalized blowup equations, we compute defect partition functions in the $\Omega$-background and show that, in the Nekrasov-Shatashvili limit, they satisfy certain difference equations that encode the quantization of classical Seiberg-Witten curves. Furthermore, we explore novel transitions in the defect partition functions and their relation to coordinate transformations of quantum Seiberg-Witten curves, with a focus on SL(2,$\mathbb{Z}$) transformations and Hanany-Witten transitions. These findings provide new insights into the interplay between codimension-2 defects, quantum curves, and the geometric structure of 5d supersymmetric QFTs.
}

\begin{document}
\preprint{KIAS-Q25003, USTC-ICTS/PCFT-25-09}
\maketitle
\section{Introduction}

Supersymmetric quantum field theories (QFTs) in five and six dimensions are a rich subject to explore. Their classification has made remarkable progress over the past decades through gauge-theoretic methods and Calabi-Yau compactifications in M-/F-theory \cite{Seiberg:1996bd, Douglas:1996xp, Intriligator:1997pq, Heckman:2013pva, Heckman:2015bfa, Bhardwaj:2015oru, Bhardwaj:2015xxa, DelZotto:2017pti, Xie:2017pfl, Jefferson:2017ahm, Jefferson:2018irk, Bhardwaj:2019fzv, Apruzzi:2019vpe, Apruzzi:2019opn, Apruzzi:2019enx, Apruzzi:2019kgb, Bhardwaj:2020gyu}. Many non-perturbative aspects, such as UV dualities and symmetry enhancements at special points in moduli space, have been explored along the lines of the classification program, brane construction in string theory \cite{Aharony:1997ju, Aharony:1997bh, Bergman:2013aca, Bergman:2014kza, Hayashi:2015fsa, Zafrir:2015rga, Hayashi:2018lyv}, and explicit computations of supersymmetry-protected observables, such as instanton partition functions \cite{Nekrasov:2002qd, Nekrasov:2003rj, Mitev:2014jza, Hayashi:2014wfa, Hayashi:2016abm, Kim:2020hhh} and superconformal indices \cite{Bhattacharya:2008zy, Kim:2012gu}, which have provided deeper insights in the field.

One of the key objects that characterizes the vacuum moduli space of 5d supersymmetric QFTs is  the Seiberg-Witten (SW) curve, an algebraic curve defined by $ H(x,p)=0 $, where $ x $ and $ p $ are complex coordinates. It was originally introduced in 4d $ \mathcal{N}=2 $ supersymmetric QFTs, whose low energy effective theory is governed by a holomorphic function known as the prepotential. The prepotential determines the effective gauge coupling, which receives both 1-loop perturbative and non-perturbative instantonic corrections. The prepotential, as well as the central charges of the BPS particles, is derived from the period integrals of the meromorphic 1-form $ \lambda_{\mathrm{SW}} = p\,dx $ on a Riemann surface defined by the Seiberg-Witten curve \cite{Seiberg:1994rs, Seiberg:1994aj}. Geometrically, 4d $ \mathcal{N}=2 $ theories can be realized via string compactifications on local Calabi-Yau threefolds, where the Seiberg-Witten curve describes the mirror dual of the Calabi-Yau manifold \cite{Katz:1996fh, Chiang:1999tz}. This picture generalizes to 5d and 6d supersymmetric theories, formulated on the spacetime manifold $ M_4 \times S^1 $ and $ M_4 \times T^2 $, respectively, where $ M_4 $ is a 4-manifold.

The Seiberg-Witten description is closely related to the classical integrable system \cite{Gorsky:1995zq, Martinec:1995by, Donagi:1995cf}. The SW curve can be interpreted as the classical phase space of an M5-brane wrapping a special Lagrangian 3-cycle in the Calabi-Yau threefold. The SW differential $ \lambda_{\mathrm{SW}} $ defines the symplectic structure $ \omega=d\lambda_{\mathrm{SW}} $ of the phase space \cite{Aganagic:2003qj}. This M5-brane configuration is associated with a codimension-2 defect in the field theory and  is also linked to open topological strings in the topological string theory framework \cite{Bershadsky:1993cx}. As an example, the SW curves of 5d $ \mathcal{N}=1 $ and $ \mathcal{N}=2 $ $ \mathrm{SU}(N) $ gauge theories correspond to the spectral curve of the classical $ N $-body relativistic Toda lattice and the elliptic Ruijsenaars-Schneider model, respectively \cite{Nekrasov:1996cz}.

A fundamental problem is the quantization of such integrable systems, which has a natural formulation in supersymmetric field theory. Localization techniques offer a powerful method for studying this problem.  In the case of 5d and 6d supersymmetric QFTs, localization is implemented by introducing the $ \Omega $-background on $ \mathbb{C}^2_{\epsilon_1,\epsilon_2} \times S^1 $ and $ \mathbb{C}^2_{\epsilon_1,\epsilon_2} \times T^2 $ with two $ \mathrm{SO}(4) $ rotational parameters $ \epsilon_1 $ and $ \epsilon_2 $ \cite{Nekrasov:2002qd, Nekrasov:2003rj}. In the limit $ \epsilon_1, \epsilon_2 \to 0 $, the BPS partition function $ Z $ defined on the $ \Omega $-background is identified with the prepotential, and the moduli space of the field theory describes the phase space of a classical integrable system. On the other hand, the Nekrasov-Shatashvili (NS) limit $ \epsilon_1 \to \hbar $, $ \epsilon_2 \to 0 $ reveals the quantization of the integrable system, characterized by the canonical commutation relation $ [x,p]=\hbar $ \cite{Nekrasov:2009rc}. In this framework, the Seiberg-Witten curve $ H(x,p)=0 $ is promoted to an operator equation $ H(x,p) \Psi(x) = 0 $, known as the quantum curve, where $ \Psi(x) $  is the wavefunction obtained from the NS-limit of the codimension-2 defect partition function.

In this paper, we study codimension-2 defect partition functions and quantum curves in 5d rank-1 supersymmetric QFTs, including 5d superconformal field theories (SCFTs) and Kaluza-Klein (KK) theories arising from the circle compactification of 6d SCFTs. We first generalize the blowup equations for the BPS partition functions of 5d theories on $ \mathbb{C}^2 \times S^1 $ in the presence of codimension-2 defect. Recent developments of the blowup formalism \cite{Nakajima:2003pg, Nakajima:2005fg, Gottsche:2006bm} enable us to compute BPS partition functions on the $ \Omega $-background, also known as Nekrasov's instanton partition functions for 5d gauge theories and elliptic genera of self-dual strings in 6d theories, for generic 5d/6d supersymmetric QFTs \cite{Huang:2017mis, Kim:2019uqw, Gu:2018gmy, Gu:2019dan, Gu:2019pqj, Gu:2020fem, Kim:2020hhh, Kim:2023glm}. The blowup formalism has also been generalized to include the partition functions in the presence of the half-BPS Wilson loop operators and codimension-4 defects in 5d/6d theories \cite{Kim:2021gyj}, as well as the surface defects in 4d $ \mathcal{N}=2 $ theories \cite{Jeong:2020uxz, Nekrasov:2020qcq}. A key point of the blowup equation is that the partition function on the one-point blow-up of $ \mathbb{C}^2 $ factorizes into contributions from the two fixed points of the blown-up $ \mathbb{P}^1 $, and it captures the partition function on $ \mathbb{C}^2 $ via a smooth blow-down procedure. Notably, this factorization structure remains unchanged even when defects are introduced. In this study, we investigate the structure of the blowup equations in the presence of codimension-2 defects and determine their solutions. To verify our results, we compare these results with the defect partition functions computed from other methods, such as ADHM construction of the instanton moduli space \cite{Hwang:2014uwa, Gaiotto:2014ina}, Higgsing \cite{Gaiotto:2012xa} and freezing \cite{Hayashi:2023boy, Kim:2023qwh, Kim:2024vci}. Furthermore, we establish the quantization of classical Seiberg-Witten curves in 5d rank-1 theories using explicit expressions for codimension-2 defect partition functions. Our results apply to a broad class of theories, including toric and non-toric models in geometry, and non-Lagrangian theories.

We also examine how the quantum curves and codimension-2 defect partition functions transform under the coordinate changes of the SW curve $ H(x,p)=0 $. Specifically, we focus on two types of coordinate transformations with concrete physical significance. One is the $ \mathrm{SL}(2,\mathbb{Z}) $, which acts on the position and momentum variables of the SW curve. This transformation originates from the $ \mathrm{SL}(2,\mathbb{Z}) $ symmetry in IIB string theory picture. At the level of defect partition functions, it corresponds to a combination of shifting background Chern-Simons level for the global symmetry introduced by the defect and applying a Fourier transformation. The other transformation is the Hanany-Witten (HW) transition \cite{Hanany:1996ie}. Although two theories related by HW transition have equivalent bulk 5d physics, they can be distinguished by the codimension-2 defects. For instance, 5d $ \mathrm{SU}(2) $ theory admits $ \mathbb{Z}_2 $-valued discrete theta angle $ \theta $, where $ \theta\to \theta+2\pi $ can be understood as a HW transition. Two theories with $ \theta=0 $ and $ \theta=\pi $ are distinguished by their instanton corrections. On the other hand, even though bulk 5d observables, such as instanton partition functions and superconformal indices, do not differentiate between $ \theta=0 $ and $ \theta=2\pi $, our analysis using blowup equations reveals that the codimension-2 defect partition functions for $ \theta=0 $ and $ 2\pi $ are distinct, but are instead related through non-trivial transformations.

The remainder of this paper is organized as follows. Section~\ref{sec:5dSCFTHW} begins with a review of 5d $ \mathcal{N}=1 $ supersymmetric QFTs and their brane constructions in IIB string theory. We then discuss classical SW curves, their quantizations, and transitions of curves under $ \mathrm{SL}(2,\mathbb{Z}) $ and HW transitions. In Section~\ref{sec:codim2-hw-blowup}, we introduce codimension-2 defects in 5d theories and generalize the blowup equations in the presence of defects. We apply blowup equations to our main examples, 5d $ \mathrm{SU}(2) $ gauge theories, and examine the transitions of curves and defect partition functions under the coordinate transformations. Section~\ref{sec:example} presents additional examples of rank-1 5d field theories, including 5d KK-theories and non-Lagrangian theories. We then conclude with subtle issues and interesting future directions in Section~\ref{sec:conclusion}. Appendix~\ref{app:special} summarizes the definitions and properties of special functions used in the paper. Appendix~\ref{app:ADHM} discusses ADHM construction of instanton partition functions in the presence of codimension-2 defects, while Appendix~\ref{app:piadj} considers partition functions obtained by freezing. Appendix~\ref{app:hw} provides detailed computations for the transformations of codimension-2 defect partition functions under Hanany-Witten transitions.


\section{5d SCFTs and Hanany-Witten transitions}\label{sec:5dSCFTHW}
In this section, we review Type IIB 5-brane construction of 5d  $\mathcal{N}=1$ superconformal field theories (SCFTs) \cite{Seiberg:1996bd,Intriligator:1997pq} and discuss how to construct Seiberg-Witten curves for SCFTs based on the corresponding 5-brane webs.

Many 5d SCFTs admit mass deformations leading to gauge theory descriptions at low energy. Given a gauge group $G$, supersymmetric multiplets of the $\mathcal{N}=1$ gauge theory consist of the vector multiplet and charged matter hypermultiplets. Restricting to bosonic fields, the vector multiplet is comprised of a vector field $A_\mu$ and a real scalar field $\phi$. The hypermultiplets are complex scalar fields forming an $\mathrm{SU}(2)_R$ doublet which transform in a representation of $G$. On the Coulomb branch of the moduli space, where gauge group $G$ is completely broken to $\mathrm{U}(1)^{\text{rank}(G)}$, the scalar field in the vector multiplet takes the expectation values, denoted as $\phi_i$, in the Cartan of $G$, where $i=1,\cdots, \text{rank}(G)$.  The Coulomb branch is hence parameterized by $\phi_i$ and the corresponding low energy theory is described by an effective theory of the Abelian gauge groups, characterized by a cubic prepotential $\mathcal{F}(\phi)$ \cite{Seiberg:1996bd, Intriligator:1997pq},
\begin{align}\label{eq:prepotential}
    \mathcal{F}(\phi) = \frac{m}{2} h_{ij}\phi_i\phi_j-\frac{\kappa}{6} d_{ijk} \phi_i\phi_j\phi_k+\frac1{12}\bigg( 
    \sum_{e\in\bf R}|e\cdot \phi|^3-\sum_f\sum_{w\in {\bf w}_f}|w\cdot\phi+m_f|^3
    \bigg)\ .
\end{align}
Here, $m=1/{g_0^2}$ is the inverse of the gauge coupling $g_0$ squared, $h_{ij}={\rm Tr}(T^G_iT^G_j)$  with the generators $T^G_i$ in the fundamental representation of $G$, and the coefficient associated with the classical Chern-Simons (CS) term of level $\kappa$ is defined by $d_{ijk}=\frac{1}{2}{\rm Tr}(T^G_i\{T^G_j,T^G_k\})$, which is nonzero only for $G=\mathrm{SU}(N)$ with $N\ge 3$. The remaining terms are the one-loop contributions where $\mathbf{R}$ are the roots of the Lie algebra of $\mathfrak{g}$ associated with $G$ and $\mathbf{w}_f$ are the weights of the $f$-th hypermultiplet with mass $m_f$.

\subsection{5-brane webs and 5d SCFTs} \label{subsec:5dscft}

A large class of 5d $\mathcal{N}=1$  supersymmetric gauge theories are engineered by 5-brane webs in Type IIB string theory. 5-branes have charges denoted by ($p,q$) and form a charge conserving configuration called 5-brane web diagram. D5-branes correspond to ($1,0$) charges, NS5-branes to ($0,1$) charges, and their bound states can have ($p,q$) charges, where $p$ and $q$ are coprime \cite{Aharony:1997bh}. The worldvolume spanned by 5-branes are given in Table \ref{tab:5-braneWB}. The common directions ($x^0, x^1, x^2, x^3, x^4$) comprise the  worldvolume of 5d theories, whereas the remaining directions ($x^5, x^6$) provide a 2-dimensional plane, called the ($p,q$)-plane, on which 5-brane web configurations realize various  properties of 5d $\mathcal{N}=1$ supersymmetric gauge theories. Rotations in the ($x^7, x^8, x^9$) directions correspond to SU(2)$_R$ R-symmetry of 5d $\mathcal{N}=1$ theories. 

\begin{table}
    \centering
    \begin{tabular}{c|ccccc|cc|ccc}
    \hline
    5-/7-branes &0&1&2&3&4&5&6&7&8&9\\
    \hline
    NS5 $(0,1)$& $\times$&$\times$&$\times$&$\times$&$\times$&$\times$ &&&&\\
    D5 $(1,0)$& $\times$&$\times$&$\times$&$\times$&$\times$& &$\times$&&&\\
    $(p,q)$& $\times$&$\times$&$\times$&$\times$&$\times$&$\vartheta $&$\vartheta$ &&&\\
    7-brane& $\times$&$\times$&$\times$&$\times$&$\times$&&&$\times$&$\times$&$\times$\\
    \hline
\end{tabular}
\caption{Type IIB brane configuration. The directions that each brane is extended are denoted by the symbols $\times$ (fully extended) or $\vartheta$ (partially extended). } 
\label{tab:5-braneWB}
\end{table}
\begin{figure}
    \centering
    \includegraphics[scale=0.5]{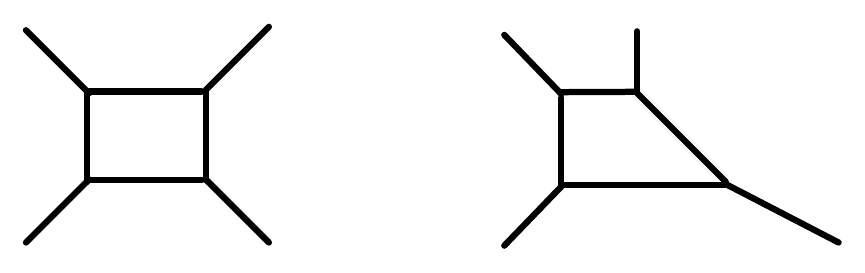}
    \caption{5-brane webs for 5d SU(2)$_\theta$ gauge theories without matter: SU(2)$_0$ on the left and SU(2)$_\pi$ on the right. }
    \label{fig:SU2s}
\end{figure}

A typical example of 5-brane webs  is given in Figure \ref{fig:SU2s}, where two distinctive BPS configurations of 5d $\mathcal{N}=1$ SU($2$) gauge theories are depicted on the ($x^6, x^5$)-plane. In this setup, the separation of two D5-branes along the vertical ($x^5$) direction represents a Coulomb phase.  The positions of D5-branes correspond to the VEV of the scalars in the vector multiplet. Fundamental strings stretched between D5-branes represent W-bosons, while D1 strings stretched between two NS5-branes correspond to instanton particle. Magnetic monopole strings in the 5d theory arise from D3-branes wrapping the compact surface in the brane web, and their tension, expressed as $\frac{\partial\mathcal{F}}{\partial{\phi}}$, is directly determined by the area of the surface.

The two SU(2) gauge theories shown in Figure~\ref{fig:SU2s}, representing distinct SU(2) theories without hypermultiplets, are known as the SU(2)$_{\theta=0, \pi}$ theories, where the discrete theta angle $\theta$ takes values in $\mathbb{Z}_2$, arising from  $\pi_4(\mathrm{SU}(2))=\mathbb{Z}_2$. While these theories are perturbatively identical, they differ non-perturbatively. Specifically, their BPS spectra, which include instanton particles, are distinct. Indeed, they exhibit different flavor symmetries at their UV fixed points and have distinct RG flows under mass deformations \cite{Morrison:1996xf,Bergman:2013ala}. For instance, the SU(2)$_\pi$ theory has a U(1) flavor symmetry, while the SU(2)$_0$ theory has an enhanced SO(3) flavor symmetry  at its UV fixed point. Additionally, the SU(2)$_\pi$ theory allows an RG flows to a non-Lagrangian theory called local $\mathbb{P}^2$ (or also frequently referred to $E_0$) theory, whereas the SU(2)$_0$ theory has no RG flow to an interacting IR theory. 

In the 5-brane web, a fundamental hypermultiplet can be introduced by attaching a semi-infinite D5-brane. Hypermultiplets in the antisymmetric or symmetric representation can also be realized by introducing orientifolds. For instance, see \cite{Bergman:2015dpa} for an explicit construction for antisymmetric and symmetric hypermultiplets. Some hypermultiplets in higher dimensional representations are also possible and discussed in \cite{Hayashi:2018lyv,Hayashi:2019yxj}.

It is often convenient to introduce 7-branes with worldvolumes as described in Table \ref{tab:5-braneWB}, where each 7-brane appears as a point in the $(p,q)$-plane. External 5-branes can be attached to 7-branes carrying the same charge. A fundamental hypermultiplet is represented by a (1,0) 7-brane (D7-brane), whose vertical distance from the center of the Coulomb branch determines the corresponding flavor mass. Thus, shifting the vertical position of a flavor D7-brane corresponds to a mass deformation for the fundamental hypermultiplet.

\begin{figure}
    \centering
    \includegraphics[width=15.1cm]{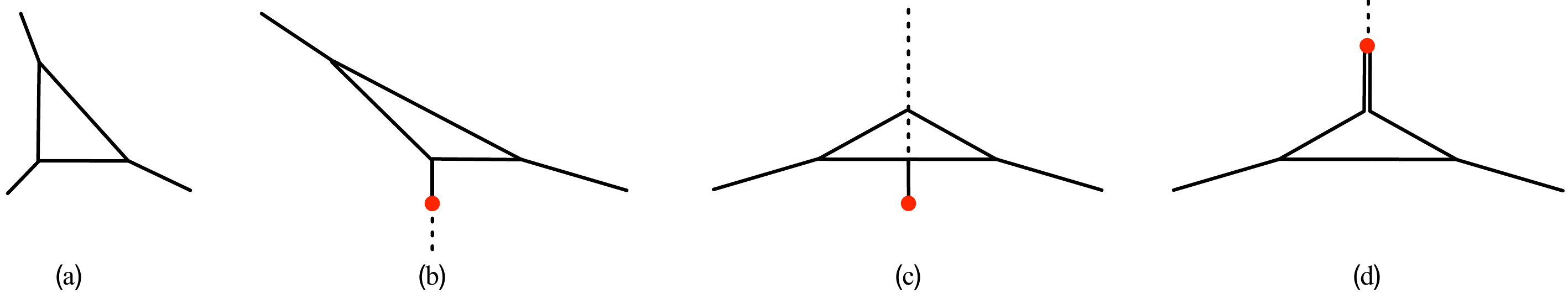}
    \caption{5-branes web for local $\mathbb{P}^2$, related by Hanany-Witten moves. After performing an SL$(2,\mathbb{Z})$ transformation, the monodromy cut (dashed line) of (0,1) 7-brane in red is rotated in the clockwise direction and the (0,1) 7-brane is moved upward.}
    \label{fig:localP2s}
\end{figure}

It is important to note that the horizontal placement of a 7-brane does not affect the physics of the 5d theory. As a result, a flavor 7-brane can be shifted freely in the horizontal direction. Similarly, 5d physics remains unchanged when a $(p,q)$ 7-brane along is moved along its charge direction $(p,q)$. Many aspects of the theory can be understood by appropriately moving 7-branes. A trivial example is the Hanany-Witten transition with a flavor D7-brane as shown in Figure~\ref{fig:localP2s}. When a Hanany-Witten move is performed, the shape of the 5-brane web changes due to the monodromy cut of a $(p, q)$ 7-brane has a monodromy cut. This cut causes the bending of branes when a 5-brane or another 7-brane crosses it. Hanany-Witten moves can also be performed with 7-branes carrying different charges, successive applications of these moves can lead to the enhancement of flavor symmetry. The enhanced flavor symmetry can be read off by putting external 7-branes inside a closed surface of the 5-brane web, where 7-brane configurations correspond to an ADE-type symmetry. Hanany-Witten transitions often result in 5-brane configurations where one 5-brane jumps over another, which leads to multiple 5-branes attaching to a single 7-brane.

\begin{figure}
    \centering
    \includegraphics[width=15.1cm]{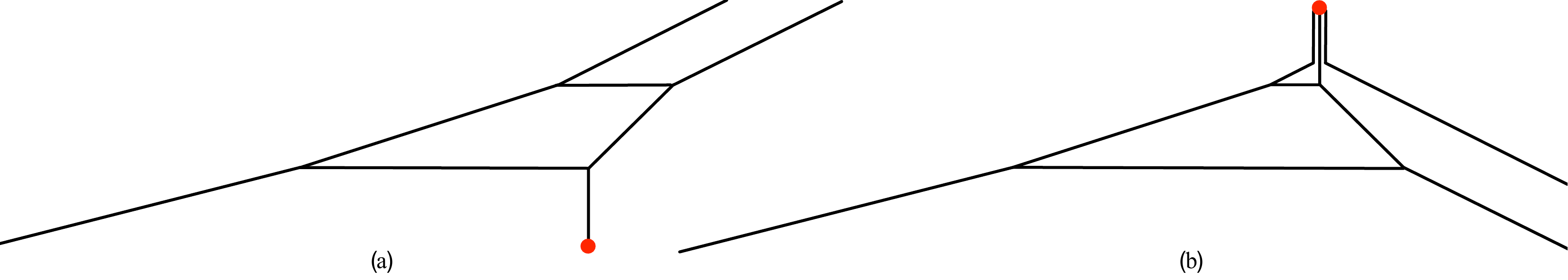}
    \caption{(a) A 5-brane web for the SU(2)$_{2\pi}$ theory and (b) A 5-brane web for  the SU(2)$_{4\pi}$ theory. Hanany-Witten move with a (0,1) 7-brane in red connects the two theories.}
    \label{fig:su2-2pi4pi}
\end{figure}

A simple example is the 5-brane web for the local $\mathbb{P}^2$ theory, as shown in Figure~\ref{fig:localP2s}. The well-known 5-brane web for local $\mathbb{P}^2$ is presented on the left in Figure~\ref{fig:localP2s}(a), while the same configuration, modified through Hanany-Witten moves, is shown on the right in Figure~\ref{fig:localP2s}(d). How these two 5-brane webs are related by the Hanany-Witten transition can be seen as follows: One first applies an SL($2,\mathbb{Z}$) transformation on the first configuration to get Figure~\ref{fig:localP2s}(b) and then rotate the monodromy cut of the (0,1) 7-brane in the clockwise direction so that it points upward as in Figure~\ref{fig:localP2s}(c). In doing so, the shape of the 5-brane configuration deforms as the monodromy cut changes $(p,q)$ brane charges. Finally, one brings the (0,1) 7-brane upward, which leads to successive brane annihilation and creation processes while preserving the charge conservation for 5-branes. Notice that in the resulting 5-brane web shown in Figure~\ref{fig:localP2s}(d), two NS5-branes are attached to a single (0,1) 7-brane. While this is not a conventional configuration, it is required to maintain charge conservation. Another example, which we will use later, is a 5-brane web for the SU(2)$_{4\pi}$ theory which can be realized by a Hanany-Witten transition from a 5-brane web of the SU(2)$_{2\pi}$ theory by applying a similar Hanany-Witten moves for the local $\mathbb{P}^2$ case. See Figure~\ref{fig:su2-2pi4pi}. We note that multiplet 5-branes are attached to a single 7-brane in this configuration. In this case, there are three NS5-branes attached to a (0,1) 7-brane in red as shown Figure~\ref{fig:su2-2pi4pi}, where two of these NS5-branes are created through the same process as in the local $\mathbb{P}^2$ theory, and the third NS5-brane is created as the (0,1) 7-brane crosses a color D5-brane. 

\subsection{Classical Seiberg-Witten curves}\label{sec:classicalSW}
One of characteristics capturing non-perturbative aspects of the theory that can be obtained from 5-brane webs is the 5d Seiberg-Witten (SW) curves which describe an M5-brane configuration. The M5-brane configuration is obtained after the circle compactification along $x^4$ of the type IIB 5-brane webs,  taking T-duality along this circle, and uplifting to the M-theory  on a M-circle along $x^{11}$. The curve lives on $\mathbb{R}^2\times T^2$ where $\mathbb{R}^2$ is along the 5-brane $(x^5,x^6)$-directions and $T^2$ is along the $(x^4, x^{11})$-directions in Table \ref{tab:5-braneWB}.   The coordinates on $\mathbb{R}^2\times T^2$ are, hence, given by two complex coordinates $t$ and $w$:
\begin{align}
    t = e^{-\frac{1}{R_M}(x^6\,+\,i\,x^{11})}, \qquad w = e^{-\frac{1}{R_5}(x^5\,+\,i\,x^{4})},
\end{align}
where $x^4= x^4+ 2\pi R_5$ and $x^{11}= x^{11}+ 2\pi R_M$. The radius of the compactification of the original 5d SCFT is  $\ell_s^2/R_5$.

\begin{figure}
    \centering
    \includegraphics[width=8.1cm]{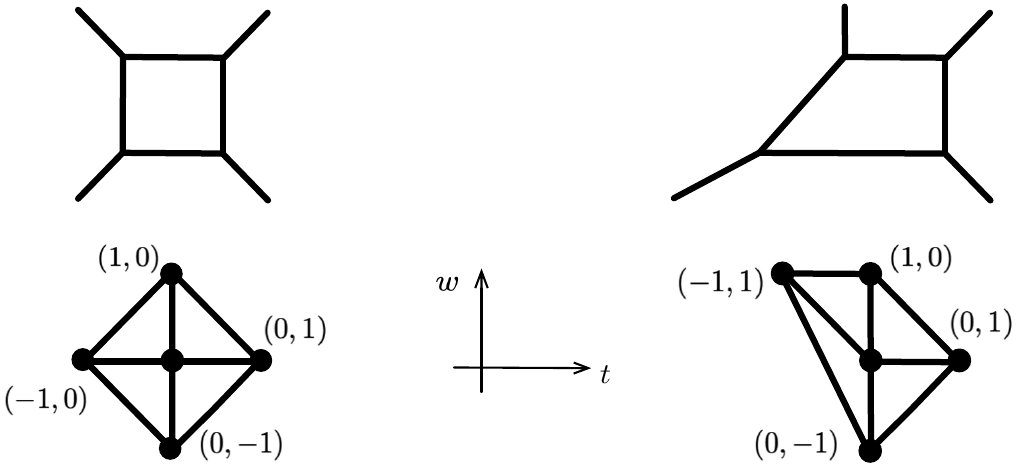}
    \caption{5-brane webs and (dual) toric diagrams for SU(2)$_0$ and SU(2)$_\pi$. Here, we choose the origin to represent the closed surface associated with the Coulomb branch moduli and the vertices $(m,n)$ are labelled accordingly.}
    \label{fig:toric-su2}
\end{figure}

The 5d SW curves can be obtained in a straightforward manner by considering the characteristic polynomial in a (dual) toric diagram of a given $(p,q)$ 5-brane web. As a Poincar\'e dual, the toric diagrams for the 5d SU(2)$_\theta$ theories are given as Figure~\ref{fig:toric-su2}.
The SW curve is obtained from the characteristic polynomial
\begin{align}\label{eq:classicalSW}
    H(t,w,c_{m,n})=\sum_{(m,n)\in {\rm vertices}} c_{m,n}\ t^m w^n =0\ ,
\end{align}
where $(m,n)$ are the positions of vertices or dots in dual toric diagram, where the origin can be chosen freely, and the coefficients $c_{m,n}$ are determined by boundary conditions up to three redundancies for $t,w$-axis shifts and an overall rescaling. As a representative example, consider the toric diagrams for SU(2)$_\theta$ given in Figure~\ref{fig:toric-su2}. Each diagram has five dots and one needs to fix the corresponding coefficients. Out of these five coefficients $c_{m,n}$, three can be set to unity due to the three redundancies from toric actions, while remaining two coefficients correspond to the physical parameters, the coupling $ \mathfrak{q} $ and the Coulomb branch parameter $u$, of the theory. The classical SW curves are then given as the following forms:  
\begin{align}\label{eq:classical-SW1}
    \begin{aligned}
        \mathrm{SU}(2)_0 ~ : &\qquad \mathfrak{q}\,t^{-1}+ t+ w+w^{-1}+u =0 \ , \\
        \mathrm{SU}(2)_\pi \,:&\qquad \mathfrak{q}\,t^{-1}w+ t+ w+w^{-1}+u =0 \ .
    \end{aligned}
\end{align}
It is worth noting that although SW curves are obtained from a specific 5-brane web corresponding to a particular range of the Coulomb branch parameter or a chosen Weyl chamber, the obtained SW curve itself covers all parameter ranges. This means that for computing SW curves, one can only use the vertices of the dual toric diagram. The resulting SW curve encodes all possible internal triangulations of the toric diagram and therefore captures non-perturbative configurations. 

\begin{figure}[t]
    \centering
    \includegraphics[width=13.1cm]{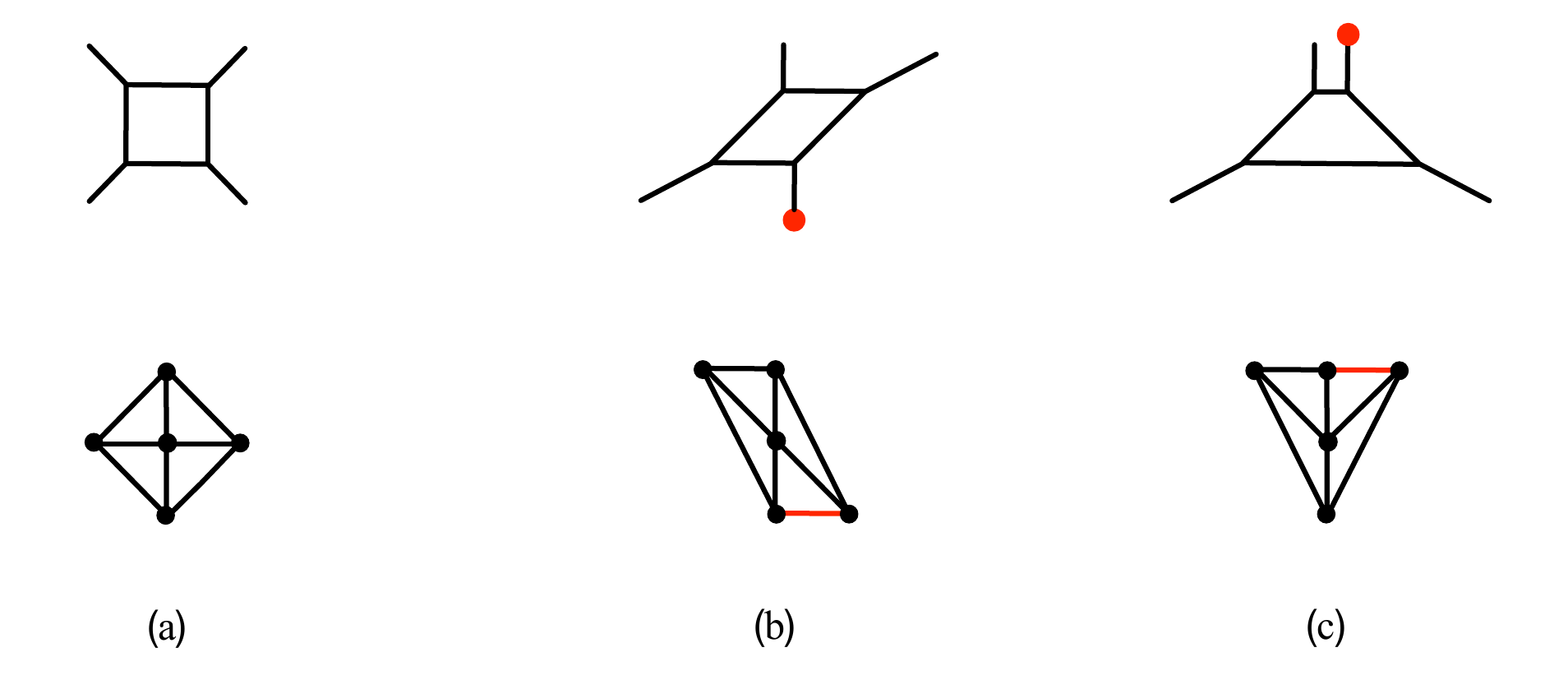}
    \caption{Toric diagrams for the SU(2)$_\theta$ theory ($\theta=0,2\pi$). (a) A toric diagram for SU(2)$_0$. (b) A toric diagram after an SL($2,\mathbb{Z}$) T-transformation is performed on the diagram (a), where the edge in red is dual of the 5-brane connected to the (0,1) 7-brane in red. (c) A toric diagram after 7-brane in red is pushed upward by a Hanany-Witten move, which leads to SU(2)$_{2\pi}$.}
    \label{fig-SU2-HW}
\end{figure}
SL($2,\mathbb{Z}$) transformations and Hanany-Witten (HW) transitions of a 5-brane web are realized as coordinate transformations of the SW curves. Consider, for instance, the 5-brane web for the SU(2)$_0$ theory above and let us take an SL($2,\mathbb{Z}$) $T$-transformation followed by a HW move, which leads to a 5-brane web for the SU(2)$_{2\pi}$ theory as depicted in Figure~\ref{fig-SU2-HW}. A $T$-transformation is done by $t\to tw^{-1} $ while $w$ is untouched, and the HW move of the $(0,-1)$ 7-brane upward is achieved by $w\to w(1+t)$. The resulting curves is 
\begin{align}\label{eq:classical-SW2}
    w^{-1}+w(1+t)(1+\mathfrak{q} \, t^{-1})+u =0\ ,
\end{align}
which precisely is that for the SU(2)$_{2\pi}$ theory, as expected. 

\begin{figure}[t]
    \centering
    \includegraphics[width=13.1cm]{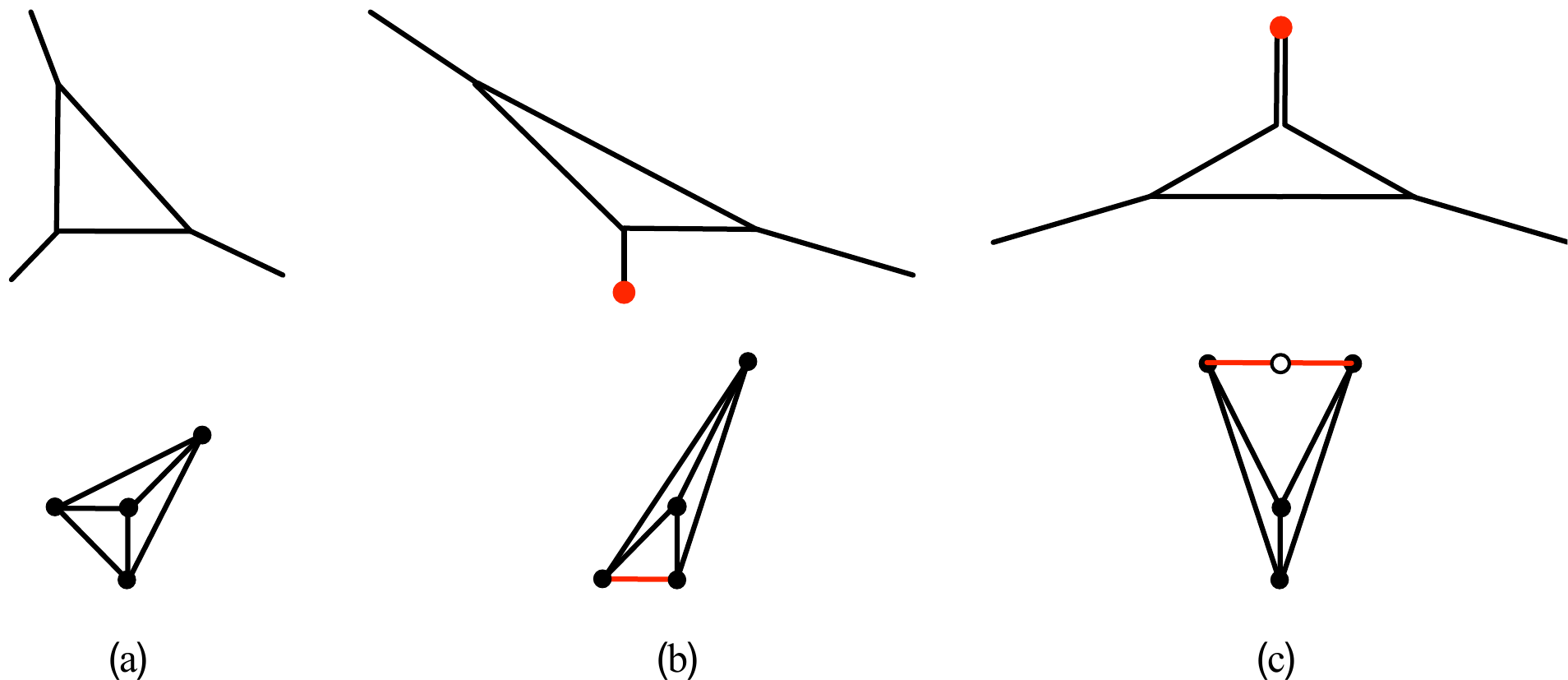}
    \caption{Various dual toric diagrams for local $\mathbb{P}^2$. (a) A standard toric diagram for the local $\mathbb{P}^2$. (b) An SL($2, \mathbb{Z}$) transformed diagram where a red edge is dual of 5-brane connected to the red 7-brane. (c) A toric diagram after a Hanany-Witten move creating a white dots.}
    \label{fig-localP2-toric}
\end{figure}

Another illuminating yet distinct example is the SW curve for the local $\mathbb{P}^2$ theory whose brane configurations are given in Figure~\ref{fig-localP2-toric}. First of all, the SW curve corresponding to Figure~\ref{fig-localP2-toric}(a) is simply 
\begin{align}
    w^{-1}+t^{-1}+ t w +u =0\ ,
\end{align}
as there is only one physical parameter $u$ for the local $\mathbb{P}^2$.
An SL($2,\mathbb{Z}$) transformation leading to Figure~\ref{fig-localP2-toric}(b) is done by $t\to t w^{-1}$ and the HW move leading to Figure~\ref{fig-localP2-toric}(c) is achieved by $w\to w(1+t^{-1})$. The resulting curve of these coordinate transformations is then given by
\begin{align}
    w^{-1}+ w^2(t+t^{-1})^2+u =0\ .
\end{align}
Notice that the $w^2$ term is given by a degenerate polynomial which describes multiple 5-branes bound to a single 7-brane and such configuration is distinguished by white dots in a dual toric diagram \cite{Benini:2009gi, Kim:2014nqa}. 
Such white dots appear as one considers SU$(2)_{\theta> 2 \pi} $, SU$(N)_{\kappa> N}$, and by increasing the number of hypermultiplets or by various Hanany-Witten moves.  

\subsection{Quantum curves and Hanany-Witten transitions}\label{sec:qcurve_HW}
The classical SW curve for a \emph{toric diagram} can be canonically quantized by promoting the variables $x=-\log w $, $y= -\log t$ to the operators $\hat{x}, \hat{y}$ that satisfy the commutation relation
\begin{align}
    \left[\hat{x},\hat{y}\right]=\epsilon_1 \, ,
\end{align}
where we will identify the Planck constant $ \epsilon_1 $ as the $ \Omega $-deformation parameter of the 5d supersymmetric QFTs on $ \mathbb{C}^2 \times S^1 $ in Sections~\ref{subsec:su2} and \ref{sec:example}. The operators $ \hat{x} $ and $ \hat{y} $ can be represented as
\begin{align}
    \hat{x}=x\ ,\qquad \hat{y}=-\epsilon_1 \partial_{x} \, ,
\end{align}
in an appropriate basis, and the exponentiated variables $X=e^{-\hat{x}}$ and $Y=e^{-\hat{y}}$ satisfy the relation
\begin{align}
    Y X=q_1 X Y \, ,
\end{align}
where $q_1=e^{-\epsilon_1}$. From here, we use $x,y$ to denote the operators $\hat{x},\hat{y}$ for simplicity.

Under the canonical quantization, the classical SW curve \eqref{eq:classicalSW} is promoted to an operator. For the toric cases, the quantized SW curve can be written as
\begin{align}\label{eq:quantumSW}
    \hat{H}=\sum_{(m,n)\in {\rm vertices}} c_{m,n}\ q_1^{\frac{1}{2}mn}X^m Y^n \, ,
\end{align}
where the $q_1$ factor is written from the Baker-Campbell-Hausdorff formula
\begin{align}
    e^{x}e^{y}=e^{x+y+\frac{1}{2}[x,y]+\frac{1}{12}[x,[x,y]]-\frac{1}{12}[y,[x,y]]+\cdots}.
\end{align}
Here ``$\cdots$" indicates terms involving higher commutators of $x$ and $y$. Note that for toric cases, the $q_1$ factors in \eqref{eq:quantumSW} can be adjusted via toric actions.

The effect of the HW moves on the classical curves has been discussed in the last subsection. These moves may transform a classical curve corresponding to a toric diagram into a \emph{generalized toric diagram} with white dots. To study their effect on the quantum curves, we consider the eigenvalue problem of the operator \eqref{eq:quantumSW} given as
\begin{align}\label{eq:qCurve}
    \hat{H}\Psi(x)=\sum_{(m,n)\in {\rm vertices}} c_{m,n}\ q_1^{\frac{1}{2}mn}X^m \Psi(x+n\epsilon_1 )=0\ ,
\end{align}
where $ \Psi(x) $ is the eigenfunction and the $Y$ operator acts on the eigenfunction as the shift operator. Then the $ \mathrm{SL}(2,\mathbb{Z}) $ transformations and HW moves described in the last subsection can be formally triggered by a sequence of transformations on $\Psi(x)$:
\begin{itemize}
    \item $ T $-transformation
        \begin{equation}\label{eq:Ttrans}\begin{split}
            \Psi(x ) \,&\rightarrow\, \widetilde{\Psi}(x)= e^{\frac{1}{2\epsilon_1}x^2}\Psi(x )\\
            X^m Y^n \,&\rightarrow\, q_1^{\frac{1}{2}n^2} X^{m+n} Y^n
        \end{split}\end{equation}
    \item $ S $-transformation
        \begin{equation}\label{eq:Strans}\begin{split}
            \Psi(x )  \,&\rightarrow\, \widetilde{\Psi}(x)= \int d x^{\prime} e^{-\frac{xx^{\prime}}{\epsilon_1}}\Psi(x^{\prime} ) \\
            X^m Y^n \,&\rightarrow\, q_1^{-mn} X^{-n} Y^m
        \end{split}\end{equation}
    \item Inverse $ S $-transformation
        \begin{equation}\begin{split}
            \Psi(x )  \,&\rightarrow\, \widetilde{\Psi}(x)= \int d x^{\prime} e^{\frac{xx^{\prime}}{\epsilon_1}}\Psi(x^{\prime} ) \\
            X^m Y^n \,&\rightarrow\, q_1^{-mn} X^{n} Y^{-m}
        \end{split}\end{equation}
    \item Similarity transformation
        \begin{equation}\label{eq:SimTrans}\begin{split}
            \Psi(x )  \,&\rightarrow\, \widetilde{\Psi}(x)= (-q_1^{1/2}X^{-1};q_1)_{\infty}^{-1}\Psi(x )\\
            X^m Y^n \,&\rightarrow\, (-q_1^{-n+1/2}X^{-1}; q_1)_n X^m Y^n\ ,
        \end{split}\end{equation}
        where $(\bullet;\bullet)_{n/\infty}$ is the q-Pochhammer symbol defined in Appendix \ref{app:special}.
\end{itemize}
In Sections~\ref{sec:codim2-hw-blowup} and \ref{sec:example}, we relate the wavefunction $ \Psi(x) $ to the codimension-2 defect partition function of the 5d theory.

For instance, in the $\mathrm{SU}(2)$ theories, the classical SW curves described in \eqref{eq:classical-SW1} are quantized according to \eqref{eq:quantumSW} as
\begin{align}
    \mathrm{SU}(2)_0 ~ : &\qquad X+X^{-1} +Y+\mathfrak{q}\,Y^{-1}= E \, , \label{eq:SU2-0-qSW} \\
    \mathrm{SU}(2)_\pi \,:&\qquad X+X^{-1} +Y+\mathfrak{q}\,q_1^{-1/2}XY^{-1}=E \, ,
\end{align}
where $ E $ is an eigenvalue. Let's first focus on the transformation of the quantum curve of $\mathrm{SU}(2)_0$ under HW moves. Applying $ T^{-1} $ to the quantum curve yields
\begin{equation}
    X+X^{-1} +q_1^{-1/2}X^{-1}Y+\mathfrak{q}\,q_1^{-1/2}XY^{-1}=E \, ,
\end{equation}
which corresponds to the transformation from Figure~\ref{fig-SU2-HW}(a) to (b). From Figure~\ref{fig-SU2-HW}(b) to (c), the classical transformation $w\to w(1+t)$ indicates a similarity transformation on the $ X $ variable. At the quantum curve level, we first apply the $ S $-transformation to make $ X $ as the shift operator, leading to
\begin{align}
    Y + (1 + q_1^{1/2} X^{-1}) Y^{-1} + \mathfrak{q}\, q_1^{1/2} X Y = E \, .
\end{align}
Next, we perform the similarity transformation \eqref{eq:SimTrans} as
\begin{align}
    (1 + q_1^{-1/2} X^{-1}) Y + Y^{-1} + \mathfrak{q} ( 1 + q_1^{1/2} X) Y = E \, .
\end{align}
This is a special type of birational transformation discussed in \cite{Arias-Tamargo:2024fjt, Ghim:2024asj}. We then apply the the inverse $ S $-transformation to make the $ Y $ operator back to the shift operator, resulting in
\begin{equation}
    X+X^{-1} +q_1^{1/2}XY+\mathfrak{q}\,(X+q_1^{-1/2}XY^{-1})=E \, .
\end{equation}
By further performing the inverse $T$-transformation, we obtain the standard quantum curve for $\mathrm{SU}(2)_{2\pi}$, which is also the quantum mirror curve for the local Hirzebruch surface $\mathbb{F}_2$:
\begin{equation}\label{eq:SU2-2-qSW}
    X+X^{-1} +Y+\mathfrak{q}\,(X+q_1^{-1}X^2Y^{-1})=E \, .
\end{equation}
Similarly, if we repeat the process above for the quantum curve of $\mathrm{SU}(2)_{\pi}$ theory, we obtain the quantum curve of $\mathrm{SU}(2)_{3\pi}$ theory as:
\begin{align}
    X+X^{-1}+Y+\mathfrak{q}\,\left(q_1^{1/2}XY+(q_1^{1/2}+q_1^{-1/2})X^2+q_1^{-3/2}X^3Y^{-1}\right)=E \, .
\end{align}

In general, one can repeat the process above for a quantum curve of $\mathrm{SU}(2)_{\theta}$ theory, to obtain the quantum curve for $\mathrm{SU}(2)_{\theta+2\pi}$. At the level of eigenfunction, we have
\begin{align}\label{eq:psitheta}
    \Psi_{\theta+2\pi}(x)=e^{-\frac{x^2}{2\epsilon_1}}\int d{s_1}d{s_2}e^{\frac{s_2x}{\epsilon_1}}e^{-\frac{s_1s_2}{\epsilon_1}}e^{-\frac{s_1^2}{2\epsilon_1}}(-q_1^{\frac{1}{2}}\sigma_2^{-1};q_1)_{\infty}^{-1}\Psi_{\theta}(s_1),
\end{align}
where we have used the notation $\sigma_i=e^{-s_i}$. In the next section, we will review that the eigenfunction $\Psi_{\theta}$ can be realized as the codimension-2 defect partition function of the 5d $\mathrm{SU}(2)_{\theta}$ theory, thus  \eqref{eq:psitheta} provides a recursive relation to the defect partition functions for theories with different discrete theta angles.

\section{Codimension-2 defects and Hanany-Witten transitions} \label{sec:codim2-hw-blowup}

In this section, we introduce codimension-2 defects in 5-dimensional theories from Higgsing, and discuss the blowup equation in the presence of the defects. We then consider 5d $ \mathrm{SU}(2) $ gauge theory as a working example. We compute the codimension-2 defect partition function of $ \mathrm{SU}(2) $ gauge theory on $ \Omega $-deformed $ \mathbb{R}^4 \times S^1 $ via Higgsing and by solving blowup equations. It turns out that blowup equations can capture various defect partition functions. Since the codimension-2 defect partition function is an eigenfunction of the quantum SW curves, we can identify each solution of blowup equations with a defect probing different theta angles related by Hanany-Witten transitions, and determine the quantization of the classical SW curve derived from the toric diagram.

\subsection{Codimension-2 defects} \label{subsec:codim2}

The supersymmetric field theories in five dimensions have half-BPS defect operators. These defects have codimension two or four to preserve supersymmetry. In type IIB 5-brane webs, the defects are realized by D3-branes perpendicular to the $ (p,q) $ 5-branes. The codimension-2 defect is a D3-brane stretched along the $ (x^0,x^1,x^2,x^7) $ directions, while the codimension-4 defect is a $\mathrm{D3}'$-brane placed along the $(x^0,x^7,x^8,x^9) $ directions. The resulting brane configuration is summarized in Table~\ref{table:braneDefect}. We will focus on the codimension-2 defects.

\begin{table}
    \centering
    \begin{tabular}{c|ccccc|cc|ccc}
        \hline
        & 0 & 1 & 2 & 3 & 4 & 5 & 6 & 7 & 8 & 9 \\ \hline
        NS5 $(0,1)$& $\times$&$\times$&$\times$&$\times$&$\times$&$\times$ &&&&\\
        D5 $(1,0)$& $\times$&$\times$&$\times$&$\times$&$\times$& &$\times$&&&\\
        $(p,q)$& $\times$&$\times$&$\times$&$\times$&$\times$&$\odot $&$\odot$ &&&\\
        7-brane& $\times$&$\times$&$\times$&$\times$&$\times$&&&$\times$&$\times$&$\times$\\
        $ \mathrm{D3} $-brane & $ \times $ & $ \times $ & $ \times $ & & & & & $ \times $ \\
        $ \mathrm{D3}' $-brane & $ \times $ & & & & & & & $ \times $ & $ \times $ & $ \times $ \\ \hline
    \end{tabular}
    \caption{Brane configuration for 5d theories with defects. $ \mathrm{D3} $- and $ \mathrm{D3}' $-branes correspond to codimension-2 and codimension-4 defects, respectively. The directions that each brane extends are marked with $\times$ or $\odot$.} \label{table:braneDefect}
\end{table}

\begin{figure}
    \centering
    \begin{subfigure}[b]{0.9\textwidth}
        \centering
        \includegraphics[align=c,scale=1]{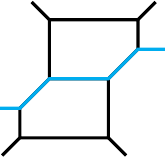}
        {\boldmath$ \to $}\hspace{1ex}
        \includegraphics[align=c,scale=1]{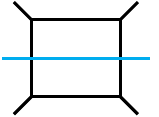}
        {\boldmath$ \to $}\hspace{1ex}
        \includegraphics[align=c,scale=1]{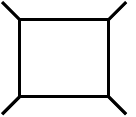}
        \caption{Mesonic Higgsing from SU($3)_0+2\mathbf{F}$ to SU($2)_0$.}
    \end{subfigure}
    \begin{subfigure}[b]{0.9\textwidth}
        \centering
        \includegraphics[align=c,scale=1]{figures/fig-SU3-0-2F.pdf}
        {\boldmath$ \to $}\hspace{1ex}
        \includegraphics[align=c,scale=1]{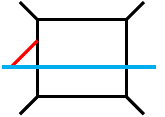}
        {\boldmath$ \to $}
        \includegraphics[align=c,scale=1]{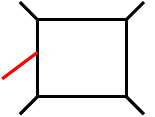}
        \caption{Defect Higgsing, where the red line represents a defect D3-brane.}
    \end{subfigure}
    \caption{Mesonic Higgs branch deformation from $ \mathrm{SU}(3)_0 + 2\mathbf{F} $ theory to $ \mathrm{SU}(2)_0 $} \label{fig:SU2-0-Higgs}
\end{figure}

A useful method to realizing such a defect D3-brane is to generalize the conventional Higgsing procedure on 5-brane webs, which can be achieved by giving constant vacuum expectation values (VEVs) to scalar fields in the theory. On the other hand, one can consider position-dependent VEVs that not only reduces the gauge symmetry but also introduces a codimension-2 defect \cite{Gaiotto:2012xa}. To illustrate this, we examine the 5d $ \mathrm{SU}(2)_0 $ gauge theory and its defect discussed in the previous section. This theory arises through a Higgsing of the 5d $ \mathrm{SU}(3)_0 $ gauge theory coupled to two fundamental hypermultiplets. Before the Higgsing, the UV theory have mesonic operators built from the fundamental and anti-fundamental chiral scalar fields in the hypermultiplets. Mesonic Higgsing is achieved by giving non-zero constant VEV to the mesonic operator, resulting in the $ \mathrm{SU}(2)_0 $ theory through an RG-flow at low energy. Moreover, one can also assign a position-dependent VEV to the mesonic operator which leads to the codimension-2 defect in the IR $ \mathrm{SU}(2)_0 $ theory. The 5-brane web pictures of these Higgsings are given in Figure~\ref{fig:SU2-0-Higgs}. The 5-brane configuration of $ \mathrm{SU}(3)_0+2\mathbf{F} $ theory is given in the leftmost diagrams of Figure~\ref{fig:SU2-0-Higgs}, and the mesonic Higgs branch opens up when the positions of two flavor D5-branes are aligned to the internal D5-brane. Higgsing is realized by moving the D5-brane to the perpendicular direction of the other 5-branes. The Higgsing with a position-dependent VEV turns out to introduce an additional D3-brane ending on a NS5-brane and the moved D5-brane. This D3-brane will become the codimension-2 defect in the 5d theory at low energy.

There is a $ \mathrm{U}(1) $ symmetry acting on the defect D3-brane. The 3d field theories with a $ \mathrm{U}(1) $ global symmetry have a natural $ \mathrm{SL}(2,\mathbb{Z}) $ action \cite{Witten:2003ya} generated by $ T $- and $ S $-transformations. The $ T $-transformation shifts the Chern-Simons level of the background $ \mathrm{U}(1) $ gauge field $ A $. The $ S $-transformation amounts to gauging the $ \mathrm{U}(1) $ symmetry which regards $ A $ as a dynamical gauge field, and adds mixed Chern-Simons term $ B dA $ for the background gauge field $ B $ of a new $ \mathrm{U}(1) $ global symmetry. This new $ \mathrm{U}(1) $ symmetry acts on monopole operators, and its conserved quantity is the magnetic flux. This $ \mathrm{SL}(2,\mathbb{Z}) $ action on a 3d theory uplifts to the $ \mathrm{SL}(2,\mathbb{Z}) $ transformation in $ (p,q) $ 5-brane construction of a 5d-3d coupled system, defining a more general type of codimension-2 defects coupled to 5d SCFTs.

\begin{figure}
    \centering
    \includegraphics[align=c,scale=1]{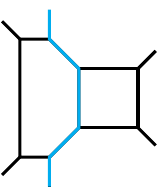}
    {\boldmath $ \to $}
    \includegraphics[align=c,scale=1]{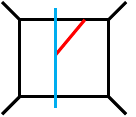}
    {\boldmath $ \to $}
    \includegraphics[align=c,scale=1]{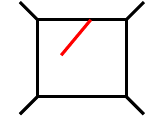}
    \caption{Baryonic Higgs branch deformation from $ \mathrm{SU}(2)_\pi \times \mathrm{SU}(2)_\pi $ theory to $ \mathrm{SU}(2)_0 $. The red line represents the defect D3-brane.} \label{fig:SU2-0-Higgs2}
\end{figure}

For instance, an $ S $-transformation in a 5-brane web rotates the diagram by 90 degrees. This is nothing but the S-duality of the type IIB theory that exchanges D5-branes and NS5-branes. The S-dual of the 5d $ \mathrm{SU}(3)_0+2\mathbf{F} $ theory considered in Figure~\ref{fig:SU2-0-Higgs} has low energy gauge theory description given by the theory with a gauge group $ \mathrm{SU}(2)_\pi \times \mathrm{SU}(2)_\pi $ coupled to a bifundamental hypermultiplet. This theory admits a baryonic Higgs branch deformation to the $ \mathrm{SU}(2)_0 $ theory by giving nonzero VEV to the baryonic operator built from the chiral scalar fields in the bifundamental hypermultiplet. The codimension-2 defect is introduced by adding an additional D3-brane orthogonal to the 5-brane system, which amounts to giving a position dependent VEV to the baryonic operator. This Higgsing procedure is given in Figure~\ref{fig:SU2-0-Higgs2}. The codimension-2 defect in Figure~\ref{fig:SU2-0-Higgs2} is related to the defect in Figure~\ref{fig:SU2-0-Higgs} via the S-transformation of the 3d theory and its uplift to the 5d-3d coupled system \cite{Gaiotto:2014ina}. See \cite{Jeong:2024onv} for a related discussion in the perspective of the bispectral duality.

\begin{figure}
    \centering
    \includegraphics[align=c,scale=1]{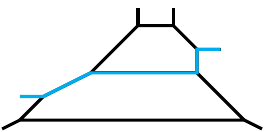}
    {\boldmath$ \to $}
    \includegraphics[align=c,scale=1]{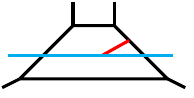}
    {\boldmath$ \to $}
    \includegraphics[align=c,scale=1]{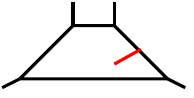}
    \caption{Codimension-2 defect in $ \mathrm{SU}(2)_{2\pi} $ theory obtained from mesonic Higgs branch deformation of $ \mathrm{SU}(3)_2 + 2\mathbf{F} $ theory. The red line represents the defect D3-brane.} \label{fig:SU2-2-Higgs}
\end{figure}

Another type of codimension-2 defects is related to HW transitions of 5-brane web. Without a defect, the HW transition does not change the 5d physics. Various physical observables such as partition functions and superconformal indices are unchanged, up to some simple extra factors. However, codimension-2 defects can distinguish two brane systems related by HW moves. One example is the $ \mathrm{SU}(2)_{2\pi} $ theory, which is equivalent to the $ \mathrm{SU}(2)_0 $ theory up to a HW transition when there is no defect, as discussed in Section~\ref{sec:5dSCFTHW}. A codimension-2 defect of the $ \mathrm{SU}(2)_{2\pi} $ theory can be introduced via the mesonic Higgs branch deformation of the $ \mathrm{SU}(3)_2 + 2\mathbf{F} $ theory as shown in Figure~\ref{fig:SU2-2-Higgs}, similar to the defect in the $ \mathrm{SU}(2)_0 $ theory. As we will see in Section~\ref{subsec:su2defect}, the partition functions of two theories in the presence of codimension-2 defects are different, and they serve as eigenfunctions of the quantum SW curves in \eqref{eq:SU2-0-qSW} and \eqref{eq:SU2-2-qSW}.

\subsection{Blowup equations review}

In this paper, we consider the BPS spectrum of 5d/6d supersymmetric field theories in the presence of codimension-2 defect operators. The BPS spectrum without a defect is counted by the partition function on $ \Omega $-deformed $ \mathbb{C}^2 \times S^1 $, which is the Witten index defined as \cite{Nekrasov:2002qd}
\begin{align}\label{eq:Z}
    Z(\phi, m, \epsilon_1, \epsilon_2) = \operatorname{Tr} \left[ (-1)^F e^{-\beta\{Q,Q^\dagger\}} e^{-\epsilon_1(J_1+J_R)} e^{-\epsilon_2(J_2 + J_R)} e^{-\phi \cdot \Pi} e^{-m\cdot H}  \right] \, ,
\end{align}
where $ F $ is the fermion number, $ Q, Q^\dagger $ are the supercharge and its conjugates; $ J_1 $, $ J_2 $ are the Cartan generators of the $ \mathrm{SO}(4)$ Lorentz rotation, $ J_R $ is the Cartan generator of the $ \mathrm{SU}(2)_R $ R-symmetry; $ \Pi $ and $ H $ are respectively gauge and flavor charges. The $ Q $ and $ Q^\dagger $ commute with $ J_1+J_R $ and $ J_2+J_R $. We denote by $ \beta $ the radius of $ S^1 $, by $ \epsilon_{1,2} $  the $ \Omega $-deformation parameters, and by $ \phi $ and $ m $, the chemical potentials of the gauge and flavor symmetries. The Witten index counts the BPS states annihilated by the supercharges $ Q $ and $ Q^\dagger $, so it is independent of $ \beta $.

The partition function is factorized into the product of the regularization factor $e^{\mathcal{E}}$ and the index factor $ Z_{\mathrm{GV}} $ as
\begin{align}\label{eq:index}
    Z(\phi,m,\epsilon_1,\epsilon_2) &= e^{\mathcal{E}(\phi,m,\epsilon_{1,2})} Z_{\mathrm{GV}}(\phi,m,\epsilon_1,\epsilon_2) \, .
\end{align}
Here, the index part $ Z_{\mathrm{GV}} $ captures the spectrum of BPS states and is given by
\begin{align}\label{eq:ZGV}
    Z_{\mathrm{GV}} &= \PE\left[ \sum_{j_l,j_r,d} (-1)^{2(j_l+j_r)} N_{j_l,j_r}^{d} \frac{\chi_{j_l}(\epsilon_-) \chi_{j_r}(\epsilon_+)}{2\sinh(\epsilon_1/2) \cdot 2\sinh(\epsilon_2/2)} e^{-d \cdot t} \right]  ,
\end{align}
where $ \PE[\cdot] $ is the plethystic exponential defined in Appendix~\ref{app:special}, $ t $ and $ d $ collectively denote the chemical potentials $ (\phi,m) $ and the associated charges respectively, $ \epsilon_\pm = \frac{\epsilon_1 \pm \epsilon_2}{2} $, and $ \chi_j $ is the $ \mathrm{SU}(2) $ character of spin $ j $-representation. The non-negative integer $ N_{j_l,j_r}^d $ is the refined Gopakumar-Vafa invariants \cite{Gopakumar:1998ii,Gopakumar:1998jq} counting the degeneracies of a BPS state on $ \Omega $-background with charge $ d $ and spin $ (j_l,j_r) = (\frac{J_1-J_2}{2},\frac{J_1+J_2}{2}) $ for the $ \mathrm{SO}(4) $ Lorentz rotation. 

The prefactor $ \mathcal{E} $ in \eqref{eq:index} is the effective prepotential \cite{Kim:2020hhh} arising from the classical contributions and regularizations of infinite products in the 1-loop part. On the $ \Omega $-background, one finds
\begin{align}
    \mathcal{E} = \frac{1}{\epsilon_1 \epsilon_2} \left( \mathcal{F} + \frac{\epsilon_1^2 + \epsilon_2^2}{48} C_i^G \phi_i + \frac{\epsilon_+^2}{2} C_i^R \phi_i \right) \, ,
\end{align}
where $ \mathcal{F} $ is the cubic prepotential \eqref{eq:prepotential} on the Coulomb branch. The remaining two terms are contributions from mixed Chern-Simons terms. Here, $ C_i^G $ and $ C_i^R $ are the mixed gauge/gravitational and gauge/$ \mathrm{SU}(2)_R $ CS coefficients given by \cite{Bonetti:2013ela, Grimm:2015zea, BenettiGenolini:2019zth}
\begin{align}
    C_i^G = -\partial_i \bigg( \sum_{e\in\mathbf{R}} |e \cdot \phi| - \sum_f |w \cdot \phi + m_f| \bigg) \, , \quad
    C_i^R = \frac{1}{2} \partial_i \sum_{e\in\mathbf{R}} |e \cdot \phi| \, .
\end{align}
For a more detailed discussion about the effective prepotential, as well as numerous intriguing examples, see \cite{Kim:2020hhh}. 

The blowup equation provides a powerful method for computing the partition function~$ Z $. Consider a blowup geometry $ \hat{\mathbb{C}}^2 $ obtained by blowing up the origin in the $ \mathbb{C}^2 $ with a 2-sphere $ \mathbb{P}^1 $. The partition function $ \hat{Z} $ defined on $ \hat{\mathbb{C}}^2 $ is factorized into two local partition function contributions around the north and south poles of the $ \mathbb{P}^1 $ under the supersymmetric localization. This was initially observed in 4d and 5d $ \mathrm{SU}(N) $ super Yang-Mills theories by G\"ottsche, Nakajima and Yoshioka \cite{Nakajima:2003pg, Nakajima:2005fg, Gottsche:2006bm} and later generalized to generic 5d/6d field theories \cite{Huang:2017mis, Kim:2019uqw, Gu:2018gmy, Gu:2019dan, Gu:2019pqj, Gu:2020fem, Kim:2020hhh, Kim:2023glm}. This relation is called the blowup equation given by
\begin{align}\label{eq:blowup}
    \Lambda(m, \epsilon_1, \epsilon_2) \hat{Z}(\phi,m, \epsilon_1, \epsilon_2)
    = \sum_{n} (-1)^{|n|} \hat{Z}^{(N)}(n, B) \hat{Z}^{(S)}(n, B) \, .
\end{align}
Here, two magnetic fluxes are denoted by $n$ and $B$, where $ n=(n_i) $ is the magnetic fluxes on $ \mathbb{P}^1 $ for the gauge symmetries, with $ |n|=\sum n_i $, while $ B=(B_j) $ is the background magnetic fluxes for the global symmetries. Two partition functions $ \hat{Z}^{(N)} $ and $ \hat{Z}^{(S)} $ are localized partition functions on the north pole and the south pole of the $ \mathbb{P}^1 $, respectively. They have the same form as $ \hat{Z} $ up to shifting the chemical potentials as
\begin{align}
    \begin{aligned}\label{eq:ZNZS}
        \hat{Z}^{(N)}(n,B) &= \hat{Z}(\phi_i + n_i \epsilon_1, m_j + B_j \epsilon_1, \epsilon_1, \epsilon_2-\epsilon_1) \, , \\
        \hat{Z}^{(S)}(n,B) &= \hat{Z}(\phi_i + n_i \epsilon_2, m_j + B_j \epsilon_2, \epsilon_1-\epsilon_2, \epsilon_2) \, .
    \end{aligned}
\end{align}
The prefactor $ \Lambda(m,\epsilon_1,\epsilon_2) $ does not depend on the dynamical parameter $ \phi $. The partition function $ \hat{Z} $ on $ \hat{\mathbb{C}}^2 $ is related to the original partition function $ Z $ on $ \mathbb{C}^2 $ by replacing $ (-1)^F \to (-1)^{2J_R} $ in \eqref{eq:Z}. This replacement is equivalent to the redefinition $ \epsilon_1 \to \epsilon_1+2\pi i $ in the index part $ Z_{\mathrm{GV}} $ of the partition function:
\begin{align}
    \hat{Z}(\phi,m,\epsilon_1, \epsilon_2) = e^{\mathcal{E}(\phi,m,\epsilon_1,\epsilon_2)} Z_{\mathrm{GV}}(\phi,m,\epsilon_1+2\pi i, \epsilon_2) \, .
\end{align}

The magnetic fluxes $ (n, B) $ on $ \mathbb{P}^1 $ must be correctly quantized. The proper quantization condition is \cite{Huang:2017mis}
\begin{align}\label{eq:blowup-flux}
    (n, B) \cdot e \text{ is integral/half-integral } \Leftrightarrow\ 2(j_l+j_r) \text{ is odd/even,}
\end{align}
for all BPS particles of charge $ e $ and spin $ (j_l,j_r) $. The blowup equation \eqref{eq:blowup} is true for special sets of magnetic fluxes, called the consistent magnetic fluxes, among the properly quantized fluxes. After identifying the effective prepotential and consistent magnetic fluxes, one can compute the BPS spectrum $ Z_{\mathrm{GV}} $ using the blowup equations. See \cite{Kim:2020hhh} for the details.

The blowup formalism can be extended in the presence of Wilson loop operators and codimension-4 defects in 5d field theories. The half-BPS Wilson loop in a 5d $ \mathcal{N}=1 $ gauge theory is a gauge invariant half-BPS operator defined by \cite{Young:2011aa, Assel:2012nf},
\begin{align}
    W_{\mathbf{r}}[C] = \operatorname{Tr}_{\mathbf{r}} \mathcal{P} \exp\bigg(\int_C (i A_\mu \dot{x}^\mu + |\dot{x}| \phi) ds \bigg) \, ,
\end{align}
where $ \mathbf{r} $ denotes its representation under the gauge group, $ \mathcal{P} $ is the path-ordering, $ A_\mu $ is the gauge field, $ \phi $ is the real scalar in the vector multiplet, $ x^\mu(s) $ is the worldline $ C $ of the loop operator, and $ \dot{x}^\mu = dx^\mu/ds $. It is a codimension-4 defect in 5d, which is located at the origin of $ \mathbb{C}^2 $ but stretched along $ S^1 $. More generally, the Wilson loop operators can be understood as an insertion of a heavy BPS particle at the origin of $ \mathbb{C}^2 $. In the geometric construction of 5d field theories from M-theory compactification on local Calabi-Yau threefolds, such heavy particles are realized by M2-branes wrapping holomorphic non-compact 2-cycles in the Calabi-Yau manifold \cite{Kim:2021gyj}. This perspective enables us to define loop operators in more general 5d field theories, including non-Lagrangian theories that do not admit a gauge theory description.

One can consider the partition function in the presence of a loop operator. The vacuum expectation value (VEV) of $ W_{\mathbf{r}} $ is given by 
\begin{align}
    \langle W_{\mathbf{r}} \rangle (\phi, m, \epsilon_1, \epsilon_2) = \frac{Z_{W_{\mathbf{r}}}(\phi,m,\epsilon_1,\epsilon_2)}{Z(\phi,m,\epsilon_1,\epsilon_2)} \, ,
\end{align}
and it counts the bound states of the loop operator. Here, $ Z_{W_{\mathbf{r}}} $ is the partition function with an insertion of $ W_{\mathbf{r}} $ on $ \Omega $-background. Since the 1d defect states arising from the heavy particles cannot move from the origin of $ \mathbb{C}^2 $, Wilson loop VEV has an index form as \eqref{eq:ZGV} without the momentum factor on the transverse $ \mathbb{C}^2 $ in the denominator \cite{Kim:2021gyj}:
\begin{align}\label{eq:WGV}
    \langle W_{\mathbf{r}} \rangle = \sum (-1)^{2(j_l+j_r)} \tilde{N}_{j_l,j_r}^d \chi_{j_l}(\epsilon_-) \chi_{j_r}(\epsilon_+) e^{-d \cdot t} \, .
\end{align}
Here, a non-negative integer $ \tilde{N}_{j_l,j_r}^d $ counts the degeneracy of the 1d BPS states with charge $ d $ for the gauge and global symmetries of the theory and with spin $ (j_l,j_r) $ for $ SO(4) $ Lorentz rotation.

\begin{figure}
    \centering
    \includegraphics[scale=1]{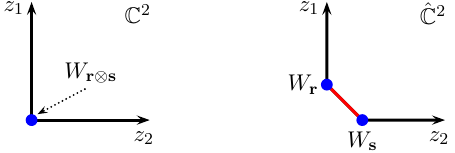}
    \caption{Wilson loop operator $ W_{\mathbf{r} \otimes \mathbf{s}} $ is inserted at the origin of $ \mathbb{C}^2 $. After replacing the origin to $ \mathbb{P}^1 $ (red line), $ W_{\mathbf{r} \otimes \mathbf{s}} $ is split into two Wilson loops $ W_{\mathbf{r}} $ and $ W_{\mathbf{s}} $ at north pole and south pole of $ \mathbb{P}^1 $.} \label{fig:blowupW}
\end{figure}

On the blowup geometry $ \hat{\mathbb{C}}^2 $, the bare partition function $ \hat{Z} $ has two contributions, coming from the north/south pole of $ \mathbb{P}^1 $. Similarly, the partition function in the presence of the Wilson loop operator receives two such contributions and is factorized into two Wilson loop partition functions localized at the two poles of the $ \mathbb{P}^1 $. If there is a Wilson loop with the representation $ \mathbf{r} $ at the north pole and $ \mathbf{s} $ at the south pole of the $ \mathbb{P}^1 $, then they will merge together and form a loop operator in the product representation $ \mathbf{r} \otimes \mathbf{s} $ after the blow-down transition $ \hat{\mathbb{C}}^2 \to \mathbb{C}^2 $. This is illustrated in Figure~\ref{fig:blowupW}. Therefore, the blowup equation in the presence of the loop operator can be written as
\begin{align}\label{eq:blowupWilson}
    \Lambda(m,\epsilon_1,\epsilon_2) \hat{Z}_{W_{\mathbf{r} \otimes \mathbf{s}}}(\phi,m,\epsilon_1,\epsilon_2) = \sum_n (-1)^{|n|} \hat{Z}^{(N)}_{W_{\mathbf{r}}}(n,B) \hat{Z}^{(S)}_{W_{\mathbf{s}}}(n,B) \, ,
\end{align}
where $ \hat{Z}^{(N)}_{W_{\mathbf{r}}} $ and $ \hat{Z}^{(S)}_{\mathbf{s}} $ respectively denote the partition function at the north pole and that at the south pole in the presence of the loop operators. They are related to $ \hat{Z}_{W_{\mathbf{r}}} $ and $ \hat{Z}_{W_{\mathbf{s}}} $ by shifting the chemical potentials in the same way as \eqref{eq:ZNZS}. Similarly, the hatted partition functions are related to the unhatted partition function by shifting the parameter $ \epsilon_1 \to \epsilon_1 + 2\pi i $. Codimension-4 defect partition functions for many 5d SCFTs and KK theories are computed using this blowup equation in \cite{Kim:2021gyj}.

\subsection{Blowup equations for codimension-2 defects}

\begin{figure}
    \centering
    \includegraphics[scale=1]{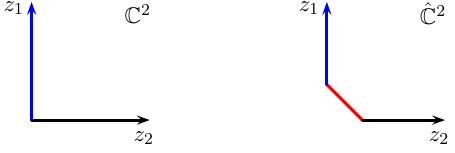}
    \caption{(Left) Codimension-2 defect (blue line) is placed at the $ z_2=0 $ plane of $ \mathbb{C}^2 $. (Right) In the blowup geometry $ \hat{\mathbb{C}}^2 $, the defect is now placed at the $ z_2=0 $ plane and attached to the north pole of $ \mathbb{P}^1 $.} \label{fig:blowupD}
\end{figure}

We now generalize the blowup equation in the presence of codimension-2 defects.\footnote{See \cite{Nekrasov:2020qcq,Jeong:2020uxz} for 4d cases.} Suppose a codimension-2 defect is placed in $ z_2=0 $ plane inside $ \mathbb{C}^2 $ and stretched along the time circle $ S^1 $. In the blowup geometry $ \hat{\mathbb{C}}^2 $, it is natural to consider the defect located at $ z_2=0 $ plane and do not contact with the south pole of the $ \mathbb{P}^1 $, as in Figure~\ref{fig:blowupD}. Blowdown of the 2-sphere $ \mathbb{P}^1 $ recovers the original defect at $ z_2=0 $ plane.
Let us denote a partition function in the presence of such a codimension-2 defect in 5d field theories as $ \Psi = \Psi(\phi,m,x,\epsilon_1,\epsilon_2) $, where $ x $ is the defect parameter. This partition function counts BPS bound states to the surface defect operator. It admits an index form 
\begin{align}\label{eq:Psi-index}
    &\frac{\Psi(\phi,m,x,\epsilon_{1,2})}{Z(\phi,m,\epsilon_{1,2})} = e^{\mathcal{E}_1}
    \PE\!\Bigg[ \!\sum_{d_1,d_2,s_l,s_r}\!\! (-1)^{2(s_l+s_r)} D_{s_l,s_r}^{d_1, d_2} \frac{e^{-2(s_l \epsilon_- + s_r \epsilon_+)}}{1-q_1} e^{-(d_1\cdot x + d_2 \cdot t)}\! \Bigg], 
\end{align}
similar to \eqref{eq:ZGV} and \eqref{eq:WGV}. Here, $ \mathcal{E}_1=\mathcal{E}_1(\phi,m,x,\epsilon_{1},\epsilon_{2}) $ is the prepotential contribution from the defect that is the regularization factor of the 1-loop part of the defect and $ D_{s_l,s_r}^{d_1,d_2} $ is the refinement of Ooguri-Vafa invariants \cite{Ooguri:1999bv,Gukov:2004hz,Aganagic:2011sg,Aganagic:2012ne,Aganagic:2012hs,Cheng:2021nex} which is conjectured to be a non-negative integer. Because the defect resides on the $\epsilon_1$-plane defined by $ z_2=0 $, the single particle index inside the plethystic exponential has a pole at $ \epsilon_1=0 $ and $ \mathrm{SU}(2)_l \times \mathrm{SU}(2)_r $ symmetry is broken.

We now propose the blowup equation in the presence of codimension-2 defect as
\begin{align}\label{eq:blowupDefect}
    \Lambda \cdot \hat{\Psi}(\phi,m,x,\epsilon_1,\epsilon_2) = \sum_n \hat{\Psi}^{(N)}(n, B) \hat{Z}^{(S)}(n,B) \, ,
\end{align}
where $ \hat{\Psi} $ is the defect partition function on $ \hat{\mathbb{C}}^2 $ that is the same as the unhatted partition function $ \Psi $ but with $ \epsilon_1 \to \epsilon_1 + 2\pi i $, $ \hat{\Psi}^{(N)} $ is the local partition at the north pole in the presence of defect and $ \hat{Z}^{(S)} $ is the bare partition function localized at the south pole. The north pole defect partition function $ \hat{\Psi}^{(N)} $ is related to $ \hat{\Psi} $ by the shifting chemical potentials as \eqref{eq:ZNZS}:
\begin{align}
    \hat{\Psi}^{(N)}(n,B) = \hat{\Psi}(\phi_i + n_i \epsilon_1, m_j + B_j \epsilon_1, x + B_x \epsilon_1, \epsilon_1, \epsilon_2-\epsilon_1) \, ,
\end{align}
where $ B_x $ is the background magnetic flux for the $ \mathrm{U}(1) $ global symmetry associated to the defect. The magnetic fluxes on the blowup $ \mathbb{P}^1 $ in the presence of a codimension-2 defect satisfy a similar quantization condition as \eqref{eq:blowup-flux}. To construct a consistent blowup equation, the magnetic fluxes $ (n, B) = (n_i, B_j, B_x) $ is quantized as \cite{Kashani-Poor:2016edc}
\begin{align}\label{eq:defect-blowup-flux}
    (n, B) \cdot e = n_i e_{\phi_i} + B_j e_{m_j} + B_x e_{x} \text{ is integral/half-integral} \, \Leftrightarrow \, s_l + s_r \text{ is even/odd,}
\end{align}
for the BPS states bound to the codimension-2 defect with charge $ e=(e_{\phi}, e_m, e_x) $ and spin $ (s_l,s_r) $. 

We remark on the $ \Lambda $ factor in the blowup equation \eqref{eq:blowupDefect}. Without a defect, $ \Lambda=\Lambda(m,\epsilon_1,\epsilon_2) $ for the consistent magnetic fluxes is a constant which does not depend on the dynamical parameter $ \phi $. However, the $ \Lambda $ factor must be generalized to an operator involving the shift operator $ \epsilon_1 \partial_x $ in the presence of codimension-2 defects, due to the various coordinate transformations explored in Section~\ref{sec:qcurve_HW}. For instance, suppose that a codimension-2 defect partition function $ \Psi $ satisfies a blowup equation with a $ \Lambda $ factor that depends on the defect parameter $ x $. For simplicity, let us consider $ \Lambda(m,x,\epsilon_1,\epsilon_2) = \lambda(m,\epsilon_1,\epsilon_2) e^{-nx} $, where a more general $ \Lambda $ can be expressed as a sum of such terms. Then the LHS and RHS of the blowup equation for the codimension-2 defect $ \widetilde{\Psi} $ obtained by the S-transformation \eqref{eq:Strans} can be written as follows:
\begin{align}
    \int \lambda(m,\epsilon_1,\epsilon_2) e^{-nx'} \Psi(x') e^{-xx'/\epsilon_1} dx'
    &= \lambda(m,\epsilon_1,\epsilon_2) Y^n \widetilde{\Psi}(x) \, , \\
    \int \sum_n \Psi^{(N)}(x'+B \epsilon_1) Z^{(S)} e^{-xx'/\epsilon_1} dx'
    &= e^{x B} \sum_n \widetilde{\Psi}^{(N)}(x) Z^{(S)} \, ,
\end{align}
where $ Y=e^{\epsilon_1 \partial_x} $ is an operator $ Y \widetilde{\Psi}(x) = \widetilde{\Psi}(x+\epsilon_1) $. The new blowup equation has $ e^{-xB} \lambda Y^n $ as a $ \Lambda $ factor in this case. Thus, even if we are able to find a blowup equation with a constant $ \Lambda $ factor, the S-transformed defect may satisfy a blowup equation with an operator
\begin{align}
    \Lambda = \Lambda(m, x, \epsilon_1 \partial_x, \epsilon_1, \epsilon_2) \, .
\end{align}
We will see the appearance of the blowup equations with such operator $ \Lambda $ factors in the next section.

For a given blowup equation in the presence of a codimension-2 defect with a constant, non-operator $ \Lambda $ factor, one may attempt to solve the equation to find the partition function $ \Psi $. Unlike the blowup equations without a defect or with a codimension-4 defect, a single blowup equation \eqref{eq:blowupDefect} is not enough to determine the codimension-2 defect partition function $ \Psi $ even if we assume the index form \eqref{eq:Psi-index}, due to the lack of $ \mathrm{SU}(2)_l \times \mathrm{SU}(2)_r $ symmetry. Nevertheless, it is possible to find $ \Psi $ using blowup equations with two different sets of consistent magnetic fluxes. The blowup equation \eqref{eq:blowupDefect} is a functional equation involving the ordinary partition function $ Z $, the defect partition function $ \Psi $, and the defect partition function $ \Psi^{(N)} $ at the north pole of $ \mathbb{P}^1 $. We first compute the partition function $ Z $ without defect using the ordinary blowup equations \eqref{eq:blowup}. If there are two different sets of consistent magnetic fluxes, then we can write two blowup equations for two unknown functions $ \Psi $ and $ \Psi^{(N)} $. By expanding the blowup equations in terms of K\"ahler parameters including the defect parameter $ x $, and solving them order by order, we can determine the unknown function $ \Psi $.

\subsection{Defects in 5d \texorpdfstring{$ \mathrm{SU}(2) $}{SU(2)} gauge theories} \label{subsec:su2}

In this subsection, we consider codimension-2 defects in the 5d $ \mathrm{SU}(2) $ gauge theory with discrete theta angle $ \theta $. We determine their partition functions on $ \Omega $-deformed $ \mathbb{R}^4 \times S^1 $ using the Higgsing and blowup equations. In particular, we discuss how the solutions of blowup equations are related to the defects for different theta angles. We then explore the quantum curves and difference equations that annihilate the defect partition functions.

\subsubsection{Defect partition functions from Higgsing} \label{subsec:su2defect}

The effective prepotential of the 5d $ \mathrm{SU}(2) $ gauge theories on the $ \Omega $-background is given by
\begin{align}\label{eq:SU2-E0}
    \mathcal{E}_0 = \frac{1}{\epsilon_1 \epsilon_2} \left( \frac{8}{6} \phi^3 + m_0\phi^2 - \frac{\epsilon_1^2 + \epsilon_2^2}{12} \phi + \epsilon_+^2 \phi \right) \, ,
\end{align}
where $ m_0=1/g^2 $ is the inverse gauge coupling squared. The partition function on $ \mathbb{R}^4 \times S^1 $ is expressed as
\begin{align}
    Z(\phi, m_0, \epsilon_1,\epsilon_2) = e^{\mathcal{E}_0} \cdot Z_{\mathrm{pert}}(\phi,\epsilon_1,\epsilon_2) \cdot \left( 1 + \sum_{k=1}^\infty \mathfrak{q}^k Z_k(\phi, \epsilon_1,\epsilon_2) \right) \, ,
\end{align}
where
\begin{align}\label{eq:SU2-pert}
    Z_{\mathrm{pert}}(\phi,\epsilon_{1,2}) = \PE \left[ -\frac{1+q_1 q_2}{(1-q_1)(1-q_2)} e^{-2\phi} \right]
\end{align}
is the perturbative partition function coming from the $ \mathrm{SU}(2) $ vector multiplet. We use a Fraktur letter $ \mathfrak{q} = e^{-m_0} $ to denote the instanton factor and $ Z_k $ to denote the $ k $-instanton partition function which can be computed by the blowup equations with the consistent magnetic fluxes \cite{Huang:2017mis,Kim:2019uqw,Kim:2020hhh}
\begin{align}\label{eq:SU2-flux}
    \left\{ \begin{array}{lll}
            \theta = 0 \ & : \ & n \in \mathbb{Z} \, , \ B_{m_0} = 0, \pm 1, \pm 2 \\
            \theta = \pi \ & : \ & n \in \mathbb{Z} \, , \ B_{m_0} = \pm 1/2, \pm 3/2 \, ,
        \end{array}\right.
\end{align}
and $ \Lambda = 1-\mathfrak{q}(q_1 q_2)^{\pm 1} $ for $ B_{m_0}=\pm 2 $ and $ \Lambda=1 $ for the other cases.

\begin{figure}
    \centering
    \begin{subfigure}[b]{0.49\textwidth}
        \centering
        \includegraphics[align=c,scale=1]{figures/fig-SU3-0-2F.pdf}
        {\boldmath$ \to $}
        \includegraphics[align=c,scale=1]{figures/fig-SU2-0.pdf}
        \caption{}
    \end{subfigure}
    \begin{subfigure}[b]{0.49\textwidth}
        \centering
        \includegraphics[align=c,scale=1]{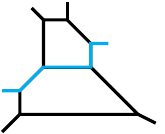}
        {\boldmath$ \to $}
        \includegraphics[align=c,scale=1]{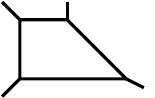}
        \caption{}
    \end{subfigure}
    \caption{5-brane web descriptions of a Higgsing: (a) from $ \mathrm{SU}(3)_0 + 2\mathbf{F} $ to $ \mathrm{SU}(2)_0 $ and (b) from $ \mathrm{SU}(3)_1 + 2\mathbf{F} $ to $ \mathrm{SU}(2)_\pi $. The 5-branes in blue are Higgsed away.}\label{fig:SU2-higgsing-usual}
\end{figure}

We now introduce a half-BPS codimension-2 defect operator extended over $ \mathbb{C} \times S^1 $, obtained by a Higgsing of the $ \mathrm{SU}(3) $ gauge theories. As we discussed in the previous section, a codimension-2 defect in the $ \mathrm{SU}(2) $ theory can be introduced via mesonic Higgs branch deformation of the 5d $ \mathrm{SU}(3)_\kappa $ gauge theory at Chern-Simons level $ \kappa $ coupled to two hypermultiplets in the fundamental representation. The Chern-Simons level can have integer values subject to $ |\kappa| \leq 6 $; otherwise the $ \mathrm{SU}(3)_\kappa + 2\mathbf{F} $ theories do not admit a UV-completion \cite{Jefferson:2017ahm, Jefferson:2018irk}. In this paper, we only consider cases with $ |\kappa| \leq 3 $. By giving nonzero VEVs to the mesonic operators, we obtain the $ \mathrm{SU}(2)_0 $ theory when $ \kappa $ is even and the $ \mathrm{SU}(2)_\pi $ theory when $ \kappa $ is odd. Representative 5-brane diagrams for the Higgsing procedures for $ \kappa=0 $ and $ 1 $ are illustrated in Figure~\ref{fig:SU2-higgsing-usual}.
This can be checked at the level of partition functions. The instanton partition function of the $ \mathrm{SU}(3)_\kappa + 2\mathbf{F} $ is computed using blowup equations as outlined in \cite{Kim:2020hhh}. We choose a chamber of the Coulomb branch as depicted in the brane diagrams Figure~\ref{fig:SU2-higgsing-usual}, where the effective prepotential is given by
\begin{align}
    \begin{aligned}
        \mathcal{E}_0^{\mathrm{SU}(3)_\kappa+2\mathbf{F}}\! &= \frac{1}{6} (8\phi_1^3\! - 3\phi_1^2\phi_2\! - 3\phi_1\phi_2^2\! + 8\phi_2^3) \!- \frac{\kappa}{2} \phi_1 \phi_2 (\phi_1\!-\!\phi_2)\! + m_0 (\phi_1^2 \!- \phi_1\phi_2\! + \phi_2^2) \\
        &\quad - \frac{1}{12} \sum_{i=1}^2 \left( (\phi_1 + s_i m_i)^3 + s_i (\phi_2-\phi_1 + s_i m_i)^3 - (-\phi_2 + s_i m_i)^3 \right) \\
        &\quad - \frac{\epsilon_1^2 + \epsilon_2^2}{48} (2\phi_1+2\phi_2-m_1-m_2) + \epsilon_+^2 (\phi_1+\phi_2) \, ,
    \end{aligned}
\end{align}
with $ s_i = (-1)^i $. One can construct solvable blowup equations using the following consistent magnetic fluxes:
\begin{align}
    n_i \in \mathbb{Z} \, , \quad
    B_{m_1} = \frac12  \, , \quad B_{m_2} = -\frac12 \, , \quad
    B_{m_0} = \left\{ \begin{array}{ll}
            0,\pm 1 & \text{ for } \kappa=1,3 \\
            \pm \frac12 , -\frac32 & \text{ for } \kappa=0,2,4 \, .
        \end{array}\right.
\end{align}

The Higgsing amounts to choosing the pole of the partition function at
\begin{align}\label{eq:SU2-higgsing-pole}
    \left( \sqrt{q_1 q_2} \cdot e^{-(\phi_1-\phi_2+m_1)} \right) \cdot \left( \sqrt{q_1 q_2} \cdot e^{-(\phi_2-\phi_1+m_2)} \right) = 1 \, .
\end{align}
At the level of the partition function, we parameterize the pole condition as
\begin{align}\label{eq:SU2-higgsing-parameter}
    2\phi_1 - \phi_2 = -\phi_1 + 2\phi_2 = \phi \, , \quad
    \phi_1-\phi_2 + m_1 = \phi_2-\phi_1+m_2 = -\epsilon_+ \, ,
\end{align}
where $ \phi $ is the Coulomb branch parameter of the Higgsed theory. Imposing \eqref{eq:SU2-higgsing-parameter} on the effective prepotential and the partition function of $ \mathrm{SU}(3)_\kappa + 2\mathbf{F} $ yields the effective prepotential \eqref{eq:SU2-E0} and the partition function of $ \mathrm{SU}(2)_\theta $ theory:
\begin{align}
    Z^{\mathrm{SU}(3)_\kappa+2\mathbf{F}} ~ \overset{\eqref{eq:SU2-higgsing-parameter}}{\longrightarrow} ~\left\{ \begin{array}{ll}
            Z^{\mathrm{SU}(2)_0} & \quad \kappa=\text{even} \\
            Z^{\mathrm{SU}(2)_\pi} & \quad \kappa=\text{odd\ .} \\
        \end{array}\right.
\end{align}

\begin{figure}
    \centering
    \begin{subfigure}[b]{0.43\linewidth}
        \centering
        \includegraphics[align=c,scale=1]{figures/fig-SU3-1-2F.pdf}
        \hspace{-2ex}{\boldmath$ \to $}
        \includegraphics[align=c,scale=1]{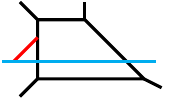}
        \caption{}
    \end{subfigure}
    \begin{subfigure}[b]{0.56\linewidth}
        \centering
        \includegraphics[align=c,scale=1]{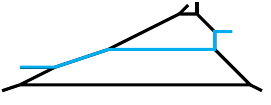}
        \hspace{-2ex}{\boldmath$ \to $}\hspace{-1ex}
        \includegraphics[align=c,scale=1]{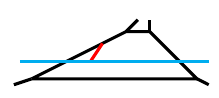}
        \caption{}
    \end{subfigure}
    \caption{Brane realization of the codimension-2 defects in (a) $ \mathrm{SU}(2)_\pi $ theory from Higgsing $ \mathrm{SU}(3)_1+2\mathbf{F} $ theory and (b) $ \mathrm{SU}(2)_{3\pi} $ theory from Higgsing $ \mathrm{SU}(3)_3+2\mathbf{F} $ theory. The red line represents a defect D3-brane.} \label{fig:SU2-1d-higgsing}
\end{figure}

A codimension-2 defect is introduced by giving a position dependent VEV to the mesonic operator. In the 5-brane diagrams, this Higgsing introduces defect D3-branes perpendicular to the $ (p,q) $ 5-branes, as depicted in Figures~\ref{fig:SU2-0-Higgs}(b), \ref{fig:SU2-2-Higgs} and \ref{fig:SU2-1d-higgsing}. Compared to \eqref{eq:SU2-higgsing-parameter}, the position dependent VEV is related to the pole
\begin{align}
    q_2 \cdot \left( \sqrt{q_1 q_2} \cdot e^{-(\phi_1-\phi_2+m_1)} \right) \cdot \left( \sqrt{q_1 q_2} \cdot e^{-(\phi_2-\phi_1+m_2)} \right) = 1 \, .
\end{align}
At the level of the partition functions, we parameterize fugacities as
\begin{alignat}{2}
    \begin{aligned}\label{eq:SU2-defH-parameter}
        &2\phi_1 - \phi_2 = \phi + x \, , \quad
        &&-\phi_1 + 2\phi_2 = \phi - x \, , \\
        &\phi_1 - \phi_2 + m_1 = -\epsilon_+ \, , \quad
        &&\phi_2 - \phi_1 + m_2 = -\epsilon_+ - \epsilon_2 \, ,
    \end{aligned}
\end{alignat}
where $ x $ is the defect parameter. We further shift $ m_0 \to m_0 + \kappa \,x/3 $ and then $ x \to x + \epsilon_2 $ for later convenience. The effective prepotential and the partition function for $\mathrm{SU}(3)_\kappa+2\mathbf{F}$ then reduce to
\begin{equation}
    Z^{\mathrm{SU}(3)_\kappa+2\mathbf{F}} \overset{\eqref{eq:SU2-defH-parameter}}{\longrightarrow} e^{\mathcal{E}_0 + \frac{\phi (2x-\epsilon_1)}{2\epsilon_1} } \cdot Z_{\mathrm{pert}} \cdot \PE\left[ \frac{e^{-x-\phi} - q_1 e^{x-\phi}}{1-q_1} \right] \cdot \Psi_{\mathrm{inst}}\ , \label{eq:SU2-Psi-h} 
\end{equation}
up to an overall extra factor which does not depend on the Coulomb parameter $ \phi $, where $ \mathcal{E}_0 $ and $ Z_{\mathrm{pert}} $ are the effective prepotential and the perturbative partition function for the $ \mathrm{SU}(2) $ theories given in \eqref{eq:SU2-E0} and \eqref{eq:SU2-pert}. We interpret the additional correction to $ \mathcal{E}_0 $ and $ Z_{\mathrm{pert}} $ as the contributions of the D3-brane defect and $ \Psi_{\mathrm{inst}} $ as the instanton correction to the codimension-2 defect partition function. This instanton contribution $ \Psi_{\mathrm{inst}} $ can also be computed from the ADHM construction of the instanton moduli space in the presence of a defect \cite{Hwang:2014uwa, Gaiotto:2014ina}, as summarized in Appendix~\ref{app:ADHM}. We will give explicit expressions of $ \Psi_{\mathrm{inst}} $ in the next subsection.

\subsubsection{Defect partition functions from blowup} \label{subsubsec:su2-blowup}

\begin{figure}
    \centering
    \includegraphics[align=c,scale=1]{figures/fig-SU2-0-def.pdf}
    \hspace{2ex}{\boldmath$ \leftrightarrow $}\hspace{1ex}
    \includegraphics[align=c,scale=1]{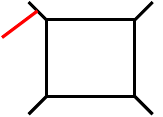}
    \caption{Moving the position of defect D3-brane (in red). Their partition functions only differ in the perturbative parts. In open topological string theory, they are called inner and outer branes, respectively.} \label{fig:SU2-inout}
\end{figure}

We now study $ \Psi_{\mathrm{inst}} $ using the blowup equation \eqref{eq:blowupDefect}. Instead of using the effective prepotential and the perturbative partition function in the presence of codimension-2 defect given in \eqref{eq:SU2-Psi-h}, we write the defect partition function as
\begin{align}\label{eq:SU2-psi}
    \Psi(\phi,m,x,\epsilon_{1,2}) &= e^{\mathcal{E}_0 + \mathcal{E}_1} \cdot \Psi_{\mathrm{pert}}(\phi,x,\epsilon_{1,2}) \cdot \left( 1 + \sum_{k=1}^\infty (\mathfrak{q} \sqrt{q_2})^k \Psi_k(\phi,x,\epsilon_{1,2}) \right) \, ,
\end{align}
where
\begin{align}\label{eq:SU2-E1}
    \mathcal{E}_1 &= \frac{\phi^2 + x^2 - x \epsilon_1 - 2\pi i x}{2\epsilon_1} \, , \quad
    \frac{\Psi_{\mathrm{pert}}}{Z_{\mathrm{pert}}} = \PE\left[ \frac{e^{-x}( e^{\phi} + e^{-\phi} )}{1-q_1}\right] = \frac{1}{(XQ^{\pm 1};q_1)_\infty} \, ,
\end{align}
for $ X=e^{-x} $, where $ (\bullet; \bullet)_\infty $ is the q-Pochhammer symbol defined in Appendix~\ref{app:special} and $ Q=e^{-\phi} $. This amounts to relocate the D3-brane defects, attached to the NS5-brane, to the external 5-brane above the top (color) D5-brane as in Figure~\ref{fig:SU2-inout}. In the context of open topological string theory, the D3-brane defect on the left of Figure~\ref{fig:SU2-inout} is called the inner brane, and the D3-brane defect on the right is called outer brane \cite{Aganagic:2001nx}.
The 5d $ \mathrm{SU}(2) $ gauge theory is now coupled to a 3d chiral multiplets which form a doublet of $ \mathrm{SU}(2) $ with $ \mathrm{U}(1) $ flavor charge $ 1 $. Note that $ \mathcal{E}_1 $ can be understood as a regularization factor of the D3-brane defect contribution in the perturbative partition function \cite{Bullimore:2014awa}.\footnote{In particular, we find $ \mathcal{E}_1 = \pi i \big(\zeta_2(0,\frac{x+\phi}{2\pi i} | 1,\frac{\epsilon_1}{2\pi i}) + \zeta_2(0,\frac{x-\phi}{2\pi i} | 1,\frac{\epsilon_1}{2\pi i})\big) $ from the regularized infinite product of the perturbative contribution $ \Psi_{\mathrm{pert}} $, where $ \zeta_2(s,z | w_1,w_2) $ is the Barnes' zeta function and we ignore the terms that depend only on $ \epsilon_1 $ but are independent of $ \phi $ and $ x $. See also Appendix~C of \cite{Bullimore:2014awa} for the further details.} We expect that the partition functions of the two defects (associated with inner/outer branes) differ only by the perturbative contributions and the effective prepotential $ \mathcal{E}_1 $, while their instanton corrections $ \Psi_k $ are the same.

Notice that the instanton factor in \eqref{eq:SU2-psi} comes with $\sqrt{q_2}$. The defect contribution to the effective prepotential $ \mathcal{E}_1 $ contains a background Chern-Simons term $ \frac{\phi^2}{2\epsilon_1} $. This term can be combined into the gauge kinetic term $ \frac{m_0 \phi^2}{\epsilon_1 \epsilon_2} $ in $ \mathcal{E}_0 $ and renormalizes the inverse gauge coupling $ m_0 $ to $ m_0 + \epsilon_2/2 $. For this reason, we factor out $ \sqrt{q_2} $ in the instanton partition function $ \Psi_k $ in \eqref{eq:SU2-psi}. In the blowup equation, this redefinition shifts the magnetic flux $ B_{m_0} $ by $ 1/2 $. When we consider the flux quantization condition \eqref{eq:defect-blowup-flux}, we need to subtract the $ 1/2 $ shift on $ B_{m_0} $ due to the renormalization effect. In the perturbative part $ \Psi_{\mathrm{pert}}/Z_{\mathrm{pert}} $, the BPS states bound to the defect have spin $ (s_l, s_r)=(0,0) $ according to comparisons between \eqref{eq:SU2-E1} and \eqref{eq:Psi-index}, so the flux quantization condition \eqref{eq:defect-blowup-flux} indicates that $ B_x $ must be an integer if the dynamical magnetic flux $ n $ for $ \phi $ is quantized to an integer. In the original blowup equations without a defect, $ B_{m_0} $ is an integer when $ \theta=0 $. Therefore, we expect $ B_{m_0} $ to become a half-integer in the presence of the defect,  but $ s_l+s_r $ still remains even for all the BPS states bound to the defect according to the quantization condition. Similarly, we expect $ s_l+s_r $ to be odd for the odd-instanton states and even for the even-instanton states in the $ \theta=\pi $ theory.

We now discuss the solution of blowup equations in the presence of a codimension-2 defect. The solutions of the usual blowup equations \eqref{eq:blowup} without a defect is uniquely determined from the effective prepotential $ \mathcal{E}_0 $ and the magnetic fluxes \eqref{eq:SU2-flux}. However, in the presence of a defect, the blowup equations with magnetic fluxes given in \eqref{eq:SU2-flux}, together with a $ 1/2 $ shift on $ B_{m_0} $ due to the mass renormalization, do not have a common solution. Instead, only the blowup equations with $ B_{m_0} $ differing by $ 1 $ share a common solution. We claim that different solutions arising from different magnetic fluxes correspond to the different discrete theta angles. We solve the blowup equation \eqref{eq:blowupDefect} using the magnetic fluxes $ n \in \mathbb{Z} $, $ B_x=0 $ together with 
\begin{align}\label{eq:su2-flux-defect}
    B_{m_0} = \left\{ \begin{array}{ll}
            -3/2 \text{ and } -1/2 & \quad \theta=-2\pi \\
            -1/2 \text{ and } 1/2 & \quad \theta=0 \\
            1/2 \text{ and } 3/2 & \quad \theta=2\pi \\
            3/2 \text{ and } 5/2 & \quad \theta=4\pi
        \end{array}\right.
    \text{ or } \quad
    B_{m_0} = \left\{ \begin{array}{ll}
            -1 \text{ and } 0 & \quad \theta=-\pi \\
            0 \text{ and } 1 & \quad \theta=\pi \\
            1 \text{ and } 2 & \quad \theta=3\pi \, . \\
        \end{array}\right.
\end{align}

Let us present explicit expressions of the solutions for each case at 1-instanton order. Denote $ \chi_j(\phi) = \sum_{k=-j}^j e^{2k\phi} $ as the spin-$ j $ character of $ \mathrm{SU}(2) $. In case of $ \theta=(\mathrm{even})\pi $, we find
\begin{align}\label{eq:su2-m2-defect}
    \Psi_1 = Z_1 - \frac{ \left( (q_1^2 q_2 \chi_{3/2}(\phi) - q_1 \chi_{1/2}(\phi)) - (q_1q_2 \chi_1(\phi)-1) e^{-x} \right) e^{-x}}{q_2(1-q_1) (1-q_1q_2 e^{-2\phi}) (1-q_1q_2 e^{2\phi}) (q_1-e^{-x-\phi}) (q_1-e^{-x+\phi})} \, ,
\end{align}
for $ \theta=-2\pi $;
\begin{align}\label{eq:su2-0-defect}
    \Psi_1 = Z_1 - \frac{q_1 \left( q_1^2 q_2 \chi_{1/2}(\phi) - (1+q_1q_2) e^{-x} \right) e^{-x}}{(1-q_1) (1-q_1q_2 e^{-2\phi}) (1-q_1q_2 e^{2\phi}) (q_1-e^{-x-\phi}) (q_1-e^{-x+\phi})} \, ,
\end{align}
for $ \theta=0 $; and
\begin{align}\label{eq:su2-2pi-psi1}
    \Psi_1 = Z_1 - \frac{q_1^2 q_2 \left( q_1 \chi_{1/2}(\phi) - ( \chi_1(\phi) - q_1q_2) e^{-x}  \right) e^{-x} }{(1-q_1)(1-q_1q_2 e^{-2\phi}) (1-q_1q_2 e^{2\phi}) (q_1-e^{-x-\phi}) (q_1-e^{-x+\phi})}\ ,
\end{align}
for $ \theta=2\pi $. Here, $ Z_1 $ is the 1-instanton partition function of the $ \mathrm{SU}(2)_0 $ theory without a defect. For these three cases, we checked that solutions of blowup equations exactly match with the Higgsing $ \mathrm{SU}(3)_{-2} + 2\mathbf{F} $, $ \mathrm{SU}(3)_0+2\mathbf{F} $ and $ \mathrm{SU}(3)_2 + 2\mathbf{F} $ theory, up to 3-instanton order. In case of $ \theta=4\pi $, the blowup equation gives
\begin{align}\label{eq:su2-4pi-psi1}
    \Psi_1 = Z_1 - \frac{q_1^3 q_2^2 \left( (q_1 \chi_{3/2}(\phi) - q_1^2 q_2 \chi_{1/2}(\phi)) - ( \chi_2(\phi) - q_1 q_2 \chi_1(\phi) )e^{-x}  \right) e^{-x} }{(1-q_1)(1-q_1q_2 e^{-2\phi}) (1-q_1q_2 e^{2\phi}) (q_1-e^{-x-\phi}) (q_1-e^{-x+\phi})}\ .
\end{align}
Although we cannot obtain this expression from the Higgsing of $ \mathrm{SU}(3)_4 + 2\mathbf{F} $, we claim that the solution of blowup equation is a sensible codimension-2 defect partition function of the $ \mathrm{SU}(2)_{4\pi} $ theory. We support this claim by considering the difference equation that annihilates the solution in Section~\ref{subsubsec:su2-diffeq}.

Similarly, we compute the 1-instanton solutions of blowup equations with integer magnetic fluxes in \eqref{eq:su2-flux-defect}, that correspond to $ \theta=(\mathrm{odd})\pi $. The blowup equations give
\begin{align}
    \Psi_1 &= Z_1 + \frac{q_1^{3/2} q_2^{-1/2} \left( q_1 q_2 \chi_1(\phi) - 1 - q_2 \chi_{1/2}(\phi) e^{-x}  \right) e^{-x} }{(1-q_1)(1-q_1q_2 e^{-2\phi}) (1-q_1q_2 e^{2\phi}) (q_1-e^{-x-\phi}) (q_1-e^{-x+\phi})}
\end{align}
for $ \theta=-\pi $ and
\begin{align}
    \Psi_1 &= Z_1 + \frac{q_1^{3/2} q_2^{1/2} \left( q_1(1+q_1q_2) - \chi_{1/2}(\phi) e^{-x}  \right) e^{-x} }{(1-q_1)(1-q_1q_2 e^{-2\phi}) (1-q_1q_2 e^{2\phi}) (q_1-e^{-x-\phi}) (q_1-e^{-x+\phi})}
\end{align}
for $ \theta=\pi $, where $ Z_1 $ is the 1-instanton partition function of the $ \mathrm{SU}(2)_\pi $ theory without a defect. We checked that solutions of the blowup equations for these two cases exactly match with the defect partition functions computed by a Higgsing of $ \mathrm{SU}(3)_{\pm 1} + 2\mathbf{F} $, up to 3-instanton order. When $ \theta=3\pi $, the blowup equation gives
\begin{align}\label{eq:su2-3-defect}
    \Psi_1 &= Z_1 + \frac{q_1^{5/2} q_2^{3/2} \left( ( q_1 \chi_1(\phi) - q_1^2 q_2 ) - (\chi_{3/2}(\phi) - q_1 q_2 \chi_{1/2}(\phi)) e^{-x}  \right) e^{-x} }{(1-q_1)(1-q_1q_2 e^{-2\phi}) (1-q_1q_2 e^{2\phi}) (q_1-e^{-x-\phi}) (q_1-e^{-x+\phi})} \, .
\end{align}
The solution differs from the Higgsing of $ \mathrm{SU}(3)_3 + 2\mathbf{F} $ up to overall factor which does not depend on the dynamical parameter $ \phi $:
\begin{align}\label{eq:SU2-3pi-higgs-blowup}
    \Psi^{\mathrm{Higgsing}} = \Psi^{\mathrm{blowup}} \cdot \PE\left[ \frac{\mathfrak{q} X \sqrt{q_1} q_2}{(1-q_1)(1-q_2)} \right] \, .
\end{align}
We checked this relation up to 3-instanton order. Since the extra factor appearing in the partition function from the Higgsing has additional pole along $ \epsilon_2=0 $, we need to factor it out. This may suggest that the solution of the blowup equation provide codimension-2 defect partition function of the $ \mathrm{SU}(2)_{3\pi} $ theory. We remark that although the defect partition functions for different theta angles appear to be very different, they are, in fact, related through the transformation given in \eqref{eq:psitheta}, which is inspired by the transition of quantum curves. A detailed computation is provided in Appendix~\ref{app:hw}.

\subsubsection{Seiberg-Witten curves and quantizations} \label{subsubsec:su2-diffeq}

We now investigate difference equations that are satisfied by the codimension-2 defect partition functions derived from the blowup equations. Let $ \langle W \rangle $ be the VEV of the Wilson loop operator in the fundamental representation of $ \mathrm{SU}(2) $ gauge algebra, which can also be computed via the blowup equation in the presence of codimension-4 defect \cite{Kim:2021gyj}. For each discrete theta angle, $ \langle W \rangle $ is given by
\begin{align}
    &\theta=0 \quad : \quad \langle W \rangle = e^{\phi} + e^{-\phi} - \frac{q_1 q_2 (e^{\phi} + e^{-\phi})}{(1-q_1q_2 e^{-2\phi}) (1-q_1 q_2 e^{2\phi})} \mathfrak{q} + \mathcal{O}(\mathfrak{q}^2) \, , \\
    &\theta=\pi \quad : \quad \langle W \rangle = e^{\phi} + e^{-\phi} + \frac{\sqrt{q_1 q_2} (1+q_1q_2)}{(1-q_1q_2 e^{-2\phi}) (1-q_1 q_2 e^{2\phi})} \mathfrak{q} + \mathcal{O}(\mathfrak{q}^2) \, .
\end{align}
The Nekrasov-Shatashvili (NS) limit of the normalized codimension-2 defect partition function
\begin{align}
    \hat{\Psi}^{\mathrm{NS}} = \lim_{\epsilon_2\to 0} \hat{\Psi} / \hat{Z} \, ,
\end{align}
where $ Z $ is the partition function without a defect, is interpreted as a wave function annihilated by a quantum Seiberg-Witten curve, whose eigenvalue is the Wilson loop expectation values $ \langle \hat{W} \rangle $. One may use either hatted or unhatted partition functions; however here we use the hatted partition functions appearing in the blowup equations. This choice makes the signs of each term in quantum SW curve positive. The replacement does not change the $ \theta=(\mathrm{even})\pi $ partition functions, and only affect to $ \mathfrak{q} \mapsto -\mathfrak{q} $ in the $ \theta=(\mathrm{odd})\pi $ partition functions.

\begin{figure}
    \centering
    \begin{subfigure}[b]{0.22\linewidth}
        \centering
        \includegraphics{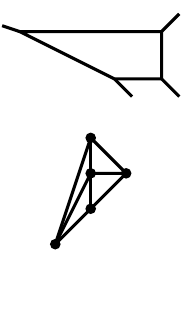}
        \caption{$ \theta=-2\pi $}
    \end{subfigure}
    \begin{subfigure}[b]{0.22\linewidth}
        \centering
        \includegraphics{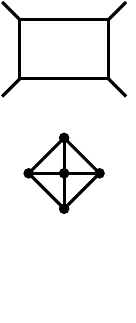}
        \caption{$ \theta=0 $}
    \end{subfigure}
    \begin{subfigure}[b]{0.22\linewidth}
        \centering
        \includegraphics{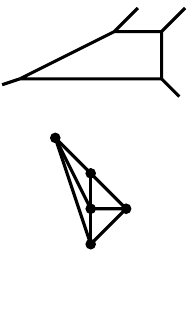}
        \caption{$ \theta=2\pi $}
    \end{subfigure}
    \begin{subfigure}[b]{0.27\linewidth}
        \centering
        \includegraphics{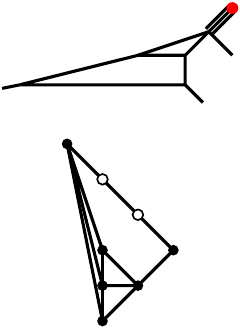}
        \caption{$ \theta=4\pi $}
    \end{subfigure}
    \caption{5-brane webs and generalized toric diagrams of $ \mathrm{SU}(2)_{\theta=(\mathrm{even})\pi} $ related with the solutions of blowup equations. They are related by HW transitions.} \label{fig:SU2-even-toric}
\end{figure}

The defect contributions of the effective prepotential $ \mathcal{E}_1 $ and the perturbative partition function $ \hat{\Psi}^{\mathrm{NS}}_{\mathrm{pert}} = \hat{\Psi}_{\mathrm{pert}} / \hat{Z}_{\mathrm{pert}} $ which are independent of the $ \theta $-angle satisfy
\begin{align}\label{eq:SU2-diff-pert}
    \left( X + X^{-1} + Y \right) e^{\mathcal{E}_1} \hat{\Psi}^{\mathrm{NS}}_{\mathrm{pert}} = \left( e^{\phi} + e^{-\phi} \right) e^{\mathcal{E}_1} \hat{\Psi}^{\mathrm{NS}}_{\mathrm{pert}} \, .
\end{align}
Here, $ X = e^{-x} $ and $ Y = e^{\epsilon_1 \partial_x} $ is the shift operator satisfying $ YX = q_1 XY $. Using the full defect partition function \eqref{eq:SU2-psi} including the instanton corrections found from the blowup equations with half-integer quantized magnetic fluxes $ B_{m_0} $, we obtain
\begin{align}
    \left( X + X^{-1} + Y + \mathfrak{q} Y^{-1} \right) \hat{\Psi}^{\mathrm{NS}} &= E \hat{\Psi}^{\mathrm{NS}} \quad \text{for } \theta=0 \, , \label{eq:SU2-0-SW} \\
    \left( X + X^{-1} + Y + \mathfrak{q} \left (X^{\pm 1} Y^{-1} X^{\pm 1} + X^{\pm 1} \right) \right) \hat{\Psi}^{\mathrm{NS}} &= E \hat{\Psi}^{\mathrm{NS}} \quad \text{for } \theta= \pm 2\pi \, , \\
    \left( X + X^{-1} + Y + \mathfrak{q} \left (Y+X + q_1 XY^{-1} (X+Y)^3 \right) \right) \hat{\Psi}^{\mathrm{NS}} &= E \hat{\Psi}^{\mathrm{NS}} \quad \text{for } \theta=4\pi \,  ,
\end{align}
where $ E = \langle \hat{W} \rangle | _{\epsilon_2=0} $ is the NS-limit of the Wilson loop expectation value. These difference equations represent the quantum Seiberg-Witten curves of $ \mathrm{SU}(2)_\theta $ with $ \theta=0,\pm 2\pi $, and $ 4\pi $, and their generalized toric diagrams are given in Figure~\ref{fig:SU2-even-toric}.

\begin{figure}
    \centering
    \begin{subfigure}[b]{0.3\linewidth}
        \centering
        \includegraphics{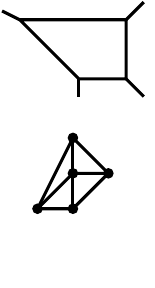}
        \caption{$ \theta=-\pi $}
    \end{subfigure}
    \begin{subfigure}[b]{0.3\linewidth}
        \centering
        \includegraphics{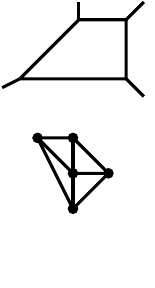}
        \caption{$ \theta=\pi $}
    \end{subfigure}
    \begin{subfigure}[b]{0.3\linewidth}
        \centering
        \includegraphics{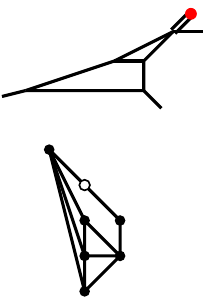}
        \caption{$ \theta=3\pi $}
    \end{subfigure}
    \caption{5-brane webs and generalized toric diagrams of the $ \mathrm{SU}(2)_{\theta=(\mathrm{odd})\pi} $ related with the solutions of blowup equations. They are related by HW transitions.} \label{fig:SU2-odd-toric}
\end{figure}

Similarly, the instanton corrections for the defect partition functions computed from the blowup equations with integer quantized magnetic fluxed $ B_{m_0} $ satisfy the difference equations
\begin{align}
    \left( X + X^{-1} + Y + \mathfrak{q} q_1^{\mp 1/2} X^{\pm 1} Y^{-1} \right) \hat{\Psi}^{\mathrm{NS}} &= E \hat{\Psi}^{\mathrm{NS}} \quad \text{for } \theta=\pm \pi \, ,\\
    \left( X + X^{-1} + Y + \mathfrak{q}\sqrt{q_1} XY^{-1} (X+Y)^2 \right) \hat{\Psi}^{\mathrm{NS}} &= E \hat{\Psi}^{\mathrm{NS}} \quad \text{for } \theta=3\pi \, . \label{eq:SU2-3pi-SW}
\end{align}
The corresponding generalized toric diagrams are shown in Figure~\ref{fig:SU2-odd-toric}. Therefore, the solutions to the blowup equations obtained in Section~\ref{subsubsec:su2-blowup} are well-defined codimension-2 defect partition functions for each discrete theta angle.

\subsubsection{S-transformation} \label{subsubsec:su2S}

In this subsection, we consider another type of codimension-2 defects in the $ \mathrm{SU}(2) $ gauge theories related to the defect studied above by $ S $-transformation. Let us first perform the $ S $-transformation to the 3d theory on the defect after decoupling 5d degrees of freedom. The $ \mathrm{U}(1) $ global symmetry parametrized by $ x $ is now gauged, and the $ S $-transformation of the 3d partition function is given by
\begin{align}
    \tilde{\Psi}^{\mathrm{3d}}(x) = \oint ds \: \frac{e^{(\phi^2 - s \epsilon_1 - 2\pi i s)/2\epsilon_1} e^{sx/\epsilon_1}}{(\sigma Q; q_1)_\infty (\sigma/Q; q_1)_\infty} \, ,
\end{align}
as in Section~\ref{sec:qcurve_HW}. Here, $ s $ and $ x $ parametrize gauged $ \mathrm{U}(1) $ symmetry and the new $ \mathrm{U}(1) $ global symmetry, respectively, and $ e^{sx/\epsilon_1} $ is the mixed Chern-Simons term introduced by the $ S $-transformation. The overall factor $ e^{(\phi^2 - s \epsilon_1 - 2\pi i s)/2\epsilon_1} $ comes from the effective prepotential $ \mathcal{E}_1 $ in \eqref{eq:SU2-E1}, while the $ e^{\frac{1}{2\epsilon_1}s^2} $ factor is eliminated by applying the $ T $-transformation \eqref{eq:Ttrans}. The denominator of the integrand is equivalent to the perturbative partition function \eqref{eq:SU2-E1} of 3d chiral multiplets, where $ \sigma=e^{-s} $, $ Q=e^{-\phi} $. For convenience, let us shift $ x \to x + \pi i + \epsilon_1/2 $ to cancel the $ e^{(-s \epsilon_1 -2\pi i s)/2\epsilon_1} $ term. The integral is evaluated by choosing residues at the poles $ \sigma = Q q_1^{-k} $ for $ k \in \mathbb{Z}_{\geq 0} $. The contour integral yields
\begin{align}\label{eq:U1-vortex}
    \tilde{\Psi}^{\mathrm{3d}}(x) = e^{\phi^2 / 2\epsilon_1} e^{x\phi/\epsilon_1} \frac{{}_0\phi_1(-; q_1 Q^{-2}; q_1, q_1^2 Q^{-2} X)}{(q_1;q_1)_\infty (Q^2; q_1)_\infty} \, ,
\end{align}
where $ X = e^{-x} $ and $ {}_0\phi_1 $ is the basic hypergeometric series defined in Appendix~\ref{app:special}. This is the vortex partition function of the 3d theory after decoupling the 5d degrees of freedom. If we choose the other poles $ \sigma = Q^{-1} q_1^{-k} $ in the contour integral, we obtain a result with $ Q \to Q^{-1} $. The vortex partition function \eqref{eq:U1-vortex} satisfies the difference equation
\begin{align}\label{eq:SU2-openToda}
    \left(Y + Y^{-1} - Y X \right) \tilde{\Psi}^{\mathrm{3d}} = \left(e^{\phi} + e^{-\phi} \right) \tilde{\Psi}^{\mathrm{3d}} \, ,
\end{align}
where $ Y = e^{\epsilon_1 \partial_{x}} $ is the momentum operator satisfying $ YX = q_1 XY $, which shifts a function $ f(x) $ to $ Y f(x) = f(x+\epsilon_1) $.\footnote{It can be proven from the property of the q-hypergeometric function \eqref{eq:qhyp-diff}.} This is the Hamiltonian of two-body open Toda system \cite{Bullimore:2014awa}.

We now promote the S-transformation to the full 5d/3d coupled system. The $ S $-transformation of the defect partition function $ \Psi $ is given by
\begin{align}\label{eq:SU2-Psitilde}
    \widetilde{\Psi}(\phi,m,x,\epsilon_{1,2}) = \oint ds \: e^{sx/\epsilon_1} \Psi(\phi,m,s,\epsilon_{1,2}) = e^{\mathcal{E}_0 + \phi^2/2\epsilon_1 + x\phi / \epsilon_1} \cdot Z_{\mathrm{pert}} \cdot \sum_{k=0}^\infty \mathfrak{q}^k \widetilde{\Psi}_k \, ,
\end{align}
where $ \mathcal{E}_0 $ and $ Z_{\mathrm{pert}} $ are the effective prepotential \eqref{eq:SU2-E0} and the 5d vector multiplet contribution \eqref{eq:SU2-pert}, respectively. In addition to the poles at $ \sigma = Q q_1^{-k} $ in the perturbative contributions, instanton corrections have poles at $ \sigma = Q q_1^k $ for the positive integers $ k $. At 1-instanton order, the contour integral gives
\begin{align}
    \widetilde{\Psi}_1 = \sum_{k=0}^\infty \frac{(1;q_1)_{-k} (Q^2; q_1)_{-k}}{(q_1;q_1)_\infty (Q^2; q_1)_\infty} X^{k} \Psi_1(s=\phi -k \epsilon_1) + \frac{1-Q^2}{(q_1;q_1)_\infty (Q^2; q_1)_\infty} \frac{1}{X} \Res_{s=\phi+\epsilon_1} \Psi_1(s)
\end{align}
for the 1-instanton defect partition function $ \Psi_1 $ given in \eqref{eq:su2-0-defect}-\eqref{eq:su2-3-defect} depending on the $ \theta $-angle. We note that it is also possible to find the same defect partition function $ \widetilde{\Psi} $ in the case of the $ \mathrm{SU}(2)_0 $ theory using the baryonic Higgs branch deformation from the 5d $ \mathrm{SU}(2)_\pi  \times \mathrm{SU}(2)_\pi $ theory as in Figure~\ref{fig:SU2-0-Higgs2}. Detailed computation is given in Appendix~\ref{app:ADHM-quiver}.

The defect partition function $ \widetilde{\Psi} $ can be also found from the blowup equations. We write the blowup equation using the effective prepotential and defect partition function, 
\begin{align}
    \widetilde{\mathcal{E}}_0 = \mathcal{E}_0 \, , \quad
    \widetilde{\mathcal{E}}_1 = \frac{2x\phi + \phi^2}{2\epsilon_1} \, , \quad
    \widetilde{\Psi} = e^{\widetilde{\mathcal{E}}_0 + \widetilde{\mathcal{E}}_1} Z_{\mathrm{pert}} \sum_{k=0}^\infty \mathfrak{q}^k \widetilde{\Psi}_k \, ,
\end{align}
together with the consistent magnetic fluxes
\begin{align}
    n \in \mathbb{Z} \, , \
    B_x = 0 \, , \
    B_{m_0} = \left\{ \begin{array}{ll}
            -3/2, -1/2 & \quad \theta=-2\pi \\
            \pm 1/2 & \quad \theta=0 \\
            1/2, 3/2 & \quad \theta=2\pi \\
            3/2, 5/2 & \quad \theta=4\pi
        \end{array}\right.
   \text{ or } \ 
   B_{m_0} = \left\{ \begin{array}{ll}
           -1, 0 & \quad \theta=-\pi \\
           0, 1 & \quad \theta=\pi \\
           1, 2 & \quad \theta=3\pi \, .
       \end{array}\right.
\end{align}
Using the vortex partition function $ \widetilde{\Psi}_0 $ obtained in \eqref{eq:U1-vortex}, we find instanton corrections for $ \widetilde{\Psi} $ from the blowup equations. The result exactly matches with the $ S $-transformation of the surface defect computed by \eqref{eq:SU2-Psitilde}. We checked the equality up to 2-instanton order.

Lastly, we find instanton corrections of the difference equation \eqref{eq:SU2-openToda} that normalized NS-limit of the defect partition function $ \widetilde{\Psi}^{\mathrm{NS}} = \lim_{\epsilon_2\to 0}\widetilde{\Psi} / Z $ is satisfied. When $ \theta=0 $, the difference equation becomes the 2-body closed Toda Hamiltonian given by
\begin{align}
    \theta=0 \ : \  E \widetilde{\Psi}^{\mathrm{NS}} = \left(Y + Y^{-1} - YX - \mathfrak{q} X^{-1} Y^{-1}\right) \widetilde{\Psi}^{\mathrm{NS}} \, ,
\end{align}
where the eigenvalue $ E = \langle W \rangle |_{\epsilon_2=0} $ is the NS-limit of the Wilson loop VEV of the $ \mathrm{SU}(2)_0 $ theory. Note that this difference equation can be also obtained by applying the $ T $- and $ S $-transformations discussed in Section~\ref{sec:qcurve_HW} to the quantum curve \eqref{eq:SU2-0-SW}, where the sign difference comes from the redefinition $ x \to x+\pi i+\epsilon_1/2 $ and use of the unhatted partition functions.

Similarly, difference equations for $ \theta=\pm 2\pi $ and $ \theta=4\pi $ are given by
\begin{align}
    \theta = 2\pi \ &: \ E\widetilde{\Psi}^{\mathrm{NS}} = \left( Y + Y^{-1} - Y X - \mathfrak{q} \left( X^{-1} Y^{-3} + Y^{-1} \right) \right) \widetilde{\Psi}^{\mathrm{NS}} \, , \\
    \theta = -2\pi \ &: \ E\widetilde{\Psi}^{\mathrm{NS}} = \left(Y + Y^{-1} - YX - \mathfrak{q} \left( Y X^{-1} + Y \right) \right) \widetilde{\Psi}^{\mathrm{NS}} \, , \\
    \theta = 4\pi \ &: \ E\widetilde{\Psi}^{\mathrm{NS}} = \left( Y + Y^{-1} - YX + \mathfrak{q} \left( -q_1^2 X^{-1} Y^{-5} + (q_1+1+q_1^{-1}) Y^{-3} \right. \right. \nonumber \\
    &\qquad \qquad \qquad \qquad + \left. \left. (1-(q_1+1+q_1^{-1})X) Y^{-1} - YX + YX^2  \right) \right) \widetilde{\Psi}^{\mathrm{NS}} \, .
\end{align}
In case of $ \theta=(\mathrm{odd})\pi $, we find
\begin{align}
    \theta=\pi \ &: \ E\widetilde{\Psi}^{\mathrm{NS}} = \left( Y + Y^{-1} - YX + \mathfrak{q} \sqrt{q_1} X^{-1} Y^{-2} \right) \widetilde{\Psi}^{\mathrm{NS}} \, , \\
    \theta=-\pi \ &: \ E\widetilde{\Psi}^{\mathrm{NS}} = \left( Y + Y^{-1} - YX + \frac{\mathfrak{q}}{\sqrt{q_1}}X^{-1} \right) \widetilde{\Psi}^{\mathrm{NS}} \, , \\
    \theta=3\pi \ &: \ E\widetilde{\Psi}^{\mathrm{NS}} = \left( Y + Y^{-1} - YX + \frac{\mathfrak{q}}{\sqrt{q_1}}\left(q_1^2 X^{-1} Y^{-4} + (1+q_1) Y^{-2} + q_1X \right) \right) \widetilde{\Psi}^{\mathrm{NS}} \, ,
\end{align}
from the solutions to the blowup equations, where $ E = \langle W \rangle |_{\epsilon_2=0} $ is the NS-limit of the Wilson loop VEV in the $ \mathrm{SU}(2)_\pi $ theory.


\section{Examples}\label{sec:example}

In this section, we explore more examples for the blowup equations and codimension-2 defects in 5d rank-1 SCFTs and KK-theories.

\subsection{\texorpdfstring{$\mathrm{SU}(2)+1{\bf F}$}{SU2+1F}}

Let us consider codimension-2 defects in the $ \mathrm{SU}(2) $ gauge theory coupled to a fundamental hypermultiplet. The effective prepotential and the perturbative partition function from the vector multiplet and fundamental hypermultiplet are
\begin{align}
    \mathcal{E}_0 &= \frac{1}{\epsilon_1 \epsilon_2} \left( \frac{8}{6} \phi^3 - \frac{1}{12} \left( (\phi+m_1)^3 + (\phi-m_1)^3 \right) + m_0 \phi^2 - \frac{\epsilon_1^2 + \epsilon_2^2}{24}\phi + \epsilon_+^2 \phi \right) \, , \label{eq:SU2-1F-E0}\\
    Z_{\mathrm{pert}} &= \PE\left[ -\frac{1+q_1q_2}{(1-q_1)(1-q_2)} e^{-2\phi} + \frac{\sqrt{q_1q_2}}{(1-q_1)(1-q_2)} \left( e^{-\phi+m_1} + e^{-\phi-m_1} \right) \right] \, ,
\end{align}
where $ m_0 $ is the inverse gauge coupling squared and $ m_1 $ is the mass parameter of the fundamental hypermultiplet. The instanton partition function is computed by using the blowup equations with a set of consistent magnetic fluxes
\begin{align}
    \left\{ \begin{array}{l}
    n \in \mathbb{Z} \, , \quad
    B_{m_0} = -7/4, -3/4, 1/4, 5/4 \, , \quad
    B_{m_1} = 1/2 \, , \\
    n \in \mathbb{Z} \, , \quad
    B_{m_0} = -5/4, -1/4, 3/4, 7/4 \, , \quad
    B_{m_1} = -1/2 \, ,
    \end{array} \right.
\end{align}
where
\begin{align}\label{eq:SU2-1F-Lambda}
    \Lambda = \left\{ \begin{array}{ll}
            1 - \mathfrak{q} e^{m_1/2} q_1^{-1} q_2^{-1} & \quad (B_{m_0}, B_{m_1}) = (-7/4, 1/2) \\
            1 - \mathfrak{q} e^{m_1/2}q_1q_2 & \quad (B_{m_0}, B_{m_1}) = (7/4, -1/2) \\
            1 & \quad \text{otherwise,}
        \end{array}\right.
\end{align}
for the instanton fugacity $ \mathfrak{q}=e^{-m_0} $. One can check these results against the partition functions computed from the ADHM construction of instanton moduli space summarized in Appendix~\ref{app:ADHM}. 

Also, we can compute the expectation values of the Wilson loop operators using the blowup equations following the method in \cite{Kim:2021gyj}. Using the effective prepotential \eqref{eq:SU2-1F-E0} and magnetic fluxes $ n \in \mathbb{Z} $, $ B_{m_0} = -3/4,1/4 $, $ B_{m_1} = 1/2 $, we calculate the Wilson loop VEV $ \langle W \rangle $. The result up to 1-instanton order is given by
\begin{align}
    \langle W \rangle = e^{\phi} + e^{-\phi} + \mathfrak{q} \frac{e^{-m_1} \sqrt{q_1q_2} (1+q_1q_2) - (e^{\phi}+e^{-\phi})q_1q_2}{e^{-m_1/2} (1-q_1q_2 e^{2\phi})(1-q_1q_2 e^{-2\phi})} + \mathcal{O}(\mathfrak{q}^2) \, .
\end{align}

Next, we consider a codimension-2 defect realized by coupling 3d chiral multiplets in the fundamental representation of the bulk $ \mathrm{SU}(2) $ gauge symmetry. As in the $ \mathrm{SU}(2)_\theta $ theory, we use the contribution of the 3d chiral multiplets to the effective prepotential and the perturbative partition function given by
\begin{align}
    \mathcal{E}_1 &= \frac{\phi^2 + x^2 - x \epsilon_1 - 2\pi i x}{2\epsilon_1} \, , \quad
    \frac{\Psi_{\mathrm{pert}}}{Z_{\mathrm{pert}}} = \PE\left[ \frac{1}{1-q_1} \left( e^{-x+\phi} + e^{-x-\phi} \right) \right] \, ,
\end{align}
where $ x $ is chemical potential for the $ U(1) $ flavor symmetry acting on the chiral multiplets. The background Chern-Simons term $ \phi^2/2\epsilon_1 $ in the effective prepotential $ \mathcal{E}_1 $ changes flux on $ m_0 $ by $ 1/2 $. We solve the blowup equation in the presence of codimension-2 defect taking this effect into account. 

\begin{figure}
    \centering
    \begin{subfigure}[b]{0.3\linewidth}
        \centering
        \includegraphics{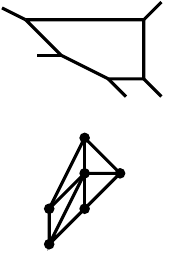}
        \caption{}
    \end{subfigure}
    \begin{subfigure}[b]{0.3\linewidth}
        \centering
        \includegraphics{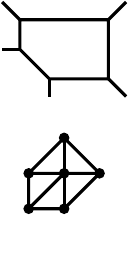}
        \caption{}
    \end{subfigure}
    \begin{subfigure}[b]{0.3\linewidth}
        \centering
        \includegraphics{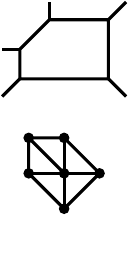}
        \caption{}
    \end{subfigure}
    \begin{subfigure}[b]{0.3\linewidth}
        \centering
        \includegraphics{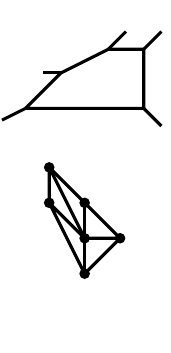}
        \caption{}
    \end{subfigure}
    \begin{subfigure}[b]{0.3\linewidth}
        \centering
        \includegraphics{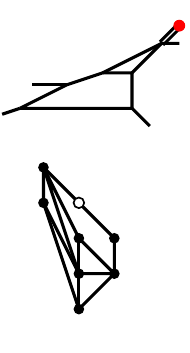}
        \caption{}
    \end{subfigure}
    \begin{subfigure}[b]{0.3\linewidth}
        \centering
        \includegraphics{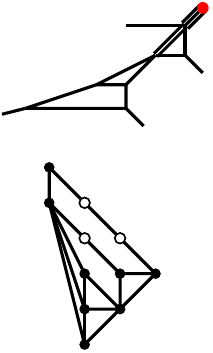}
        \caption{}
    \end{subfigure}
    \caption{Various 5-brane webs and the corresponding toric diagrams of the $ \mathrm{SU}(2)+1\mathbf{F} $ theory. All these diagrams are related by Hanany-Witten transitions. The codimension-2 defect partition functions from blowup equations probe these Hanany-Witten moves.} \label{fig:SU2-1F}
\end{figure}

We find three defect partition functions from the blowup equations with $ B_{m_1} = +1/2 $ and another three partition functions from $ B_{m_1} = -1/2 $. We propose that each defect partition function probes Hanany-Witten transitions in the 5-brane web for the $ \mathrm{SU}(2)+1\mathbf{F} $ theory given in Figure~\ref{fig:SU2-1F}. The correspondences between the solutions to the blowup equations and the web diagrams in Figure~\ref{fig:SU2-1F} are given as follows:
\begin{alignat}{2}
    \begin{aligned}\label{eq:SU2-1F-fluxd}
    &B_{m_1} = 1/2 \, , \ &&B_{m_0} = \left\{ \begin{array}{ll}
            -5/4 \text{ and } -1/4 & \quad \cdots\text{ Figure~\ref{fig:SU2-1F}(a)} \\
            -1/4 \text{ and } 3/4 & \quad \cdots\text{ Figure~\ref{fig:SU2-1F}(c)} \\
            3/4 \text{ and } 7/4 & \quad \cdots\text{ Figure~\ref{fig:SU2-1F}(e)}
    \end{array} \right.  \\
        &B_{m_1} = -1/2 \, , \ &&B_{m_0} = \left\{ \begin{array}{ll}
            -3/4 \text{ and } 1/4 & \quad \cdots\text{ Figure~\ref{fig:SU2-1F}(b)} \\
            1/4 \text{ and } 5/4 & \quad \cdots\text{ Figure~\ref{fig:SU2-1F}(d)} \\
            5/4 \text{ and } 9/4 & \quad \cdots\text{ Figure~\ref{fig:SU2-1F}(f).}
    \end{array} \right.
    \end{aligned}
\end{alignat}
In each case, the $ \Lambda $ factor for the defect blowup equation \eqref{eq:blowupDefect} is the same as that of the corresponding case in \eqref{eq:SU2-1F-Lambda}, with a shift $ m_0 \to m_0 + \epsilon_2/2 $.

To verify this correspondence, we first compute the codimension-2 defect partition functions using the blowup equations and then confirm the result by showing they satisfy the quantum difference equations. The explicit expressions for the 1-instanton solutions of the blowup equations are as follows. For the magnetic fluxes given in \eqref{eq:SU2-1F-fluxd}(a), we compute
\begin{align}
    \Psi_1 &= Z_1 - \frac{q_1 \left(q_1q_2 \chi_{3/2}(\phi) - M \sqrt{q_1q_2} (q_1q_2 \chi_{1}(\phi)-1) - \chi_{1/2}(\phi) \right)e^{-x}}{\sqrt{M} q_2 (1-q_1)(1-q_1q_2 e^{-2\phi}) (1-q_1q_2e^{2\phi}) (q_1-e^{-x-\phi}) (q_1-e^{-x+\phi})} \nonumber \\
    &\quad + \frac{\left( q_1q_2 \chi_{1}(\phi) - M (q_1q_2)^{3/2} \chi_{1/2}(\phi) - 1 \right) e^{-2x}}{\sqrt{M} q_2 (1-q_1)(1-q_1q_2 e^{-2\phi}) (1-q_1q_2e^{2\phi}) (q_1-e^{-x-\phi}) (q_1-e^{-x+\phi})} \, ,
\end{align}
where $ Z_1 $ is the 1-instanton partition function of $ \mathrm{SU}(2)+1\mathbf{F} $ without a defect, $ M=e^{-m_1} $ and $ \chi_{j}(\phi) $ is the character of the spin $ j $ representation. Similarly, we find
\begin{align}
    \Psi_1 &= Z_1 - \frac{q_1^{5/2} q_2^{1/2} \left(\sqrt{q_1q_2} \chi_{1/2}(\phi) - M(1+q_1q_2) \right) e^{-x}  }{\sqrt{M} (1-q_1)(1-q_1q_2 e^{-2\phi}) (1-q_1q_2e^{2\phi}) (q_1-e^{-x-\phi}) (q_1-e^{-x+\phi})} \nonumber \\
    &\quad - \frac{ q_1 \left(M \sqrt{q_1q_2} \chi_{1/2}(\phi) - (1+q_1q_2) \right) e^{-2x}}{\sqrt{M} (1-q_1)(1-q_1q_2 e^{-2\phi}) (1-q_1q_2e^{2\phi}) (q_1-e^{-x-\phi}) (q_1-e^{-x+\phi})} \, ,
\end{align}
for \eqref{eq:SU2-1F-fluxd}(c); and
\begin{align}
    \Psi_1 &= Z_1 + \frac{q_1^3 q_2 \left(M \sqrt{q_1q_2} (\chi_{1}(\phi)-q_1q_2) - \chi_{1/2}(\phi) \right) e^{-x}}{\sqrt{M} (1-q_1)(1-q_1q_2 e^{-2\phi}) (1-q_1q_2e^{2\phi}) (q_1-e^{-x-\phi}) (q_1-e^{-x+\phi})}  \nonumber \\
    &\quad - \frac{ q_1^2 q_2 \left(M \sqrt{q_1q_2} (\chi_{3/2}(\phi) - q_1q_2 \chi_{1/2}(\phi)) - \chi_{1}(\phi) + q_1q_2 \right) e^{-2x}}{\sqrt{M} (1-q_1)(1-q_1q_2 e^{-2\phi}) (1-q_1q_2e^{2\phi}) (q_1-e^{-x-\phi}) (q_1-e^{-x+\phi})} \, ,
\end{align}
for \eqref{eq:SU2-1F-fluxd}(e). For these three cases, we find the following difference equations that is satisfied by the normalized defect partition functions $ \hat{\Psi}^{\mathrm{NS}} = \lim_{\epsilon_2\to 0} \hat{\Psi} / \hat{Z} $ in the NS-limit:
\begin{align}
    \text{(a)} \, &: \,  E \hat{\Psi}^{\mathrm{NS}} = \left( X + X^{-1} + Y + \mathfrak{q} \frac{q_1}{\sqrt{M}} X^{-2} Y^{-1} (1 + \sqrt{q_1} M X + XY) \right) \hat{\Psi}^{\mathrm{NS}}\, , \\
    \text{(c)} \, &: \,  E \hat{\Psi}^{\mathrm{NS}} = \left( X + X^{-1} + Y + \frac{\mathfrak{q}}{\sqrt{q_1}} \left( \sqrt{M} X + \sqrt{\frac{q_1}{M}} \right) Y^{-1} \right) \hat{\Psi}^{\mathrm{NS}}  \, , \\
    \text{(e)} \, &: \,  E \hat{\Psi}^{\mathrm{NS}} = \left( X + X^{-1} + Y + \frac{\mathfrak{q}}{\sqrt{M}} X Y^{-1} (X+Y) ( 1 + \sqrt{q_1} M (X+Y)) \right) \hat{\Psi}^{\mathrm{NS}} ,
\end{align}
where $ E = \langle \hat{W} \rangle|_{\epsilon_2=0} $ is the NS-limit of the Wilson loop expectation value. We checked these difference equations up to 2-instanton order. One can see that these difference equations are the quantum SW curves associated to the toric diagrams given in Figure~\ref{fig:SU2-1F}(a), (c) and (e), respectively. 

For the other remaining fluxes, we find
\begin{align}
    \Psi_1 &= Z_1 + \frac{ q_1^{3/2} \left( M (q_1q_2 \chi_{1}(\phi)-1) - (q_1q_2)^{3/2} \chi_{1/2}(\phi) \right) e^{-x}  }{\sqrt{q_2 M} (1-q_1)(1-q_1q_2 e^{-2\phi}) (1-q_1q_2e^{2\phi}) (q_1-e^{-x-\phi}) (q_1-e^{-x+\phi})} \nonumber \\
    &\quad - \frac{ q_1 \left( M \sqrt{q_1q_2} \chi_{1/2}(\phi) - (1+q_1q_2) \right) e^{-2x}  }{\sqrt{M} (1-q_1)(1-q_1q_2 e^{-2\phi}) (1-q_1q_2e^{2\phi}) (q_1-e^{-x-\phi}) (q_1-e^{-x+\phi})}
\end{align}
for \eqref{eq:SU2-1F-fluxd}(b);
\begin{align}
    \Psi_1 &= Z_1 - \frac{ q_1^{5/2} q_2^{1/2} \left( \sqrt{q_1q_2} \chi_{1/2}(\phi) - M(1+q_1q_2) \right) e^{-x}  }{\sqrt{M} (1-q_1)(1-q_1q_2 e^{-2\phi}) (1-q_1q_2e^{2\phi}) (q_1-e^{-x-\phi}) (q_1-e^{-x+\phi})} \nonumber \\
    &\quad + \frac{ q_1^{3/2} q_2^{1/2} \left( \sqrt{q_1q_2} \chi_{1}(\phi) - M \chi_{1/2}(\phi) - (q_1q_2)^{3/2} \right) e^{-2x}  }{\sqrt{M} (1-q_1)(1-q_1q_2 e^{-2\phi}) (1-q_1q_2e^{2\phi}) (q_1-e^{-x-\phi}) (q_1-e^{-x+\phi})}
\end{align}
for \eqref{eq:SU2-1F-fluxd}(d); and
\begin{align}
    \Psi_1 &= Z_1 - \frac{ q_1^{7/2} q_2^{3/2} \left( \sqrt{q_1q_2} (\chi_{3/2}(\phi) - q_1q_2 \chi_{1/2}(\phi)) - M(\chi_{1}(\phi) - q_1q_2) \right) e^{-x}  }{\sqrt{M} (1-q_1)(1-q_1q_2 e^{-2\phi}) (1-q_1q_2e^{2\phi}) (q_1-e^{-x-\phi}) (q_1-e^{-x+\phi})} \nonumber \\
    &\quad + \frac{ q_1^{5/2} q_2^{3/2} \left( \sqrt{q_1q_2} (\chi_{2}(\phi) - q_1q_2 \chi_{1}(\phi)) - M (\chi_{3/2}(\phi) - q_1q_2 \chi_{1/2}(\phi)) \right) e^{-2x}  }{\sqrt{M} (1-q_1)(1-q_1q_2 e^{-2\phi}) (1-q_1q_2e^{2\phi}) (q_1-e^{-x-\phi}) (q_1-e^{-x+\phi})}
\end{align}
for \eqref{eq:SU2-1F-fluxd}(f). These defect partition functions satisfy the difference equations
\begin{align}
    \text{(b)} \ &: \  E \hat{\Psi}^{\mathrm{NS}} = \left( X + X^{-1} + Y + \frac{\mathfrak{q}}{\sqrt{M}} \left( 1 + \sqrt{q_1} M X^{-1} \right) Y^{-1} \right) \hat{\Psi}^{\mathrm{NS}}  \, , \\
    \text{(d)} \ &: \  E \hat{\Psi}^{\mathrm{NS}} = \left( X + X^{-1} + Y + \frac{\mathfrak{q}}{\sqrt{M}} \left( \left( X + \sqrt{q_1} M \right) Y^{-1} X + X \right) \right) \hat{\Psi}^{\mathrm{NS}} \, , \\
    \text{(f)} \ &: \  E \hat{\Psi}^{\mathrm{NS}} = \left( X + X^{-1} + Y  \phantom{+\frac{\mathfrak{q}}{\sqrt{M}}} \right. \\
    & \qquad \qquad \qquad \left. + \: \frac{\mathfrak{q}}{\sqrt{M}} \left( XY^{-1} (q_1(X+Y)+\sqrt{q_1}M)(X-Y) + 1 \right) (X+Y) \right) \hat{\Psi}^{\mathrm{NS}}\, , \nonumber
\end{align}
which are the quantization of the Seiberg-Witten curves associated to the toric diagrams in Figures~\ref{fig:SU2-1F}(b), (d), and (f). We checked these equations up to 2-instanton order.

\subsection{Local \texorpdfstring{$\mathbb{P}^2$}{P2}}

Our next example is the 5d SCFT engineered by M-theory compactified on a local $ \mathbb{P}^2 $, which is also called as $ E_0 $ theory or $ \mathcal{O}(-3) \to \mathbb{P}^2 $. Its partition function without a defect on the $ \Omega $-background can be computed using the blowup equation \cite{Huang:2017mis, Kim:2020hhh} with the effective prepotential
\begin{align}
    \mathcal{E}_0 = \frac{1}{\epsilon_1 \epsilon_2} \left( \frac{9}{6} \phi^3 - \frac{\epsilon_1^2 + \epsilon_2^2}{8} \phi + \epsilon_+^2 \phi \right) \, ,
\end{align}
and the consistent magnetic fluxes
\begin{align}\label{eq:P2-flux}
    n \in \mathbb{Z} \pm 1/6 \, .
\end{align}
The result for the first few orders is given by
\begin{align}\label{eq:P2-Z}
    Z(\phi,\epsilon_1, \epsilon_2) = e^{\mathcal{E}_0} \PE\left[ \frac{\sqrt{q_1q_2}}{(1-q_1)(1-q_2)} \left( \chi_1(\epsilon_+) e^{-3\phi} - \chi_{5/2}(\epsilon_+) e^{-6\phi} + \mathcal{O}(e^{-9\phi}) \right)  \right] \, .
\end{align}

As this theory is non-Lagrangian and does not admit a mass deformation leading to a gauge theory realization, we instead construct Wilson loop operators in M-theory, where they correspond to M2-branes wrapping non-compact 2-cycles of the local Calabi-Yau manifold. Following \cite{Kim:2021gyj}, one can compute the Wilson loop expectation value of the local $ \mathbb{P}^2 $ theory using the blowup equations, and the result is
\begin{align}\label{eq:P2-W}
    \langle W \rangle = e^\phi - \chi_{1/2}(\epsilon_+) e^{-2\phi} + \chi_2(\epsilon_+) e^{-5\phi} + \mathcal{O}(e^{-8\phi}) \, .
\end{align}
As explained in Section~\ref{subsec:5dscft}, the local $ \mathbb{P}^2 $ theory can be engineered by integrating out an instantonic hypermultiplet in the 5d $ \mathrm{SU}(2)_\pi $ gauge theory. Thus, the partition function \eqref{eq:P2-Z} and the Wilson loop VEV \eqref{eq:P2-W} can be obtained from the instanton partition function and the Wilson loop VEV of the $ \mathrm{SU}(2)_\pi $ theory through an RG-flow.

We now examine a codimension-2 defect in the local $ \mathbb{P}^2 $ theory. At leading order in the K\"ahler parameter expansion, the defect partition function is given by
\begin{align}
    \Psi / Z = e^{\mathcal{E}_1} \PE\left[ \frac{e^{-x}}{1-q_1} \left( e^\phi + \mathcal{O}(e^{0\phi} \right) + \mathcal{O}(e^{-2x}) \right] \, .
\end{align}
Here, we regard $ \mathcal{E}_1 $ as a constant that does not depend on the dynamical parameter $ \phi $. This is possible because the only term in $ \mathcal{E}_1 $ of the $ \mathrm{SU}(2)_\pi $ theory that depends on the dynamical parameter is $ \phi^2/2\epsilon_1 $, which only affects the renormalization of the mass parameter and decouples in the RG-flow limit to the local $ \mathbb{P}^2 $ theory. The constant $ \mathcal{E}_1 $ is irrelevant in the blowup equation since it can be absorbed into the $ \Lambda $ factor.

Now, we choose two sets of consistent magnetic fluxes
\begin{align}
    n \in \mathbb{Z} \pm 1/6 \, , \quad
    B_x = \pm 1/6 \, ,
\end{align}
to construct the blowup equations in the presence of the codimension-2 defect. We compute the defect partition function using the two blowup equations and the ansatz for the defect partition function \eqref{eq:Psi-index}. We present some non-vanishing $ D_{s_l,s_r}^{(d_1,d_2)} $ in Table~\ref{table:P2-invariant}.
\begin{table}
    \scriptsize
    \centering
    \begin{subtable}[b]{0.2\linewidth}
        \centering
        \begin{tabular}{|c|c|} \hline
            $ 2s_l \setminus 2s_r $ & 0  \\ \hline
            0 & 1 \\ \hline
        \end{tabular}
        \caption{$ x-\phi $}
    \end{subtable}
    \begin{subtable}[b]{0.25\linewidth}
        \centering
        \begin{tabular}{|c|ccc|} \hline
            $ 2s_l \setminus 2s_r $ & -1 & 0 & 1 \\ \hline
            0 & 1 & 0 & 1 \\ \hline
        \end{tabular}
        \caption{$ x+2\phi $}
    \end{subtable}
    \begin{subtable}[b]{0.47\linewidth}
        \centering
        \begin{tabular}{|c|ccccccccc|} \hline
            $ 2s_l \setminus 2s_r $ & -4 & -3 & -2 & -1 & 0 & 1 & 2 & 3 & 4 \\ \hline
            0 & 1 & & 1 & & 1 & & 1 & & 1 \\ \hline
        \end{tabular}
        \caption{$ x+5\phi $}
    \end{subtable}

    \begin{subtable}[b]{0.95\linewidth}
        \centering
        \begin{tabular}{|c|ccccccccccccccccc|} \hline
            $ 2s_l \setminus 2s_r $ & -8 & -7 & -6 & -5 & -4 & -3 & -2 & -1 & 0 & 1 & 2 & 3 & 4 & 5 & 6 & 7 & 8 \\ \hline
            -1 & 1 & & 1 & & 1 & & 1 & & 1 & & 1 & & 1 & & 1 & & 1 \\
            0 & & 1 & & 2 & & 2 & & 2 & & 2 & & 2 & & 2 & & 1 & \\
            1 & 1 & & 1 & & 1 & & 1 & & 1 & & 1 & & 1 & & 1 & & 1 \\ \hline
        \end{tabular}
        \caption{$ x+8\phi $}
    \end{subtable}

    \begin{subtable}[b]{0.2\linewidth}
        \centering
        \begin{tabular}{|c|c|} \hline
            $ 2s_l \setminus 2s_r $ & 0 \\ \hline
            -1 & 1 \\ \hline
        \end{tabular}
        \caption{$ 2x+\phi $}
    \end{subtable}
    \begin{subtable}[b]{0.4\linewidth}
        \centering
        \begin{tabular}{|c|ccccccc|} \hline
            $ 2s_l \setminus 2s_r $ & -3 & -2 & -1 & 0 & 1 & 2 & 3 \\ \hline
            -1 & 1 & & 1 & & 1 & & 1 \\ \hline
        \end{tabular}
        \caption{$ 2x+4\phi $}
    \end{subtable}

    \begin{subtable}[b]{0.75\linewidth}
        \centering
        \begin{tabular}{|c|ccccccccccccccc|} \hline
            $ 2s_l \setminus 2s_r $ & -7 & -6 & -5 & -4 & -3 & -2 & -1 & 0 & 1 & 2 & 3 & 4 & 5 & 6 & 7 \\ \hline
            -2 & 1 & & 1 & & 1 & & 1 & & 1 & & 1 & & 1 & & 1 \\ 
            -1 & & 1 & & 2 & & 2 & & 2 & & 2 & & 2 & & 1 & \\
            0 & & & & & & & & & & & & & & & 1 \\ \hline
        \end{tabular}
        \caption{$ 2x+7\phi $}
    \end{subtable}
    \begin{subtable}[b]{0.2\linewidth}
        \centering
        \begin{tabular}{|c|c|} \hline
            $ 2s_l \setminus 2s_r $ & -1 \\ \hline
            -2 & 1 \\ \hline
        \end{tabular}
        \caption{$ 3x $}
    \end{subtable}

    \begin{subtable}[b]{0.31\linewidth}
        \centering
        \begin{tabular}{|c|ccccc|} \hline
            $ 2s_l \setminus 2s_r $ & -2 & -1 & 0 & 1 & 2 \\ \hline
            -2 & 1 & & 1 & & 1 \\ \hline
        \end{tabular}
        \caption{$ 3x+3\phi $}
    \end{subtable}
    \begin{subtable}[b]{0.68\linewidth}
        \centering
        \begin{tabular}{|c|ccccccccccccc|} \hline
            $ 2s_l \setminus 2s_r $ & -6 & -5 & -4 & -3 & -2 & -1 & 0 & 1 & 2 & 3 & 4 & 5 & 6  \\ \hline
            -3 & 1 & & 1 & & 1 & & 1 & & 1 & & 1 & & 1 \\
            -2 & & 1 & & 2 & & 2 & & 2 & & 2 & & 1 & \\
            -1 & & & & & & & & & & & & & 1 \\ \hline
        \end{tabular}
        \caption{$ 3x+6\phi $}
    \end{subtable}
    \caption{OV-invariants $ D_{s_l,s_r}^{(d_1,d_2)} $ of codimension-2 defect in local $ \mathbb{P}^2 $ for $ d_1 \leq 3 $, $ d_2 \leq 8 $.} \label{table:P2-invariant}
\end{table}%
The same result can also be derived from the RG-flow of the codimension-2 defect partition function in the $ \mathrm{SU}(2)_\pi $ theory discussed in Section~\ref{subsec:su2}. In the limit $ \epsilon_{1,2} \to 0 $, $ \Psi $ reproduces the open topological string computation in \cite{Lerche:2001cw}.

We also find the difference equation from the defect partition function $ \Psi $. Here, we choose $ \mathcal{E}_1 = (x^2 - x \epsilon_1 - 2\pi i x) / 2\epsilon_1 $ for the effective prepotential. Then the defect partition function we obtained satisfies the difference equation
\begin{align}
    \left( X^{-1} + Y + q_1^{-1/2} X Y^{-1} \right) \hat{\Psi}^{\mathrm{NS}} = E \hat{\Psi}^{\mathrm{NS}} \, ,
\end{align}
where $ X=e^{-x} $, $ Y = e^{x \partial_x} $ and $ E = \langle \hat{W} \rangle|_{\epsilon_2=0} $ is the NS-limit of the Wilson loop VEV \eqref{eq:P2-W}. This difference equation is the quantum Seiberg-Witten curve associated to the toric diagram given in Figure~\ref{fig-localP2-toric}(a).

\subsection{M-string} \label{sec:Mstr}

In this subsection, we consider codimension-2 defects and their blowup equations in the 6d $ \mathcal{N}=(2,0) $ $ A_1 $ theory. Upon circle compactification, this theory gives rise to the 5d KK theory, which at low energy is described by the $ \mathrm{SU}(2)_0 + 1\mathbf{Adj} $ theory. The effective prepotential of this theory is 
\begin{align}\label{eq:Mstr-E0}
    \mathcal{E}_0 = \frac{1}{\epsilon_1 \epsilon_2} \left( \tau \phi^2 + \phi \left(\epsilon_+^2 - m^2 \right) \right) \, ,
\end{align}
where $ \tau = 1/R $ is the inverse radius of the 6d circle, $ \phi $ is the 6d tensor parameter, and $ m $ is the mass parameter of the 6d R-symmetry. In 5d perspective, $ \tau $ and $ m $ are identified with the inverse gauge coupling squared and mass parameter of the adjoint hypermultiplet, respectively, and $ \phi $ corresponds to the Coulomb parameter. The blowup equations can be constructed using the consistent magnetic fluxes
\begin{align}\label{eq:Mstr-flux}
    n \in \mathbb{Z} \, ,\ \mathbb{Z}+1/2 \, , \quad
    B_\tau = 0 \, , \quad  B_m = \pm 1/2 \, .
\end{align}
The blowup equations can be solved to compute the partition function of the M-string theory  \cite{Gu:2019pqj, Kim:2020hhh}, which can also be obtained using the supersymmetric localization method outlined in Appendix~\ref{app:ADHM}. The partition function of the M-string theory is given by
\begin{align}\label{eq:Mstr-Z}
    Z = e^{\mathcal{E}_0} \left( 1 + \sum_{k=0}^\infty e^{-2k\phi} Z_k \right)
    = e^{\mathcal{E}_0} \left( 1 + e^{-2\phi} \frac{\theta_1(\tau, m \pm \epsilon_+)}{\theta_1(\tau,\epsilon_{1,2})} + \mathcal{O}(e^{-4\phi}) \right) \, ,
\end{align}
where $ Z_k $ is the $ k $-string elliptic genus, $ \theta_i $ is the Jacobi theta function defined in Appendix~\ref{app:special} and we use shorthand notation $ \theta_1(\tau, m\pm \epsilon_+) \equiv \theta_1(\tau, m+\epsilon_+) \theta_1(m-\epsilon_+) $. In this expression, we omit the perturbative contribution from the tensor multiplet which does not contain dynamical tensor parameter $ \phi $. 

The VEV of the fundamental Wilson surface operator can be also computed using the blowup equations \cite{Kim:2021gyj}. Here, we present the result up to 1-string order as
\begin{align}
    \langle W \rangle = e^{\phi} + e^{-\phi} \frac{\theta_1(m \pm \epsilon_-) - \theta_1(m \pm \epsilon_+)}{\theta_1(\tau,\epsilon_{1,2})} + \mathcal{O}(e^{-3\phi}) \, .
\end{align}

\begin{figure}
    \centering
    \begin{subfigure}[b]{0.47\linewidth}
        \small
        \centering
        \begin{tabular}{c|cccccccccc}
            & 0 & 1 & 2 & 3 & 4 & 5 & 6 & 7 & 8 & 9 \\ \hline
            NS5 & $ \!\times\! $ & $ \!\times\! $ & $ \!\times\! $ & $ \!\times\! $ & $ \!\times\! $ & $ \!\times\! $  \\
            D6 & $ \!\times\! $ & $ \!\times\! $ & $ \!\times\! $ & $ \!\times\! $ & $ \!\times\! $ & $ \!\times\! $ & $ \!\times\! $  \\
            D2 & $ \!\times\! $ & & & & & $ \!\times\! $ & $ \!\times\! $ \\
            D4 & $ \!\times\! $ & $ \!\times\! $ & $ \!\times\! $ & &  & $ \!\times\! $ & & $ \!\times\! $
        \end{tabular}
        \caption{}
    \end{subfigure}
    \hfill
    \begin{subfigure}[b]{0.49\linewidth}
        \centering
        \includegraphics{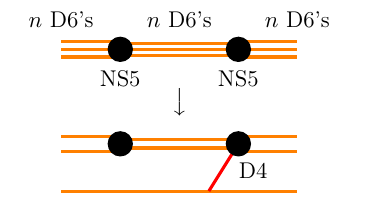}
        \caption{}
    \end{subfigure}
    \caption{(a) is the brane configurations for 6d $ \mathrm{SU}(n) + 2n\mathbf{F} $ theory. (b) represents the Higgsing of $ \mathrm{SU}(n) + 2n\mathbf{F} $ to $ \mathrm{SU}(n-1) + 2(n-1)\mathbf{F} $ theory by moving one D6-brane along the $ x^{7} $ direction. The codimension-2 defect is realized by a D4-brane between NS5-brane and D6-brane on the Higgs branch.} \label{fig:Mstr-brane}
\end{figure}

Now, consider a codimension-2 defect studied in \cite{Chen:2020jla}. This defect operator is introduced by Higgsing the 6d $ \mathcal{N}=(1,0) $ SCFT, which describes the worldvolume theory of two M5-branes probing the A-type singularty $ \mathbb{C}^2 / \mathbb{Z}_n $. The type IIA brane configuration of the worldvolume theory consists of $ 2 $ NS5-branes and $ n $ D6-branes as summarized in Figure~\ref{fig:Mstr-brane}. The NS5-branes and D6-branes become M5-branes and ALE space in the M-theory uplift, respectively. The self-dual BPS strings in the 6d theory are realized by D2-branes suspended between NS5-branes. 

\begin{figure}
    \centering
    \includegraphics[align=c]{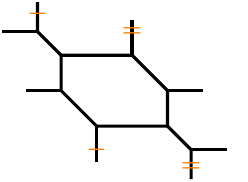}
    {\boldmath$ \to $}
    \includegraphics[align=c]{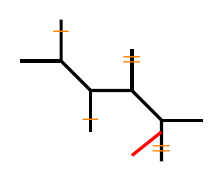}
    \caption{
    A 5-brane web for Higgsing 6d $ \mathrm{SU}(2)+4\mathbf{F} $ theory to M-string theory with a defect, which is equivalent to Figure~\ref{fig:Mstr-brane} by compactifying the theory on a circle along the $x^5$ direction and taking the T-duality along the circle direction. T-duality maps D6-branes to D5-branes and the defect D4-brane to D3-brane represented in the red line on this 5-brane web where the vertical direction represents the $ x^5 $ direction.} \label{fig:Mstr-brane2}
\end{figure}

The elliptic genera of the strings capture the BPS spectrum of the 6d theory. On the tensor branch, the worldvolume theory has an $ \mathrm{SU}(n) $ gauge theory description coupled to $ 2n $ fundamental hypermultiplets. This theory admits a mesonic Higgs branch deformation, which corresponds to shifting one D6-brane into the perpendicular direction relative to the other D6-branes and NS5-branes. As depicted in Figure~\ref{fig:Mstr-brane}(b), one may introduce an additional D4-brane between the shifted D6-brane  and NS5-brane. This D4-brane realizes a codimension-2 defect in the 6d theory. We will focus on the Higgsing of the 6d $ \mathrm{SU}(2) + 4\mathbf{F} $ theory to $ n=1 $ case, which has an enhanced $ \mathcal{N}=(2,0) $ supersymmetry with trivial gauge symmetry. The brane web for this theory, which is T-dual to Figure~\ref{fig:Mstr-brane}, is depicted in Figure~\ref{fig:Mstr-brane2}.

Let us construct the blowup equations for the codimension-2 defect using the Higgsing. The effective prepotential of the 6d $ \mathrm{SU}(2)+4\mathbf{F} $ theory is given by
\begin{align}\label{eq:6dSU2E}
    \mathcal{E}_0^{\mathrm{SU}(2)} \!=\! \frac{1}{\epsilon_1 \epsilon_2} \Bigg(\! \tau \phi_0^2 \!+\! 2\phi_0 \bigg( \epsilon_+^2 \!+\! \phi_1^2 \!-\! \frac{1}{4}\sum_{l=1}^4 m_l^2 \bigg) \!+\! \frac{4}{3}\phi_1^3 \!-\! \frac{1}{12}\sum_{l=1}^4 (\phi_1 \pm m_l)^3 \!+\! \frac{\epsilon_1^2 \!+\! \epsilon_2^2}{12} \phi_1 \!+\! \epsilon_+^2 \phi_1 \! \Bigg) ,
\end{align}
where $ \phi_1 $ is the $ \mathrm{SU}(2) $ gauge holonomy and $ m_{1,2,3,4} $ are chemical potentials for the $ \mathrm{SU}(4) $ flavor symmetry satisfying $ \sum m_i=0 $. The full partition function can be computed via localization method summarized in Appendix~\ref{app:ADHM}. The usual Higgs branch deformation that leads to the M-string theory is achieved by giving a VEV to the mesonic operator. In terms of fugacities, this Higgsing is realized by
\begin{align}\label{eq:M-usualHiggs}
    (\sqrt{q_1q_2} e^{-(\phi_1-m_3)}) \cdot (\sqrt{q_1q_2} e^{\phi_1-m_4})=1 \, ,
\end{align}
where the two factors in the LHS correspond to contributions from the chiral fields in the fundamental hypermultiplets which together form a mesonic operator. If one tunes the chemical potentials in the partition function as
\begin{align}
    \phi_0 \to \phi \, , \quad
    \phi_1 \to x \, , \quad
    m_{1,2} \to \pm m - x \, , \quad
    m_3 \to \epsilon_+ + x \, , \quad
    m_4 \to -\epsilon_+ + x \, ,
\end{align}
where $ x $ is a free parameter, then the effective prepotential \eqref{eq:6dSU2E} reduces to the M-string effective prepotential \eqref{eq:Mstr-E0} up to the terms independent of the dynamical parameter $ \phi $. One can also check that the full partition function of $ \mathrm{SU}(2)+4\mathbf{F} $ reduces to the M-string partition function up to overall Goldstone mode contributions and non-dynamical constant factor.

On the other hand, Higgsing mesonic operators by giving a position dependent VEV realizes the codimension-2 defect. At the level of partition function, we parametrize the chemical potentials as \cite{Chen:2020jla}
\begin{gather}\label{eq:Mstr-Higgs-parameter}
    \begin{gathered}
        \phi_0 \to \phi \, , \quad
        \phi_1 \to x + \epsilon_+ \, , \quad
        m_1 \to m - x - \epsilon_+ \, , \\
        m_2 \to -m - x - \epsilon_+ + \epsilon_2 \, , \quad
        m_3 \to x + 2\epsilon_+ \, , \quad
        m_4 \to x - \epsilon_2 \, ,
    \end{gathered}
\end{gather}
where $ x $ is related to the $ U(1) $ holonomy associated to the defect, and \eqref{eq:M-usualHiggs} changes to
\begin{align}
    q_2 \cdot (\sqrt{q_1q_2} e^{-(\phi_1-m_3)}) \cdot (\sqrt{q_1q_2} e^{\phi_1-m_4})=1 \, .
\end{align}
For conveninence, we rescale $ x \to x/2 $ and then shift $ x \to x - m - \epsilon_- $. The effective prepotential \eqref{eq:6dSU2E} becomes
\begin{align}\label{eq:Mstr-E1}
    \mathcal{E}^{\mathrm{SU}(2)} \to \mathcal{E}_0 + \mathcal{E}_1 \, , \quad 
    \mathcal{E}_1 = \frac{x \phi}{\epsilon_1} \, ,
\end{align}
under this parametrization up to constant terms that do not depend on $ \phi_0 $, where $ \mathcal{E}_0 $ is the M-string effective prepotential \eqref{eq:Mstr-E0}. We interpret $ \mathcal{E}_1 $ as the contribution from the codimension-2 defect to the effective prepotential.

We construct the blowup equations starting from the additional effective prepotential $ \mathcal{E}_1 $. The defect partition function of the M-string theory on the tensor branch can be written as
\begin{align}\label{eq:Mstr-defect}
    \Psi(\phi, m, x, \epsilon_1, \epsilon_2) = e^{\mathcal{E}_0 + \mathcal{E}_1} \left( 1 + \sum_{k=0}^\infty e^{-2k\phi} \Psi_k(m, x, \epsilon_1, \epsilon_2) \right) \, , 
\end{align}
where $ \Psi_k $ is the $ k $-string elliptic genus in the presence of the codimension-2 defect, and we ignore the overall perturbative contributions which does not depend on the dynamical parameter $ \phi $. For the partition function \eqref{eq:Mstr-defect}, we construct the blowup equation using the consistent magnetic fluxes
\begin{align}
    n \in \mathbb{Z} \, ,\ \mathbb{Z}+1/2 \, , \quad
    B_\tau = 0 \, , \quad  B_m = -1/2 \, , \quad
    B_x = 0 \, .
\end{align}
For each set of magnetic fluxes, the blowup equation can be written as\footnote{Here, we shift $ \epsilon_1 \to \epsilon_1+2\pi i $ in the blowup equations to express them in terms of unhatted partition functions.}
\begin{align}\label{eq:Mstr-blowup}
    \left\{\begin{array}{ll}
            \displaystyle \sum_{k_1+k_2=k} \theta_3(2\tau, x \!+\! \epsilon_+ \!-\! m \!-\! 2k_1 \!-\! 2k_2) \Psi_{k_1}^{(N)} Z_{k_2}^{(S)} = \theta_3(2\tau, x + \epsilon_+ - m) \Psi_k \quad &(n \in \mathbb{Z})  ,  \vspace{1ex}\\
            \displaystyle \sum_{k_1+k_2=k} \theta_2(2\tau, x \!+\! \epsilon_+ \!-\! m \!-\! 2k_1 \!-\! 2k_2) \Psi_{k_1}^{(N)} Z_{k_2}^{(S)} = \theta_2(2\tau, x + \epsilon_+ - m) \Psi_k \quad &(n \in \mathbb{Z} + 1/2)  .
        \end{array}\right.
\end{align}
The theta function factor on the LHS comes from the effective prepotential, while the theta function in RHS is the $ \Lambda $ factor. The 1-string order of the blowup equations using this result yields
\begin{align}\label{eq:Mstr-def-Psi1}
    \Psi_1 = \frac{\theta_1(\tau, m + \epsilon_-) \theta_1(\tau, m - \epsilon_+) \theta_1(\tau, x - m - \epsilon_+)}{\theta_1(\tau, \epsilon_1) \theta_1(\tau, \epsilon_2) \theta_1(\tau, x-m-\epsilon_-)} \, .
\end{align}
Note that Higgsing the 1-string elliptic genus of the 6d $ \mathrm{SU}(2)+4\mathbf{F} $ theory leads to 
\begin{align}\label{eq:Mstr-def-Psi1H}
    \Psi_1 = \frac{\theta_1(\tau, m\!+\!\epsilon_+) \theta_1(\tau, m\!-\!\epsilon_+\!-\!\epsilon_2) \theta_1(\tau, x\!-\!m\!+\!\epsilon_+)}{\theta_1(\tau, \epsilon_1) \theta_1(\tau, \epsilon_2) \theta_1(\tau, x\!-\!m\!+\!\epsilon_+\!+\!\epsilon_2)} + \frac{\theta_1(\tau, 2\epsilon_+) \theta_1(\tau, x) \theta_1(\tau,x \!-\! 2m \!+\! \epsilon_2)}{\theta_1(\tau, \epsilon_1) \theta_1(\tau, x\!-\!m \!\pm\! \epsilon_+ \!+\! \epsilon_2)} \, .
\end{align}
We checked that two results \eqref{eq:Mstr-def-Psi1} and \eqref{eq:Mstr-def-Psi1H} agree with each other up to order $ e^{-10\tau} $. We further verified that the two 2-string elliptic genera $ \Psi_2 $ in the presence of the codimension-4 defect, computed using both the blowup equations and Higgsing, match perfectly up to order $ e^{-5\tau} $ order.

From the solutions to the blowup equations, one checked that the normalized defect partition function $ \Psi^{\mathrm{NS}} = \lim_{\epsilon_2\to 0} \Psi/Z $ of the M-string theory in the NS-limit satisfies the difference equation
\begin{align}\label{eq:Mstr-diff1}
    \left(Y + \frac{\theta_1(\tau, x) \theta_1(\tau, x - 2m)}{\theta_1(\tau, x - m + \frac{\epsilon_1}{2}) \theta_1(\tau, x - m - \frac{\epsilon_1}{2})} Y^{-1}\right) \Psi^{\mathrm{NS}} = E \Psi^{\mathrm{NS}} \, ,
\end{align}
where $ Y $ is the momentum operator satisfying $ YX = q_1 XY $ and $ E = \langle W \rangle_{\epsilon_2=0} $ is the NS-limit of the Wilson loop VEV. This elliptic difference equation is related to the integrable model called the 2-body elliptic Ruijsenaars-Schneider system \cite{Bullimore:2014awa}.

\paragraph{S-transformation}

Another type of codimension-2 defects can be constructed by coupling a 3d hypermultiplet to the 5d $ \mathrm{SU}(2)_0 + 1\mathbf{Adj} $ theory. This type of the defect is related to the one obtained from Higgsing the 6d $ \mathrm{SU}(2)+4\mathbf{F} $ theory through the S-transformation \cite{Bullimore:2014awa}. Here, we study the blowup equations for the 3d hypermultiplet defect. Instead of using the tensor branch expansion of the M-string partition function \eqref{eq:Mstr-Z}, let us use the 5d Coulomb branch expansion:
\begin{align}\label{eq:Mstr-5d-Z}
    Z(\phi,m,\tau,\epsilon_{1,2}) = e^{\mathcal{E}_0} Z_{\mathrm{pert}}^{\mathrm{5d}}(\phi,m,\epsilon_{1,2}) \left( 1 + \sum_{k=1}^\infty \mathfrak{q}^k Z_k^{\mathrm{5d}}(\phi,m,\epsilon_{1,2}) \right) \, .
\end{align}
Here, $ \tau $ and $ m $ in the effective prepotential \eqref{eq:Mstr-E0} are interpreted as the inverse gauge coupling and the mass parameter of the adjoint hypermultiplet in the 5d maximally supersymmetric $ \mathrm{SU}(2) $ Yang-Mills theory, respectively, and $ \mathfrak{q}=e^{-\tau} $ is the 5d instanton fugacity. The 1-loop perturbative partition function $ Z_{\mathrm{pert}} $ is the contributions of the vector and adjoint hypermultiplet given by
\begin{align}
    Z_{\mathrm{pert}}^{\mathrm{5d}} &= \PE\Bigg[-\frac{1+q_1q_2}{(1-q_1)(1-q_2)} e^{-2\phi} + \frac{\sqrt{q_1q_2}}{(1-q_1)(1-q_2)} \big( e^{-(2\phi + m)} + e^{-(2\phi-m)} \big) \Bigg] \, ,
\end{align}
and $ Z_k^{\mathrm{5d}} $ is the $ k $-instanton partition function that can be computed using the ADHM construction of the instanton moduli space summarized in Appendix~\ref{app:ADHM-Mstr}.

The 3d hypermultiplet consists of two chiral multiplets, each transforming in the fundamental and antifundamental representations of the 5d $ \mathrm{SU}(2) $ gauge symmetry. We write the partition function in the presence of the defect as
\begin{align}\label{eq:Mstr-5d-psi}
    \Psi^{\mathrm{5d}}(\phi,m,\tau,x,\epsilon_{1,2}) = e^{\mathcal{E}_0} \Psi_{\mathrm{pert}}^{\mathrm{5d}}(\phi,m,x,\epsilon_{1,2}) \left( 1 + \sum_{k=1}^\infty \mathfrak{q}^k \Psi_k^{\mathrm{5d}}(\phi,m,x,\epsilon_{1,2}) \right) \, ,
\end{align}
where the perturbative contribution from the 3d hypermultiplet is given by
\begin{align}\label{eq:Mstr-def2-pert}
    \frac{\Psi_{\mathrm{pert}}}{Z_{\mathrm{pert}}} &= \PE\Bigg[ \bigg(\frac{ e^{-(m+\epsilon_+)/2}}{1-q_1} \!-\! \frac{e^{-(3\epsilon_+-m)/2}}{1-q_1} \bigg) e^{-x}(e^{\phi}\!+\!e^{-\phi}) \Bigg] = \frac{(M^{-1/2}(q_1q_2)^{3/4} XQ^{\pm 1}; q_1)_\infty}{(M^{1/2}(q_1q_2)^{1/4} XQ^{\pm 1}; q_1)_\infty} .
\end{align}
Here, $ x $ is the chemical potential for the $ \mathrm{U}(1) $ flavor symmetry associated to the defect and $ X=e^{-x} $. The instanton partition function $ \Psi_k^{\mathrm{5d}} $ can be computed via ADHM construction \cite{Gaiotto:2014ina, Bullimore:2014awa} given in Appendix~\ref{app:ADHM-Mstr}. 

One can perform the S-transformation by gauging the $ \mathrm{U}(1) $ flavor symmetry for the defect as
\begin{align}
    \Psi(x') = \oint ds \, e^{-sx'/\epsilon_1} \Psi^{\mathrm{5d}}(x=s) \, .
\end{align}
The residue integral, similar to that considered for the 5d $ \mathrm{SU}(2)_\theta $ gauge theory in Section~\ref{subsubsec:su2S}, reproduces the defect partition function $ \Psi $ derived from the Higgsing of the 6d $ \mathrm{SU}(2)+4\mathbf{F} $ theory, up to constant factors that do not contain the dynamical parameter $ \phi $.

However, the blowup equations for this defect is rather different from the usual blowup equations. Since the regularization factor of the infinite product in the 1-loop part \eqref{eq:Mstr-def2-pert} does not depend on the dynamical parameter $ \phi $, it suffices to consider only $ \mathcal{E}_0 $ when constructing the blowup equations. We consider two sets of magnetic fluxes
\begin{align}
    B_\tau = 0 \, , \quad B_m = -1/2 \, , \quad
    \left\{\begin{array}{ll}
            n \in \mathbb{Z} \, , & B_x = 1/2 \\
            n \in \mathbb{Z}+1/2 \, ,\ & B_x = 0 \, .
        \end{array}\right.
\end{align}
Using the instanton partition function $ \Psi_k^{\mathrm{5d}} $ computed by the ADHM construction, we find that the $ \Lambda $ factor in the blowup equation \eqref{eq:blowupDefect} is not a simple constant, but an operator that includes the shift operator $ Y = e^{\epsilon_1 \partial_x} $. For the first set of magnetic fluxes, $ n \in \mathbb{Z} $ and $ B_x=1/2 $, the blowup equation is given by
\begin{align}
    &\sum_{n\in \mathbb{Z}} \mathfrak{q}^{n^2} e^{-n(m-\epsilon_+)} \Psi_{\mathrm{pert}}^{\mathrm{5d},(N)} Z_{\mathrm{pert}}^{\mathrm{5d},(S)} \Psi_{\mathrm{inst}}^{\mathrm{5d},(N)} Z_{\mathrm{inst}}^{\mathrm{5d},(S)} = \sum_{n \in \mathbb{Z}} \mathfrak{q}^{n^2} e^{-n(m-\epsilon_+)} Y^{-n} \Psi_{\mathrm{pert}}^{\mathrm{5d}} \Psi_{\mathrm{inst}}^{\mathrm{5d}} \, ,
\end{align}
whereas for the second set of magnetic fluxes, when $ n\in \mathbb{Z}+1/2 $ and $ B_x=0 $, the blowup equation is
\begin{align}
    &\sum_{n\in \mathbb{Z}+\frac{1}{2}} \mathfrak{q}^{n^2} e^{-n(m-\epsilon_+)} \Psi_{\mathrm{pert}}^{\mathrm{5d},(N)} Z_{\mathrm{pert}}^{\mathrm{5d},(S)} \Psi_{\mathrm{inst}}^{\mathrm{5d},(N)} Z_{\mathrm{inst}}^{\mathrm{5d},(S)}  = \sum_{n \in \mathbb{Z}+\frac{1}{2}} \mathfrak{q}^{n^2} e^{-n(m-\epsilon_+)} Y^{-n-1/2} \Psi_{\mathrm{pert}}^{\mathrm{5d}} \Psi_{\mathrm{inst}}^{\mathrm{5d}} \, .
\end{align}
Here, we denote $ \Psi_{\mathrm{inst}} $ and $ Z_{\mathrm{inst}} $ as the instanton part in \eqref{eq:Mstr-5d-psi} and \eqref{eq:Mstr-5d-Z}. 

The first factor on LHS comes from the effective prepotential, while the terms involving $ Y $ on RHS are interpreted as the $ \Lambda $ factor. We checked these blowup equations up to 2-instanton order. Using the Jacobi theta functions, the $ \Lambda $ factors on RHS can be recasted as
\begin{align}
    \Lambda = \left\{\begin{array}{ll}
            \displaystyle \theta_3(2\tau, m-\epsilon_+ + \epsilon_1 \partial_x) & \quad (n \in \mathbb{Z}, \ B_x=1/2) \\
            \displaystyle e^{-(\epsilon_1/2) \partial_x} \theta_2(2\tau, m-\epsilon_+ + \epsilon_1 \partial_x) & \quad (n \in \mathbb{Z}+1/2,\ B_x=0) \, .
        \end{array}\right.
\end{align}
We remark that these $ \Lambda $ factors are related to the $ \Lambda $'s in \eqref{eq:Mstr-blowup} by exchanging $ x $ and the shift operator $ \epsilon_1 \partial_x $. This is a consequence of the S-transformation.

The difference equation \eqref{eq:Mstr-diff1} is also changed due to the S-transformation. The  normalized defect partition function $ \Psi^{\mathrm{NS}}_{\mathrm{5d}}=\lim_{\epsilon_2\to 0} \Psi^{\mathrm{5d}}/Z $ in the NS-limit satisfies
\begin{align}\label{eq:Mstr-diff2}
    \left( X \frac{\theta_1(\tau, y - 2m + \epsilon_1)}{\theta_1(\tau, y - m + \frac{3}{2}\epsilon_1)} + X^{-1} \frac{\theta_1(\tau, y)}{\theta_1(\tau, y - m - \frac{\epsilon_1}{2})} \right) \Psi^{\mathrm{NS}}_{\mathrm{5d}} = E \Psi^{\mathrm{NS}}_{\mathrm{5d}} \, ,
\end{align}
where $ Y=e^{-y}=e^{\epsilon_1 \partial_x} $ is the momentum operator satisfying $ YX = q_1 XY $ and $ E = \langle W \rangle_{\epsilon_2=0} $ is the NS-limit of the fundametal Wilson loop VEV. This curve is related to the one in \eqref{eq:Mstr-diff1} by coordinate transformations. In the classical limit $ \epsilon_1 \to 0 $, redefining $ Y \to Y \frac{\theta_1(x-2m)}{\theta_1(x-m)} $ and then exchanging $ x $ and $ y $ in \eqref{eq:Mstr-diff1} yields the curve in \eqref{eq:Mstr-diff2}.

\subsection{E-string}

In this subsection, we consider the codimension-2 defect and its blowup equations for the E-string theory, which is UV-dual to the 5d $ \mathrm{Sp}(1) $ gauge theory with 8 fundamental hypermultiplets. The effective prepotential of the E-string theory is
\begin{align}\label{eq:Estr-E0}
    \mathcal{E}_0 = \frac{1}{\epsilon_1 \epsilon_2} \left( \frac{\tau}{2}\phi^2 + \phi \bigg( \frac{\epsilon_1^2 + \epsilon_2^2}{4} + \epsilon_+^2 - \frac{1}{2}\sum_{l=1}^8 m_l^2 \bigg) \right) \, ,
\end{align}
where $ \phi $ is the 6d tensor parameter and $ m_l $ are eight chemical potentials for the $ E_8 $ global symmetry. The partition function is given by
\begin{align}\label{eq:Estr-Z}
    Z = e^{\mathcal{E}_0} \left( 1 + \sum_{k=0}^\infty e^{-k \phi} Z_k \right) \, ,
\end{align}
where $ Z_k $ is the $ k $-string elliptic genus that can be computed from the ADHM construction \cite{Kim:2014dza}, and we omit the perturbative contribution from the tensor multiplet, which does not depend on the dynamical parameter $ \phi $. This partition function satisfies the blowup equation with the set of the consistent magnetic fluxes
\begin{align}\label{eq:Estr-flux}
    n \in \mathbb{Z} + 1/2 \, , \quad
    B_\tau = 0 \, , \quad
    B_{m_i} \in \Phi_{E_8} \, ,
\end{align}
where $ \Phi_{E_8} $ is the $ E_8 $ root system in the orthogonal basis \cite{Gu:2019pqj, Kim:2020hhh}. These blowup equations are also useful for studying the partition function in the presence of the codimension-4 defect \cite{Kim:2021gyj}. We summarize explicit formulae for the elliptic genus $ Z_k $ and the VEV of the codimension-4 Wilson surface defect in the fundamental representation in Appendix~\ref{app:ADHM-Sp}.

\begin{figure}
    \centering
    \begin{subfigure}[b]{0.51\linewidth}
        \small
        \centering
        \begin{tabular}{c|cccccccccc}
            & 0 & 1 & 2 & 3 & 4 & 5 & 6 & 7 & 8 & 9 \\ \hline
            NS5 & $ \!\times\! $ & $ \!\times\! $ & $ \!\times\! $ & $ \!\times\! $ & $ \!\times\! $ & $ \!\times\! $  \\
            D6/O6 & $ \!\times\! $ & $ \!\times\! $ & $ \!\times\! $ & $ \!\times\! $ & $ \!\times\! $ & $ \!\times\! $ & $ \!\times\! $ \\
            D4 & $ \!\times\! $ & $ \!\times\! $ & $ \!\times\! $ & $ \!\times\! $ &  & & & $ \!\times\! $
        \end{tabular}
        \caption{}
    \end{subfigure}
    \hfill
    \begin{subfigure}[b]{0.45\linewidth}
        \centering
        \includegraphics{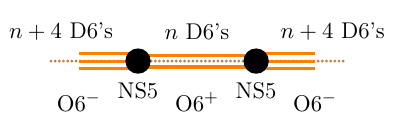}
        \caption{}
    \end{subfigure}

    \begin{subfigure}[b]{0.8\linewidth}
        \centering
        \includegraphics{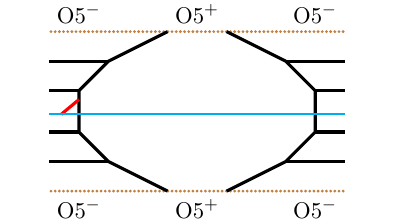}
        \caption{}
    \end{subfigure}
    \caption{(a) and (b) are IIA brane configurations for 6d $ \mathrm{Sp}(n) + (2n+8)\mathbf{F} $ theory. Higgsing to $ \mathrm{Sp}(n-1) + (2n+6)\mathbf{F} $ theory can be performed by moving one D6-brane along the $ x^{7} $ direction. The codimension-2 defect is a D4-brane between NS5-brane and the D6-brane on the Higgs branch. (c) is the dual IIB brane configuration after Higgsing, where the red line denotes the D3-brane realizing the codimension-2 defect.} \label{fig:Estr-brane}
\end{figure}

We are interested in the codimension-2 defect of the E-string theory obtained by Higgsing the 6d $ \mathrm{Sp}(1) $ gauge theory coupled to 10 fundamental hypermultiplets \cite{Kim:2020npz,Chen:2021ivd}. In type IIA picture, the $ \mathrm{Sp}(N) $ gauge theory is described by the NS5/D6/O6 brane system summarized in Figures~\ref{fig:Estr-brane}(a) and (b). The Higgsing process is realized by moving D6-brane along the orthogonal direction to Figure~\ref{fig:Estr-brane}(b). One may introduce a defect D4-brane between the NS5-brane and the Higgsed D6-brane, and it realizes the codimension-2 defect of the Higgsed theory. We also present type IIB brane picture after performing T-duality in Figure~\ref{fig:Estr-brane}(c).

The effective prepotential of the 6d $ \mathrm{Sp}(1) + 10\mathbf{F} $ theory is
\begin{align}\label{eq:sp1-10F-E0}
    \mathcal{E}_0^{\mathrm{Sp}(1)} = \frac{1}{\epsilon_1\epsilon_2} & \Bigg( \frac{\tau}{2}\phi_0^2 + \phi_0 \bigg(\frac{\epsilon_1^2+\epsilon_2^2}{4} + 2\epsilon_+^2 + \phi_1^2 - \frac{1}{2}\sum_{l=1}^{10} m_l^2 \bigg) \\
    &\quad + \frac{1}{6} \bigg( 8\phi_1^3 - 2\tau\phi_1^2 - \frac{1}{2}\sum_{l=1}^{10} (\phi_1 \pm m_l)^3 - \tau\phi_1^2 \bigg) + \frac{\epsilon_1^2+\epsilon_2^2}{3}\phi_1 + \epsilon_+^2 \phi_1 \Bigg) \, , \nonumber
\end{align}
where $ \phi_0 $ is the tensor parameter, $ \phi_1 $ is the $ \mathrm{Sp}(1) $ gauge holonomy, and $ m_l $ are the chemical potentials for ten hypermultiplets. The elliptic genus can be obtained from the ADHM computation, and we summarize it in Appendix~\ref{app:ADHM-Sp}. The usual Higgsing from the $ \mathrm{Sp}(1) $ gauge theory to the E-string theory without a defect is achieved by
\begin{align}\label{eq:Estr-Higgs-usual}
    \phi_0 \to \phi \, , \quad
    \phi_1 \to x \, , \quad
    m_9 \to x + \epsilon_+ \, , \quad
    m_{10} \to -x + \epsilon_+ \, ,
\end{align}
where $ x $ is a free parameter. The effective prepotential \eqref{eq:sp1-10F-E0} becomes \eqref{eq:Estr-E0} up to the terms independent of the dynamical parameter $ \phi $, under these identifications. It can be also verified that the full partition function of the 6d $ \mathrm{Sp}(1)+10\mathbf{F} $ theory reduces to the partition function of the E-string theory up to an overall constant factor including the Goldstone boson modes \cite{Kim:2020npz,Chen:2021ivd}.

To obtain a codimension-2 defect, we give a position dependent VEV to the mesonic operator built from the chiral fields in the hypermultiplet. At the level of partition function, we parameterize the chemical potentials as
\begin{align}\label{eq:Estr-defectHiggs}
    \phi_0 \to \phi \, , \quad
    \phi_1 \to x + \frac{\epsilon_2}{2} \, , \quad
    m_9 \to x + \epsilon_+ + \frac{\epsilon_2}{2} \, , \quad
    m_{10} \to -x + \epsilon_+ + \frac{\epsilon_2}{2} \, .
\end{align}
Unlike the usual Higgsing, the partition function now depends on $ x $, which is interpreted as the defect parameter. Using these identifications, we obtain the effective prepotential of the E-string theory in the presence of the codimension-2 defect as
\begin{align}\label{eq:Estr-defectE}
    \mathcal{E}_0^{Sp(1)} \to \mathcal{E}_0 + \mathcal{E}_1 \, , \quad
    \mathcal{E}_1 = \frac{(x-\epsilon_+)\phi}{\epsilon_1} \, ,
\end{align}
up to constant terms that do not depend on $ \phi $. 

The defect partition function of E-string on the tensor branch is expressed by
\begin{align}\label{eq:Estr-Psi}
    \Psi(\phi,m_l,x,\epsilon_1,\epsilon_2)
    = e^{\mathcal{E}_0+\mathcal{E}_1} \left( 1 + \sum_{k=1}^\infty e^{-k\phi} \Psi_k(m_l,x,\epsilon_1,\epsilon_2) \right) \, ,
\end{align}
where $ \Psi_k $ is the $ k $-string elliptic genus in the presence of the codimension-2 defect, and we omit the perturbative part which is independent of the dynamical parameter $ \phi $. Higgsing the 6d $ \mathrm{Sp}(1) + 10\mathbf{F} $ partition functions using \eqref{eq:Estr-defectHiggs} yields, for instance,
\begin{align}
    \Psi_1 = \sum_{I=1}^4 \frac{\prod_{l=1}^8 \theta_I(\tau, m_l)}{2\eta(\tau)^6 \theta_1(\tau, \epsilon_1) \theta_1(\tau,\epsilon_2)} \frac{\theta_I(\tau, x- \epsilon_1/2- \epsilon_2)}{\theta_I(\tau,x-\epsilon_1/2)} \, ,
\end{align}
at 1-string order. The explicit expression for the 2-string elliptic genus is given in Appendix~\ref{app:ADHM-Sp}. Using this result, we find two blowup equations for the defect partition function \eqref{eq:Estr-Psi} constructed from the effective prepotential \eqref{eq:Estr-defectE} and the following sets of magnetic fluxes:
\begin{align}\label{eq:Estr-def-flux}
    \left\{\begin{array}{l}
            \displaystyle n \in \mathbb{Z} + 1/2 \, , \quad
            B_{\tau} = 0 \, , \quad
            B_x = 0 \, , \quad
            B_{m_i} = \delta_{ij} \ \text{ for } 1 \leq j \leq 8 \, , \\
            \displaystyle n \in \mathbb{Z} + 1/2 \, , \quad
            B_\tau = 0 \, , \quad
            B_x = -\frac{1}{2} \, , \quad
            B_{m_i} = (-1)^{1+\delta_{ij}} \frac{1}{2} \ \text{ for } 1 \leq j \leq 8 \, .
    \end{array}\right.
\end{align}
Here, $ \delta_{ij} $ denotes the Kronecker delta. The blowup equation from the first set of magnetic fluxes at the $ k $-string order is given by
\begin{align}\label{eq:Estr-def-blowup1}
    \sum_{k_1+k_2=k} \theta_1(\tau, x + B_m \cdot m + \tfrac{\epsilon_1}{2} - k_1 \epsilon_1 - k_2 \epsilon_2) \Psi_{k_1}^{(N)} Z_{k_2}^{(S)}
    = \theta_1(\tau, x+B_m \cdot m + \tfrac{\epsilon_1}{2}) \Psi_k \, ,
\end{align}
where $ B_m \cdot m = \sum_l B_{m_l} m_l $ denotes the inner product. The theta function on the LHS arises from the effective prepotential \eqref{eq:Estr-defectE}, while $ \Lambda = \theta_1(\tau,x + B_m \cdot m + \frac{\epsilon_1}{2}) $ from the RHS.

On the other hand, the blowup equation coming from the second set of magnetic fluxes in \eqref{eq:Estr-def-flux} is more complicated. In this case, $ \Lambda $ is not a simple constant, but an operator given by
\begin{equation}
    \Lambda = \theta_1(\tau, x + B_m \cdot m + \epsilon_+) -  \frac{\theta_1(\tau, x - B_m \cdot m - \epsilon_1 - \epsilon_+) \prod_{l=1}^8 \theta_1(\tau,x + 2B_{m_l} m_l - \frac{\epsilon_1}{2})}{\eta(\tau)^6 \theta_1(\tau,2x-2\epsilon_1) \theta_1(\tau,2x-\epsilon_1)} Y^{-1} ,
\end{equation}
where $ Y $ is the shift operator $ Yf(x) = f(x+\epsilon_1) $. At the $ k $-string order, this blowup equation can be expressed as
\begin{align}\label{eq:Estr-def-blowup2}
    \sum_{k_1+k_2=k} & \theta_1(\tau, x + B_m\cdot m + \epsilon_+ - k_1 \epsilon_1 - k_2 \epsilon_2) \Psi_{k_1}^{(N)} Z_{k_2}^{(S)}
    = \theta_1(\tau, x + B_m \cdot m + \epsilon_+) \Psi_k(x) \nonumber \\
    & \quad - \frac{\theta_1(\tau, x - B_m \cdot m - \epsilon_+ - \epsilon_1) \prod_{l=1}^8 \theta_1(\tau, x + 2B_lm_l - \tfrac{\epsilon_1}{2})}{\eta(\tau)^6 \theta_1(\tau,2x-2\epsilon_1) \theta_1(\tau,2x-\epsilon_1)} \Psi_{k-1}(x-\epsilon_1) \, , 
\end{align}
where the $ (k-1) $-string contribution in the second line appears due to the action of $ Y^{-1} $ on $ \mathcal{E}_1 $. We check that $ \Lambda $ depends solely on the non-dynamical parameters and the shift operator $ Y $ up to 3-string order.

Unlike the blowup equations for M-string theory, we were unable to determine the defect partition function $ \Psi_k $ using these blowup equations. There are two major difficulties in solving them. First, the blowup equation \eqref{eq:Estr-def-blowup1} is not sufficient to fully determine the defect partition function, in contrast to the partition functions without a defect or with a codimension-4 defect. The index form \eqref{eq:Psi-index} is also insufficient to constrain the defect partition function, due to the lack of $ \mathrm{SU}(2)_l \times SU(2)_r $ symmetry. Second, we do not know how to determine the $ \Lambda $ factor involving $ Y $ in the other blowup equation \eqref{eq:Estr-def-blowup2} without prior knowledge of the defect partition function. These difficulities also arise in the blowup equations considered in Section~\ref{subsec:su2pi-adj} and \ref{subsec:P2adj}.

Nevertheless, it is possible to find a difference equation that annihilates the normalized defect partition function $ \Psi^{\mathrm{NS}} = \lim_{\epsilon_2 \to 0} \Psi/Z $ in the NS-limit:
\begin{equation}\label{eq:Estr-diffeq}
    \left( Y + V_0(x) + \frac{\prod_{l=1}^8 \theta_1(\tau,x\pm m_l -\epsilon_1/2)}{\eta(\tau)^{12} \theta_1(\tau,2x) \theta_1(\tau,2x-\epsilon_1)^2 \theta_1(\tau,2x-2\epsilon_1)} Y^{-1} \right) \Psi^{\mathrm{NS}} = E \Psi^{\mathrm{NS}} .
\end{equation}
Here, $ E = \langle W \rangle|_{\epsilon_2=0} $ is the VEV of the codimension-4 Wilson surface defect and $ V_0(x) $ is a constant factor depending on the non-dynamical parameters and $ x $. Their expressions are provided in Appendix~\ref{app:ADHM-Sp}. This eigensystem is related to the van Diejen integrable model \cite{Diejen:1994, Nazzal:2018brc, Chen:2021ivd} which is a generalization of the relativistic Calogero-Moser model.

\subsection{\texorpdfstring{$\mathrm{SU}(2)_{\pi}+1\mathbf{Adj}$}{SU(2)π+1Adj}} \label{subsec:su2pi-adj}

It was argued in \cite{Hayashi:2023boy,Kim:2023qwh,Kim:2024vci} that the 5d $ \mathrm{Sp}(N) $ gauge theory coupled to 1 antisymmetric and 8 fundamental hypermultiplets is related to the 5d $ \mathrm{Sp}(N)_\pi + 1\mathbf{Adj} $ theory by tuning the mass parameters to special values. When $ N=1 $, the former theory is the KK-theory of E-strings and the latter is UV dual to the $ \mathbb{Z}_2 $ twisted compactification of the 6d $ \mathcal{N}=(2,0) $ $ A_2 $ theory. In this subsection, we study the defect partition function and blowup equations for the $ \mathrm{SU}(2)_\pi + 1\mathbf{Adj} $ theory using this relation.

Let us begin with the partition function of the E-string theory in  \eqref{eq:Estr-Z} and consider the following map, which is called the \emph{freezing}:
\begin{align}\label{eq:mapsthetapi6d}
    \begin{aligned}
        &m_{1} \mapsto \frac{m+\epsilon_+}{2},\quad
        m_{2} \mapsto \frac{m-\epsilon_+}{2},\quad
        m_{3} \mapsto \frac{m+\epsilon_-}{2},\quad
        m_{4} \mapsto \frac{m-\epsilon_-}{2},\\
        &m_{5}\mapsto\frac{m+\epsilon_+}{2}+\pi i,\quad
        m_{6}\mapsto\frac{m-\epsilon_+}{2}+\pi i,\quad
        m_{7}\mapsto\frac{m+\epsilon_-}{2}+\pi i,\\
        &m_{8}\mapsto \pm \frac{m-\epsilon_-}{2}+\pi i+\tau, \quad
        \phi \mapsto \phi \pm \frac{m - \epsilon_- + \tau}{2} \, .
    \end{aligned}
\end{align}
If we choose the plus sign in the last line, the E-string partition function becomes M-string partition function, which is equivalent to the instanton partition function of the 5d $ \mathrm{SU}(2)_0 + 1\mathbf{Adj} $ theory. Here, $ m $ is interpreted as the mass parameter of the adjoint hypermultiplet. Additionally, it is possible to reproduce the Wilson loop VEV in the fundamental representation of $ \mathrm{SU}(2) $, starting from the codimension-4 defect partition function of the E-string theory. On the other hand, choosing the minus sign in the last line of the mapping \eqref{eq:mapsthetapi6d} yields the partition function of $ \mathbb{Z}_2 $ twist of the rank-2 M-string theory. 

The effective prepotential becomes
\begin{align}
    \mathcal{E}_0 = \frac{1}{\epsilon_1 \epsilon_2} \left( \frac{\tau}{2}\phi^2 + \phi \left( \epsilon_+^2 - m^2 \right) \right) \, ,
\end{align}
where $ \phi $ and $ m $ are identified to the chemical potential for $ \mathrm{SU}(2) $ gauge symmetry and the mass parameter of the adjoint hypermultiplet in dual 5d theory. Its partition function is given by
\begin{align}\label{eq:su2piadj-Z}
    Z = e^{\mathcal{E}_0} \left( 1 + \sum_{k=1}^\infty e^{-k\phi} Z_k \right) \, .
\end{align}
where we find
\begin{align}\label{eq:su2piadj-Z1}
    Z_1 = -\frac{\theta_4(2\tau,0)^4 \theta_1(2\tau, m \pm \epsilon_+) \theta_1(2\tau, m \pm \epsilon_-)}{\eta(\tau)^6 \theta_1(\tau,\epsilon_1) \theta_1(\tau,\epsilon_2)} \, ,
\end{align}
at 1-string order. The 2-string elliptic genus is given in Appendix~\ref{app:piadj}. This result matches  the GV-invariant computation given in \cite{Kim:2020hhh}. The mapping \eqref{eq:mapsthetapi6d} can be also applied to obtain the codimension-4 defect partition function, and its explicit expression is also summarized in Appendix~\ref{app:piadj}. The partition functions with/without the codimension-4 defect satisfy the blowup equations with sets of the consistent magnetic fluxes
\begin{align}
    n \in \mathbb{Z} + 1/2 \, , \quad
    B_\tau = 0 \, , \quad
    B_m = \pm 1/2 \, .
\end{align}

We now consider the codimension-2 defect in the $ \mathrm{SU}(2)_\pi + 1\mathbf{Adj} $ theory. Given that the mapping \eqref{eq:mapsthetapi6d} correctly reproduces the elliptic genus and the VEV of the codimension-4 defect, we expect that the SW curve, and consequently, the codimension-2 defect partition function of the E-string theory will also map to those of the $ \mathrm{SU}(2)_\pi+1\mathbf{Adj} $ theory. In addition to \eqref{eq:mapsthetapi6d}, we perform
\begin{align}\label{eq:xmappi}
    x \to x + \frac{\epsilon_2}{2} + \pi i 
\end{align}
to the codimension-2 defect partition function \eqref{eq:Estr-Psi} of the E-string theory. The $ \pi i $ shift in $ x $ is necessary to establish the correct pole structure for the defect partition function, and $ \epsilon_2 $ shift is performed for convenience. The contribution of the defect to the effective prepotential is
\begin{align}
    \mathcal{E}_1 = \frac{(x - \epsilon_1/2)\phi}{\epsilon_1} \, ,
\end{align}
and the defect partition function is given by
\begin{align}
    \Psi(\phi,\tau,m,x,\epsilon_1,\epsilon_2) = e^{\mathcal{E}_0+\mathcal{E}_1} \left( 1 + \sum_{k=1}^\infty e^{-k\phi} \Psi_k(\tau, m,x,\epsilon_1,\epsilon_2) \right) \, ,
\end{align}
where the 1-string elliptic genus in the presence of the codimension-2 defect is
\begin{align}
    \Psi_1 &= \sum_{I=1}^4 (-1)^{1+\delta_{I4}} \frac{\theta_4(2\tau,0)^4 \theta_{\mu(I)}(2\tau, m \pm \epsilon_+) \theta_{\mu(I)}(2\tau, m \pm \epsilon_-) \theta_I(x - \epsilon_+)}{2\eta(\tau)^6 \theta_1(\tau,\epsilon_{1}) \theta_1(\tau, \epsilon_2) \theta_I(x - \epsilon_-)} \, .
\end{align}
Here, $ \delta_{I4} $ is the Kronecker delta symbol, $ \mu(1)=\mu(2) = 1 $ and $ \mu(3)=\mu(4)=4 $. The expression for the 2-string elliptic genus is provided in Appendix~\ref{app:piadj}.

We then construct the blowup equation for the codimension-2 defect partition function using the magnetic flux
\begin{align}
    n \in \mathbb{Z} \, , \quad
    B_m = \pm 1/2 \, , \quad
    B_x = 0 \, .
\end{align}
This blowup equation is related to the E-string blowup equation \eqref{eq:Estr-def-blowup2} through the mappings \eqref{eq:mapsthetapi6d} and \eqref{eq:xmappi}. The $ \Lambda $ factor involves the operator $ Y $ as
\begin{align}\label{eq:su2-pi-adj-lambda}
    \Lambda = \theta_3(\tau,x+2B_mm+\tfrac{\epsilon_2}{2}) &- \frac{\theta_4(2\tau,0)^4 \theta_1(2\tau, 2x \!\pm\! m \!-\! \epsilon_- \!-\! \epsilon_1) \theta_1(2\tau, 2x \!+\! 2B_mm \!-\! 2\epsilon_- \!\pm\! \epsilon_-)}{\eta(\tau)^6 \theta_1(\tau,2x - \epsilon_1 + \epsilon_2) \theta_1(\tau, 2x - 2\epsilon_1 + \epsilon_2)} \nonumber \\
    &\qquad \cdot Y^{-1} \theta_3(\tau, x-2B_mm+\tfrac{\epsilon_2}{2}) \, .
\end{align}
The blowup equation at $ k $-string order is given by
\begin{equation}\label{eq:piadj-blowup}
    \sum_{k_1+k_2=k}\theta_3(\tau, x+2B_m m+\tfrac{\epsilon_2}{2}-k_1\epsilon_1-k_2\epsilon_2) \Psi_{k_1}^{(N)} Z_{k_2}^{(S)} = \theta_3(\tau,x+2B_mm+\tfrac{\epsilon_2}{2}) \Psi_k - \tilde{\Lambda} \Psi_{k-1} \, ,
\end{equation}
where the theta function in the LHS is obtained from the effective prepotential, and $ \tilde{\Lambda} $ is the second term of \eqref{eq:su2-pi-adj-lambda}. The $ (k-1) $-string partition function $ \Psi_{k-1} $ arises because the action of $ Y^{-1} $ on $ \mathcal{E}_1 $ generates an additional $ e^{-\phi} $ factor. We verified this blowup equation up to 2-string order. However, similar to the blowup equations in E-string theory, we do not have a systematic way to solve it.

From the E-string difference equation \eqref{eq:Estr-diffeq} and the defect partition function of $ \mathrm{SU}(2)_\pi + 1\mathbf{Adj} $ theory, we checked that the normalized defect partition function $ \Psi^{\mathrm{NS}} = \lim_{\epsilon_2 \to 0} \Psi/Z $ in the NS-limit satisfies the difference equation
\begin{align}\label{eq:su2pi-adj-diff}
    \left( Y + V_0(x) + \frac{\theta_4(2\tau,0)^8 \theta_1(2\tau,2x \pm m - \frac{\epsilon_1}{2})^2 \theta_1(2\tau,2x \pm m - \frac{3\epsilon_1}{2})^2}{\eta(\tau)^{12} \theta_1(\tau,2x) \theta_1(\tau,2x-\epsilon_1)^2 \theta_1(\tau,2x-2\epsilon_1)} Y^{-1} \right) \Psi^{\mathrm{NS}} = E \Psi^{\mathrm{NS}} \, ,
\end{align}
where $ E = \langle W \rangle|_{\epsilon_2\to0} $ is the NS-limit of the Wilson loop VEV and $ V_0 $ is the constant term independent of the dynamical parameter $ \phi $. The expression for $ V_0 $ is provided in Appendix~\ref{app:piadj}.

\subsection{\texorpdfstring{$\mathbb{P}^2$ + $1 {\bf Adj}$}{P2+1Adj}} \label{subsec:P2adj}

\begin{figure}
    \centering
    \includegraphics[align=c]{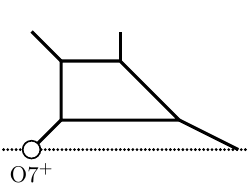}
    \hspace{-2ex}{\boldmath$ \rightarrow $}\hspace{0ex}
    \includegraphics[align=c]{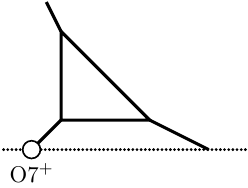}
    \hspace{-2ex}{\boldmath$ \leftrightarrow $}\hspace{-1ex}
    \includegraphics[align=c]{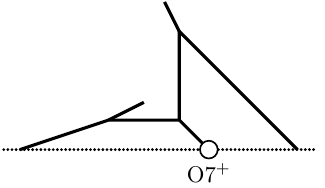}
    \caption{The left diagram is the 5-brane web for $ \mathrm{SU}(2)_\pi + 1\mathbf{Adj} $ theory. By integrating out the instantonic hypermultiplet, we obtain the 5-brane webs for local $ \mathbb{P}^2 + 1\mathbf{Adj} $ in the middle. The right diagram is obtained by flopping the ``adjoint'' hypermultiplet in the middle diagram.} \label{fig:P2adj}
\end{figure}

We now consider another rank-1 non-Lagrangian theory called the local $ \mathbb{P}^2 + 1\mathbf{Adj} $ theory \cite{Bhardwaj:2019jtr, Kim:2020hhh}. This theory can be obtained by integrating out an instantonic hypermultiplet in the $ \mathrm{SU}(2)_\pi + 1\mathbf{Adj} $ theory. A 5-brane web diagram for the $ \mathrm{SU}(2)_\pi + 1\mathbf{Adj} $ theory is engineered by attaching an orientifold $ \mathrm{O7}^+ $-plane to the web diagram of $ \mathrm{SU}(2)_\pi $ theory, as shown on the left in Figure~\ref{fig:P2adj}.\footnote{See also \cite{Kim:2023qwh} for another 5-brane web diagram of  $\mathbb{P}^2 + 1\mathbf{Adj}$ that is obtained from a Higgsing of SU($3)_\frac12+1\mathbf{Sym}$.} The $ (p,q) $-string between the lower D5-brane and the $ \mathrm{O7}^+ $-plane represents a hypermultiplet state in the adjoint representation. To obtain the local $ \mathbb{P}^2 + 1\mathbf{Adj} $ theory, we flop the instantonic hypermultiplet and integrate it out, similar to the RG-flow from $ \mathrm{SU}(2)_\pi $ to local $ \mathbb{P}^2 $ theory. As a result, we obtain the 5-brane web given in the middle of Figure~\ref{fig:P2adj}. The ``adjoint'' hypermultiplet coming from the $ (p,q) $-string between the D5-brane and the $ \mathrm{O7}^+ $-plane remains under this RG-flow, which distinguishes this theory from the local $ \mathbb{P}^2 $ theory. If we flop this adjoint hypermultiplet, we obtain the 5-brane web on the right in Figure~\ref{fig:P2adj}.

To derive its partition function on $ \Omega $-background, it is convenient to parametrize the partition function \eqref{eq:su2piadj-Z} in terms of
\begin{align}
    t_1 = 3\tilde{\phi} + \frac{\tau}{2} - 2 \tilde{m} \, , \quad
    t_2 = \tilde{m} - 2 \tilde{\phi}  \, , \quad
    t_3 = -\tilde{\phi} - \frac{\tau}{2} + 2 \tilde{m} \, .
\end{align}
where we used the notation $ \tilde{\phi} $ and $ \tilde{m} $ to distinguish the parameters in the local $ \mathbb{P}^2 + 1\mathbf{Adj} $ theory. In this parametrization, one can find a hypermultiplet with charge $ -t_3 $ in the 1-string partition function \eqref{eq:su2piadj-Z1} of the $ \mathrm{SU}(2)_\pi + 1\mathbf{Adj} $ theory. We flop this hypermultiplet and take the limit $ t_3 \to \infty $, while keeping $ t_1 $ and $ t_2 $ finite. Then we rewrite $ t_1 = 3\phi - 4m $ and $ t_2 = 3m-2\phi $, where $ m $ is the mass parameter of the non-Lagrangian theory. In this parametrization, the effective prepotential of the local $ \mathbb{P}^2 + 1\mathbf{Adj} $ theory is
\begin{align}\label{eq:P2adj-E0}
    \mathcal{E}_0 = \frac{1}{\epsilon_1 \epsilon_2} \left( \frac{\phi^3}{6} - m^2 \phi - \frac{\epsilon_1^2 + \epsilon_2^2}{24}\phi + \epsilon_+^2 \phi \right) \, ,
\end{align}
and the partition function is given by
\begin{align}\label{eq:P2adj-Z}
    Z &= e^{\mathcal{E}_0} \left(1 + \sum_{k=1}^\infty e^{-k\phi} Z_k \right)
\end{align}
where we find
\begin{align}
    Z_1 = \frac{(q_1+q_2)(1+q_1q_2)}{(1-q_1)(1-q_2)\sqrt{q_1q_2}} -\frac{(1+q_1)(1+q_2) (M+M^{-1})}{(1-q_1)(1-q_2)} \, ,
\end{align}
at $ e^{-\phi} $ order, with $ M=e^{-m} $. Similarly, we obtain the Wilson loop VEV as
\begin{align}
    \langle W \rangle
    &= e^\phi + \bigg( \! \big(2 + q_1^{\pm 1} + q_2^{\pm 1}\big) (M^2 + M^{-2}) - \frac{(1\!+\!q_1)(1\!+\!q_2)(q_1\!+\!q_2)(1\!+q_1q_2)}{q_1 q_2 \sqrt{q_1q_2}} (M+M^{-1}) \nonumber \\
    &\ + \frac{(q_1 \!+\! q_2 \!+\! q_1q_2 \!+\! q_1^2q_2 \!+\! q_1q_2^2) (q_1 \!+\! q_2 \!+\! 2q_1q_2 \!+\! q_1^2q_2 \!+\! q_1q_2^2)}{q_1^2q_2^2} \bigg) e^{-\phi} +  \mathcal{O}(e^{-2\phi}) .
\end{align}
These results can also be computed from the blowup equation with the effective prepotential \eqref{eq:P2adj-E0} and magnetic fluxes
\begin{align}
    n \in \mathbb{Z} + 1/2 \, , \quad
    B_m = \pm 1/2 \, .
\end{align}

We now discuss codimension-2 defects in this theory. We obtain the codimension-2 defect partition function from the defect partition function of $ \mathrm{SU}(2)_\pi + 1\mathbf{Adj} $ theory in a similar manner, but with an additional shift $ \phi \to \phi-\epsilon_2/2 $. The contribution from the defect to effective prepotential is given by
\begin{align}
    \mathcal{E}_1 = \frac{(x-\epsilon_1/2)(\phi-\epsilon_2/2)}{\epsilon_1} \, ,
\end{align}
and the defect partition function is
\begin{align}
    \Psi = e^{\mathcal{E}_0 + \mathcal{E}_1} \left( 1 + \sum_{k=1}^\infty e^{-k\phi} \Psi_k(m, x, \epsilon_1, \epsilon_2) \right) \, ,
\end{align}
where the defect partition function $ \Psi_1 $ at the first order can be written as
\begin{align}
    \Psi_1 &= \frac{-(1+q_1)(1+q_2)(M+M^{-1})}{(1-q_1)(1-q_2)} + \frac{q_1^2 - X^4q_2}{(1-q_1)\sqrt{q_1q_2} X^2} + \frac{(1+q_1)^2 \sqrt{q_2}}{(1-q_1)(1-q_2) \sqrt{q_1}} \\
    &\ + \frac{(q_1q_2(M^2+M^{-2}) + q_2+q_1^2q_2 + q_1(1+q_2)^2)X}{(1-q_1)(1-q_1^{-1}q_2X^2)q_1\sqrt{q_2}} - \frac{X(1+q_1)(1+q_2)(M+M^{-1})}{(1-q_1)(1-q_1^{-1}q_2X^2)\sqrt{q_1}}. \nonumber
\end{align}

This defect partition function serves as a solution to the blowup equations constructed from the magnetic fluxes
\begin{align}\label{eq:P2adj-def-flux}
    n \in \mathbb{Z} + 1/2 \, , \quad
    B_x = 0 \, , \quad
    B_m = \pm 1/2 \, .
\end{align}
These blowup equations are descendants of the $ \mathrm{SU}(2)_\pi + 1\mathbf{Adj} $ blowup equations \eqref{eq:piadj-blowup}. The $ \Lambda $ factor is an operator given as
\begin{align}\label{eq:P2adj-Lambda}
    \Lambda &= \frac{2\cosh(\frac{x+2B_m m+\epsilon_2/2}{2})}{(XM^{2B_m}\sqrt{q_2})^{1/2}} \\
    &\qquad + \frac{8\cosh(\frac{x-2B_mm-\epsilon_1 + \epsilon_2/2}{2})\sinh(\frac{2x \pm m - \epsilon_- - \epsilon_1)}{2}) \sinh(\frac{2x+2B_mm-2\epsilon_- \pm \epsilon_-}{2})}{(XM^{2B_m}\sqrt{q_2})^{1/2} \sinh(\frac{2x-\epsilon_1+\epsilon_2}{2}) \sinh(\frac{2x-2\epsilon_1+\epsilon_2}{2})} Y^{-1} \, , \nonumber
\end{align}
and the $ k $-th order of the blowup equation in $ e^{-\phi} $ expansion is given by
\begin{align}
    \sum_{\frac{l(l-1)}{2}+k_1+k_2=k} (-1)^{\frac{l(l-1)}{2}} \frac{(q_1q_2)^{\frac{l(l-1)(2l-1)}{12}}}{(X M^{2B_m}\sqrt{q_2})^{l} } \Psi_{k_1}^{(N)} Z_{k_2}^{(S)} 
    = \left( 1 + \frac{M^{-2B_m}}{X\sqrt{q_2}} \right) \Psi_k + \tilde{\Lambda} \Psi_{k-1} \, ,
\end{align}
where $ \tilde{\Lambda} $ is the second term of \eqref{eq:P2adj-Lambda} and $ l=n+1/2 \in \mathbb{Z} $ represents the dynamical magnetic flux in \eqref{eq:P2adj-def-flux}. We checked this blowup equation up to $ e^{-2\phi} $ order. However, as in the E-string and $ \mathrm{SU}(2)_\pi + 1\mathbf{Adj} $ theories, a systematic method for determining the $ \Lambda $ factor and solving this blowup equation is still lacking.

Lastly, we find the difference equation for the defect partition function. The normalized codimension-2 defect partition function $ \Psi^{\mathrm{NS}} = \lim_{\epsilon_2 \to 0} \Psi / Z $ in the NS-limit satisfies the trigonometric version of the difference equation \eqref{eq:su2pi-adj-diff} given by
\begin{align}\label{eq:P2adj-diff}
    \left( Y + V_0(x) + \frac{16\sinh^2(\frac{2x\pm m - \epsilon_1}{2}) \sinh^2(\frac{2x\pm m -3\epsilon_1/2}{2})}{\sinh(x) \sinh^2(\frac{2x-\epsilon_1}{2}) \sinh(\frac{2x-2\epsilon_1}{2})} Y^{-1} \right) \Psi^{\mathrm{NS}} = E \Psi^{\mathrm{NS}} \, ,
\end{align}
where $ E = \langle W \rangle|_{\epsilon_2=0} $ is the NS-limit of the Wilson loop VEV and
\begin{align}
    V_0(x) = 4\cosh\left(\frac{\epsilon_1}{2}\right) \cosh(2x) - \frac{8\sinh^2(\frac{m + \epsilon_1/2}{2}) \sinh^2(\frac{m - \epsilon_1/2}{2})}{\sinh(\frac{2x + \epsilon_1}{2}) \sinh(\frac{2x - \epsilon_1}{2})} \cosh(x)
\end{align}
is independent of the dynamical parameter $ \phi $. One can compare this result with the classical SW curve computed in \cite{Hayashi:2023boy}. In the classical limit $ \epsilon_1 \to 0 $, the difference operator reduces to the classical curve given by
\begin{align}\label{eq:classical limit of Local P^2}
    t + \left( 4\cosh(2x) - \frac{8\sinh^4(\tfrac{m}{2}) \cosh(x)}{\sinh^2(x)} \right) + \frac{16 \sinh^4(x\pm\frac{m}{2})}{\sinh^4(x)} \frac{1}{t} = u\ ,
\end{align}
where we define $ Y \to t $ and $ X \to w $ in this limit and $ u $ is the genus zero part of the Wilson loop VEV, $ u= \langle W \rangle|_{\epsilon_1=\epsilon_2=0} $. By rescaling $ t \to -t/(2\sinh x)^2 $ we obtain
\begin{align}
    \begin{aligned}
    t^2 + \left(\frac{(1-M)^4}{M^2} (w+w^{-1}) - 2(w-w^{-1})^2 \left(w^2+w^{-2} - \frac{u}{2}\right) \right) t \\
     + \frac{(w^2-M)^4 (w^{-2} - M)^4}{M^4} = 0 \, ,
    \end{aligned}
\end{align}
which is the same result as in \cite{Hayashi:2023boy}. Thus, our difference equation \eqref{eq:P2adj-diff} provides the quantization of the classical SW curve of the local $\mathbb{P}^2+1{\bf Adj}$ theory in  \cite{Hayashi:2023boy}.


\section{Conclusion}\label{sec:conclusion}

In this paper, we studied BPS partition functions in the presence of codimension-2 defects in 5d and 6d supersymmetric theories, focusing on their properties under various transitions in the moduli space. To achieve this, we first established blowup equations for the partition functions of rank-1 5d and 6d SCFTs with codimension-2 defects. We demonstrated how the codimension-2 defects arise from the Higgsing, and formulated the blowup equations that defect partition functions satisfy. Unlike in cases without a defect or with codimension-4 defects, a single blowup equation is not sufficient to uniquely determine the codimension-2 defect partition function. Instead, the blowup equation admits multiple solutions, which depend on the magnetic fluxes on the blown-up $ \mathbb{P}^1 $, and we were able to uniquely fix the defect partition function when blowup equations coming from two different magnetic fluxes share a common solution. These multiple solutions correspond to different defect partition functions related via the Hanany-Witten transition. This can be verified by the fact that the codimension-2 defect partition function is an eigenfunction of a difference equation, which represents the quantization of the classical Seiberg-Witten curve. Using this property, we present the quantum curves of rank-1 5d/6d SCFTs and how they change under the Hanany-Witten transitions.

There are several intriguing directions for future research and open questions that need further investigation. A natural extension of our work is to explore codimension-2 defects in higher-rank theories using the blowup equations. For example, the blowup equation provides a framework to determine codimension-2 defect partition functions and quantum curves for 5d SCFTs with SU(3) gauge symmetry and Chern-Simons level greater than 3, which correspond to non-toric theories. It would be particularly interesting to find new quantum curves and defect partition functions that remain unknown, such as 6d minimal $ \mathrm{SU}(3) $ theory, with the help of blowup equations.  We leave these generalizations for future work.

The study of quantum curves and $ \Omega $-deformed codimension-2 defect partition functions is deeply related to integrable systems. In particular, there have been recent developments for understanding tau functions of Painlev\'e integrable systems through the quantum SW curves and blowup equations of 4d $ \mathcal{N}=2 $ $ \mathrm{SU}(2) $ gauge theories and Argyres-Douglas theories \cite{Nekrasov:2020qcq, Bonelli:2024wha, Bonelli:2025owb}. Since 5d $ \mathrm{SU}(2) $ gauge theories are related to the $ q $-Painlev\'e equations \cite{Bershtein:2016aef,Bershtein:2017swf,Bonelli:2017gdk,Gamayun:2012ma,Gavrylenko:2016zlf}, it would be interesting to explore their tau functions using the blowup equations with a defect.

Another important open problem is the general structure of the blowup equations in the presence of codimension-2 defects. A single (unity) blowup equation with a constant $ \Lambda $ factor is sufficient to determine the BPS spectra of 5d/6d SCFTs, as well as the partition functions in the presence of codimension-4 defects \cite{Kim:2020hhh, Kim:2021gyj}. In these cases, the factor $ \Lambda $ can be obtained from the large Coulomb parameter limit $ \phi \to \infty $ of the equation. However, when a codimension-2 defect is present, a single blowup equation is not enough to determine the partition function. One reason is that the codimension-2 defect breaks $ \mathrm{SU}(2)_l \times \mathrm{SU}(2)_r $ symmetry, making the index form assumption \eqref{eq:Psi-index} insufficient to provide enough constraints to solve the blowup equation. Another issue is the ambiguity in whether $ \Lambda $ is a simple constant or an operator involving the shift operator $ Y $, as well as how to determine such an operator $ \Lambda $ without prior knowledge of the codimension-2 defect partition function. Resolving the issue with the $ \Lambda $ factor will be crucial for understanding the general structure of the blowup equation and for systematically determining codimension-2 defect partition functions in more general theories.

The reason why the factor $ \Lambda $ is a constant, or a combination of Wilson loop VEVs for a more general magnetic fluxes, follows from the modularity of the closed topological string partition functions \cite{Aganagic:2006wq} and the holomorphic anomaly equations \cite{Huang:2017mis, Wang:2023zcb}. In the unrefined limit, the modular property and the holomorphic anomaly equation for the open topological string amplitudes are derived in \cite{Fang:2018ett,Ruan:2019aug} from the topological recursion \cite{Bouchard:2007ys,Eynard:2007kz}. Since the codimension-2 defect partition function on the $ \Omega $-background is a refined generalization of the open topological string partition function, we expect its modular properties can be used to study the properties of the factor $ \Lambda $ in the blowup equation in the presence of the defect. It would be intriguing to identify the modularity of the open topological string amplitude and determine the factor $ \Lambda $ involving the operator $ Y $ based on the modularity.

In this work, we consider BPS partitions in the presence of codimension-2 defects, as formal solutions to the quantum Seiberg-Witten curves, which are difference operators. However, they are generally not analytic square-integrable solutions to these difference operators. As observed in \cite{Kashani-Poor:2016edc}, one important reason is that the defect partition function develops poles when $\hbar=\epsilon_1=2\pi \frac{r}{s}$ for coprime integers $r$ and $s$. To resolve this issue, square-integrable solutions have been constructed in \cite{Marino:2016rsq,Francois:2025wwd} by incorporating non-perturbative corrections that cancel these poles, with subsequent generalizations to higher-rank cases \cite{Marino:2017gyg}. It would be interesting to investigate how Hanany-Witten transitions and blowup equations influence these square-integrable solutions.

\acknowledgments
We would like to thank Yuji Sugimoto for useful discussions. The research of HK is supported by Samsung Science and Technology Foundation under Project Number SSTF-BA2002-05 and by the National Research Foundation of Korea (NRF) grant funded by the Korean government (MSIT) (2023R1A2C1006542). MK is supported by a KIAS Individual Grant QP097501. The research of SK is supported by the NSFC grant No. 12250610188. KL is supported in part by BIMSA Start-up Research Fund, KIAS Individual Grant PG006904 and the National Research Foundation of Korea (NRF) Grant funded by the Korea government (MSIT) (No. 2017R1D1A1B06034369). XW is supported by the National Natural Science Foundation of China
Grants No.12247103.

\appendix 

\section{Special functions} \label{app:special}

We summarize definitions and properties of some special functions used in this paper.

\paragraph{Plethystic exponential.}
The plethystic exponential is defined as
\begin{align}
    \PE\left[ f(x) \right] = \exp\left(\sum_{n=1}^\infty \frac{1}{n} f(x^n)\right) \, .
\end{align}

\paragraph{q-Pochhammer symbol.}
The q-Pochhammer symbol $(z; q)_n$ is defined as
\begin{align}\label{eq:q-Pochysmb}
    (z; q)_n = \frac{(z; q)_\infty}{(z q^n; q)_\infty} =  \left\{ \begin{array}{ll}
            \prod_{k=0}^{n-1} (1-z q^k) & \quad (n > 0) \\
            1 & \quad (n = 0) \\
            \prod_{k=1}^{-n} \frac{1}{1-z/q^k} & \quad (n < 0) \, ,
    \end{array}\right.
\end{align}
where $n$ is an integer, and for $n$ infinity, 
\begin{align}
        (z; q)_\infty = \prod_{k=0}^\infty (1-zq^k) \, .
\end{align}
The q-Pochhammer symbol is related to the plethystic exponential via 
\begin{align}
    \PE\left[\frac{z}{1-q}\right] = \frac{1}{(z;q)_\infty} \, .
\end{align}
The residue for the inverse of q-Pochhammer symbol is
\begin{align}\label{eq:poch-residue}
    \Res_{z=q^{-k}} \left( \frac{1}{z} \frac{1}{(z;q)_{\infty}} \right) = -\frac{(1;q)_{-k}}{(q;q)_\infty} \, ,
\end{align}
where $ k $ is a non-negative integer. The  followings are useful identities to simplify the residue integral:
\begin{align}\label{eq:poch-id}
    (z; q)_{-n} = \frac{(-q/z)^n q^{n(n-1)/2}}{(q/z; q)_n} \, , \quad
    (z; q)_\infty = \sum_{n=0}^\infty \frac{(-1)^n q^{n(n-1)/2}}{(q;q)_n} z^n \, .
\end{align}

\paragraph{Basic hypergeometric series.}
The basic hypergeometric series, or q-hypergeometric series is defined by
\begin{align}
    {}_r \phi_s(a_1, \cdots, a_r; b_1, \cdots, b_s; q, z) &= \sum_{k=0}^\infty \frac{(a_1; q)_k \cdots (a_r; q)_k}{(b_1; q)_k \cdots (b_s; q)_k} \big( (-1)^k q^{k(k-1)/2} \big)^{1+s-r} \frac{z^k}{(q;q)_k} \, , \nonumber \\
 {}_0 \phi_s(-; b_1, \cdots, b_s; q, z) &= \sum_{k=0}^\infty \frac{\big( (-1)^k q^{k(k-1)/2} \big)^{1+s}}{(b_1; q)_k \cdots (b_s; q)_k}  \frac{z^k}{(q;q)_k} \, .
\end{align}
Let $ \Delta_b f(z) = bf(qz) - f(z) $ be the $ q $-difference operator. Then the basic hypergeometric function $ {}_r\phi_s $ satisfies the following difference equation:
\begin{align}\label{eq:qhyp-diff}
    \begin{aligned}
        &\left( \Delta_1 \Delta_{b_1/q} \Delta_{b_2/q} \cdots \Delta_{b_s/q}\right) {}_r\phi_s(a_1,\cdots,a_r; b_1,\cdots,b_s; q,z) \\
        &\quad= z \left( \Delta_{a_1}\cdots \Delta_{a_r} \right) {}_r\phi_s(a_1,\cdots,a_r;b_1,\cdots,b_s;q,z q^{1+s-r}) \, .
    \end{aligned}
\end{align}

\paragraph{Dedekind eta function.}
With $ q=e^{-\tau} $, the Dedekind eta function is defined as
\begin{align}
    \eta(\tau) = q^{1/24} \prod_{n=1}^\infty (1-q^n) = q^{1/24} (q;q)_\infty \, .
\end{align}

\paragraph{Jacobi theta functions.}
The Jacobi theta functions are defined as follows:
\begin{align}
    \begin{alignedat}{2}\label{Jacobi-theta}
        &\theta_1(\tau, x) = -i \sum_{n \in \mathbb{Z}} (-1)^{n} q^{\frac{1}{2}(n+1/2)^2} y^{n+1/2} \, , \quad
        &&\theta_2(\tau, x) = \sum_{n \in \mathbb{Z}} q^{\frac{1}{2}(n+1/2)^2} y^{n+1/2} \, , \\
        &\theta_3(\tau, x) = \sum_{n \in \mathbb{Z}} q^{\frac{n^2}{2}} y^n \, , \quad
        &&\theta_4(\tau, x) = \sum_{n \in \mathbb{Z}} (-1)^n q^{\frac{n^2}{2}} y^n \, ,
    \end{alignedat}
\end{align}
where $ q=e^{-\tau} $ and $ y=e^{-x} $.

\section{ADHM computations of 5d/6d partition functions}\label{app:ADHM}

In this appendix, we review instanton partition functions of 5d theories and elliptic genera of 6d theories discussed in this paper using the ADHM construction of the instanton moduli space described in \cite{Nekrasov:2002qd,Nekrasov:2003rj,Shadchin:2004yx,Hwang:2014uwa,Gaiotto:2014ina}.

\subsection{5d \texorpdfstring{$ \mathrm{SU}(N) $}{SU(N)} theories}

We start with the instanton partition function of 5d $ \mathrm{SU}(N)_\kappa $ gauge theory coupled to $ N_f $ hypermultiplets in the fundamental representation and $ N_A $ hypermultiplets in the rank-2 antisymmetric representation, where $ \kappa $ is the classical Chern-Simons level. The $ k $-instanton partition function without a defect is given by:
\begin{align}\label{eq:SU-ADHM}
    Z_k = \frac{(-1)^{N+\kappa}}{k!} \oint \left[\prod_{I=1}^k \frac{d\phi_I}{2\pi i}\right] e^{-\kappa \sum_I \phi_I} Z_{\mathrm{vec}} Z_{\mathrm{fund}} 
    Z_{\mathrm{anti}} \, ,
\end{align}
with
\begin{align}
    Z_{\mathrm{vec}} &= \frac{\prod_{I\neq J}^k 2\sinh(\frac{\phi_{IJ}}{2}) \cdot \prod_{I,J=1}^k 2\sinh(\frac{\phi_{IJ}+2\epsilon_+}{2})}{\prod_{I=1}^k \prod_{i=1}^N 2\sinh(\frac{\pm (\phi_I - a_i) + \epsilon_+}{2}) \cdot \prod_{I,J=1}^k 2\sinh(\frac{\phi_{IJ} + \epsilon_{1,2}}{2})} \label{eq:SU-vector-adhm} \ ,\\
    Z_{\mathrm{fund}} &= \prod_{b=1}^{N_f} \prod_{I=1}^k 2\sinh\left(\frac{\phi_I+m_b}{2}\right)\ ,\\
    Z_{\mathrm{anti}} &= \prod_{a=1}^{N_a} \frac{\prod_{I=1}^k \prod_{i=1}^N 2\sinh(\frac{\phi_I + a_i - m_a}{2}) \cdot \prod_{I<J}^k 2\sinh(\frac{\pm(\phi_I+\phi_J-m_a)-\epsilon_-}{2})}{\prod_{I<J}^k 2\sinh(\frac{\pm (\phi_I+\phi_J-m_a)-\epsilon_+}{2}) \cdot \prod_{I=1}^k 2\sinh(\frac{\pm(2\phi_I-m_a)-\epsilon_+}{2})}     \ , 
\end{align}
where they represent the 1-loop determinants for the vector multiplet and the hypermultiplets in the fundamental representation and antisymmetric representation, respectively, as indicated by  the subscripts. Here, we use the shorthand notation for $ \phi_{IJ} = \phi_I-\phi_J $, and denote $ a_i$ as the chemical potentials for $ \mathrm{SU}(N) $ gauge symmetry, satisfying $ \sum_i a_i=0 $. Additionally, $ m_a $ and $ m_b $ denote the mass parameters of the hypermultiplets. The contour of the integral \eqref{eq:SU-ADHM} is chosen according to the Jeffery-Kirwan (JK) residue prescription \cite{Benini:2013nda,Benini:2013xpa}.

To realize a codimension-2 defect in the 5d theory, one can couple 3d chiral multiplets to the system. The 1-loop determinant for the additional 3d fields is given by
\begin{align}\label{eq:SU-ADHM-3d}
    Z_{\mathrm{3d}} = \prod_{I=1}^k \frac{2\sinh(\frac{-\phi_I-x+\epsilon_+}{2})}{2\sinh(\frac{-\phi_I-x+\epsilon_+-\epsilon_2}{2})} \, ,
\end{align}
where $ x $ is the defect parameter. The defect partition function is then expressed as  the residue integral
\begin{align}\label{eq:SU-ADHM-Psi}
    \Psi_k = \frac{(-1)^{N+\kappa}}{k!} \oint \left[\prod_{I=1}^k \frac{d\phi_I}{2\pi i}\right] e^{-\kappa \sum_I \phi_I} Z_{\mathrm{vec}} Z_{\mathrm{anti}} Z_{\mathrm{fund}} Z_{\mathrm{3d}} \, ,
\end{align}
where the contour is chosen by the JK-residue prescription.

\paragraph{$ \mathrm{SU}(2) $ theory.}

Using the contour integral formula \eqref{eq:SU-ADHM-Psi}, we reproduce the codimension-2 defect partition function of the $ \mathrm{SU}(2)_\theta $ theory studied in Section~\ref{subsec:su2} for $ |\theta| \leq 3\pi $. In the case where $ |\theta| \leq 2\pi $, we choose the contour of the integral \eqref{eq:SU-ADHM-3d} such that the residues at
\begin{align}
    \phi_I = a_i - \epsilon_+ - (m-1) \epsilon_1 - (n-1)\epsilon_2
\end{align}
contribute to the integral. The residues are labeled by colored Young diagrams $ (Y_1, Y_2) $, and the residue integral is given by
\begin{align}
    \Psi_k = \sum_{|Y_i|=k} \prod_{i=1}^2 \prod_{s\in Y_i} \frac{2\sinh(\frac{\varphi(s)+x-\epsilon_+}{2})}{\prod_{j=1}^2 2\sinh(\frac{E_{ij}}{2}) 2\sinh(\frac{E_{i}-2\epsilon_+}{2}) 2\sinh(\frac{\varphi(s)+x-\epsilon_++\epsilon_2}{2}) }  \, .
\end{align}
Here, $ s \in Y_i $ is a box in the Young diagram $ Y_i $, and
\begin{align}
    E_{ij} = a_i - a_j + \epsilon_1 (v_j(s) + 1) - \epsilon_2 h_i(s) \, ,\quad
    \varphi(s) = a_i - \epsilon_+ - (m-1)\epsilon_1 - (n-1)\epsilon_2 \, ,
\end{align}
where $ h_i(s) $ and $ v_j(s) $ are the arm length of $ s \in Y_i $ and the leg length of $ s \in Y_j $, respectively, while $ (m,n) $ represent the vertical and horizontal positions of $ s \in Y_i $. For $ \theta=0 $ and $ \theta=\pm \pi $, this result exactly reproduces the solutions of the blowup equations studied in Section~\ref{subsec:su2defect}. We have verified  the equivalence of two results up to 3-instanton order. For $ \theta=\pm 2\pi $, as noted in \cite{Hwang:2014uwa}, the ADHM quantum mechanics develops a continuum on the Coulomb branch, corresponding to states that decouple from the 5d physics. As a consequence, the contour integral has nonzero residues at infinity, $ \phi_I = \pm \infty $. We subtract such extra factor by dividing 
\begin{align}
    \Psi_{\mathrm{extra}} = \left\{ \begin{array}{ll}
            \displaystyle \PE\left[ -\mathfrak{q}\frac{q_1 q_2 \sqrt{q_2}}{(1-q_1)(1-q_2)} \right] & \quad \theta=2\pi \\ \\
            \displaystyle \PE\left[ -\mathfrak{q}\frac{\sqrt{q_2}}{(1-q_1)(1-q_2)} \right] & \quad \theta=-2\pi
    \end{array}\right.
\end{align}
from the residue integral. We have checked that this reproduces the results from the blowup equations up to 3-instanton order.

For $ |\theta| \geq 3\pi $, the localization integral encounters higher-degree poles at the infinity,  $ \phi_I = \pm \infty $. Currently, no general systematic method exists to consistently handle such higher-degree poles in the localization framework.

However, in the specific cases of $\theta=\pm 3\pi$, it is possible to circumvent these higher-degree poles by introducing an auxiliary hypermultiplet in the antisymmetric representation. This additional antisymmetric hypermultiplet effectively reduces the degree of divergences that arise at the infinities. Once this adjustment is made and the localization integral is properly handled, the auxiliary  hypermultiplet can be integrated out at the end of the computation. To check the $ \theta=3\pi $ result from the blowup equations, we consider 
\begin{align}
    \Psi_k = \lim_{m \to \infty} \frac{1}{k!} \oint \left[\prod_{I=1}^k \frac{d\phi_I}{2\pi i}\right] e^{2 \sum_I \phi_I} Z_{\mathrm{vec}} Z_{\mathrm{anti}} Z_{\mathrm{3d}} \, ,
\end{align}
where $ m $ is the mass parameter of the antisymmetric hypermultiplet. The contour integral computes the partition function of the $ \mathrm{SU}(2)_{2\pi} + 1\mathbf{AS}$ which is coupled with 3d fields. We expect that decoupling the antisymmetric matter will shift the $ \theta $-angle by one unit, yielding $ \mathrm{SU}(2)_{3\pi} $ with a codimension-2 defect. 

\begin{figure}
    \centering
    \includegraphics{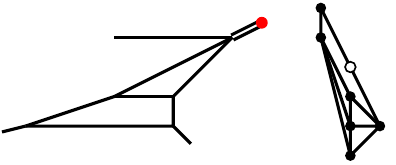}
    \caption{5-brane web and toric diagram of $ \mathrm{SU}(2)_{3\pi} $ theory related to the curve \eqref{eq:su2-3-sw-adhm}. } \label{fig:su2-3-adhm}
\end{figure}

Note that the poles from the antisymmetric hypermultiplet also contribute to the residue integral. For instance, one may choose poles at
\begin{align}
    \phi_1 - a_i + \epsilon_+ = 0 \, , \quad
    2\phi_1 - m - \epsilon_+ = 0 \, ,
\end{align}
in the 1-instanton order. The contour integral result matches the $ \theta=3\pi $ solution of the blowup equations up to an overall constant factor:
\begin{align}
    \Psi^{\mathrm{ADHM}} = \Psi^{\mathrm{blowup}} \cdot \PE\left[ \frac{ \sqrt{q_1q_2}}{1-q_1} \mathfrak{q} e^{-x}\right] \, .
\end{align}
We have verified this relation up to the 2-instanton order. It is worth noting that the extra constant factor slightly changes the quantum difference equation \eqref{eq:SU2-3pi-SW}. Assuming that \eqref{eq:SU2-3pi-SW} is true for $ \Psi^{\mathrm{blowup}} $, one can check that the normalized NS-limit of the partition function from $ \Psi^{\mathrm{ADHM}} $ satisfies
\begin{align}\label{eq:su2-3-sw-adhm}
    \left( X + X^{-1} + Y + \frac{\mathfrak{q}}{\sqrt{q_1}} \left( X^2 Y^{-1} X + (1+q_1) X^2 \right) + \mathfrak{q}^2 X^2 Y^{-1} X^2\right) \hat{\Psi}^{\mathrm{NS}} = E \hat{\Psi}^{\mathrm{NS}} \, .
\end{align}
The 5-brane diagram and the generalized toric diagram associated with this curve are shown in Figure~\ref{fig:su2-3-adhm}. These diagrams are identical to those in  Figure~\ref{fig:SU2-odd-toric}(c), except for the resolution of 7-brane monodromies in different directions.

\subsection{5d Linear quiver gauge theories} \label{app:ADHM-quiver}

Next, we consider $ \prod_{p=1}^n \mathrm{SU}(N_i) $ linear quiver gauge theory with bifundamental hypermultiplets. The $ (k_1,\cdots,k_n) $-instanton partition function of this linear quiver gauge theory is given by
\begin{align}\label{eq:quiver-adhm}
    Z_{k_1,\cdots,k_n} = \frac{1}{\prod_{p=1}^n k_p!} \oint \left[\prod_{p=1}^n \prod_{I=1}^k \frac{d\phi_I^{(p)}}{2\pi i}\right] e^{-\sum_{I,p} \kappa_p \phi_I^{(p)}} Z_{\mathrm{vec}}^{(p)} Z_{\mathrm{bifund}}^{(p)} \, ,
\end{align}
where $k_p$ and $ \kappa_p $ are respectively the instanton number and the Chern-Simons level for each gauge node, and
\begin{align}
    Z_{\mathrm{vec}}^{(p)} &= \prod_{p=1}^n \frac{\prod_{I\neq J}^k 2\sinh\big(\frac{\phi_{IJ}^{(p)}}{2}\big) \cdot \prod_{I,J=1}^k 2\sinh\big(\frac{\phi_{IJ}^{(p)}+2\epsilon_+}{2}\big)}{\prod_{I=1}^k \prod_{i=1}^N 2\sinh\big(\frac{\pm (\phi_I^{(p)} - a_i^{(p)}) + \epsilon_+}{2}\big) \cdot \prod_{I,J=1}^k 2\sinh\big(\frac{\phi_{IJ}^{(p)} + \epsilon_{1,2}}{2}\big)} \, , \\
    Z_{\mathrm{bifund}}^{(p)} &= \prod_{p=1}^{n-1} \prod_{I=1}^{k_p} \prod_{i=1}^{N_{p+1}} \! 2\sinh\bigg(\frac{\phi_I^{(p)} \!+\! a_i^{(p+1)} \!+\! m^{(p)}}{2}\bigg) \!\cdot\! \prod_{p=1}^{n-1} \prod_{J=1}^{k_{p+1}} \prod_{j=1}^{N_p} \! 2\sinh\bigg(\frac{\phi_J^{(p+1)} \!+\! a_j^{(p)} \!+\! m^{(p)}}{2}  \bigg) \nonumber \\
    &\quad \cdot \prod_{p=1}^{n-1} \prod_{I=1}^{k_p} \prod_{J=1}^{k_{p+1}} \frac{2\sinh\big(\frac{\pm(\phi_I^{(p)} - \phi_J^{(p+1)} + m^{(p)}) + \epsilon_-}{2}\big)}{2\sinh\big(\frac{\pm(\phi_I^{(p)} - \phi_J^{(p+1)} + m^{(p)}) + \epsilon_+}{2}\big)} \, ,
\end{align}
are the 1-loop determinants of the vector multiplet and the bifundamental hypermultiplets. We denote $ a_i^{(p)} $ as the chemical potentials for $ \mathrm{SU}(N_p) $ gauge symmetry satisfying $ \sum_i a_i^{(p)} = 0 $, and $ m^{(p)} $ as the mass parameter of the hypermultiplets in the bifundamental representation of $ \mathrm{SU}(N_p) \times \mathrm{SU}(N_{p+1}) $. The contour of the integral is chosen by the JK-residue prescription.

\paragraph{$ \mathrm{SU}(2) \times \mathrm{SU}(2) $ theory and baryonic Higgsing.}

We consider the quiver gauge theory $ \mathrm{SU}(2)_\pi \times \mathrm{SU}(2)_\pi $ with a bifundamental hypermultiplet and its baryonic Higgs branch deformation to the $ \mathrm{SU}(2)_0 $ theory coupled with a codimension-2 defect discussed in Section~\ref{subsubsec:su2S}. The partition function of the $  \mathrm{SU}(2)_\pi \times \mathrm{SU}(2)_\pi $ gauge theory ($\mathrm{SU}(2)^2$) is given by
\begin{align}\label{eq:SU2SU2-Z}
    Z^{\mathrm{SU}(2)^2} = e^{\mathcal{E}_0^{\mathrm{SU}(2)^2}} \cdot Z_{\mathrm{pert}}^{\mathrm{SU}(2)^2} \cdot \bigg(1 + \sum_{(k_1,k_2)\neq0} \mathfrak{q}_1^{k_1} \mathfrak{q}_2^{k_2} Z_{k_1,k_2}^{\mathrm{SU}(2)^2} \bigg) \, ,
\end{align}
where the effective prepotential and 1-loop contributions take the form
\begin{align}
    \mathcal{E}_0^{\mathrm{SU}(2)^2} &= \frac{1}{\epsilon_1 \epsilon_2} \bigg( \frac{4}{3} \left(\phi_1^3 + \phi_2^3\right) - \frac{1}{2} \left((\phi_1 + \phi_2 \pm m_3)^3 + (\phi_1 - \phi_2 \pm m_3)^3 \right)  \nonumber \\
    &\qquad \qquad + m_1 \phi_1^2 + m_2 \phi_2^2 - \frac{\epsilon_1^2 + \epsilon_2^2}{12} \phi_2 + \epsilon_+^2 (\phi_1 + \phi_2) \bigg) , \\
    Z_{\mathrm{pert}}^{\mathrm{SU}(2)^2} &= \PE\left[-\frac{1\!+\!q_1q_2}{(1\!-\!q_1)(1\!-\!q_2)} \left(e^{-2\phi_1} \!+\! e^{-2\phi_2}\right) + \frac{\sqrt{q_1q_2} (e^{-m_3} \!+\! e^{m_3})}{(1\!-\!q_1)(1\!-\!q_2)} e^{-\phi_1} (e^{-\phi_2}+e^{\phi_2}) \right] \, , \nonumber
\end{align}
and $ \mathfrak{q}_{1,2} = e^{-m_{1,2}} $ are the instanton fugacities for the two $ \mathrm{SU}(2) $ gauge theories. Here, $ \phi_{1,2} $ are two Coulomb branch parameters and $ m_3 $ is the mass parameter of the bifundamental hypermultiplet. The instanton corrections $ Z_{k_1,k_2}^{\mathrm{SU}(2)^2} $ are computed via the residue integral \eqref{eq:quiver-adhm} with $ \kappa_1=\kappa_2=1 $.

The usual baryonic Higgsing from the $ \mathrm{SU}(2)_\pi \times \mathrm{SU}(2)_\pi $ theory to the $ \mathrm{SU}(2)_0 $ theory amounts to tuning the chemical potentials as
\begin{align}
    \phi_1 = \phi_2 \to \phi \, , \quad
    m_1 + m_2 \to m_0 \, , \quad
    m_3 \to -\epsilon_+ \, ,
\end{align}
where $ \phi $ and $ m_0 $ are the Coulomb parameter and instanton chemical potential of the higgsed theory. On the other hand, we obtain a codimension-2 defect by giving position dependent VEV to the baryonic operator, which corresponds to tuning the parameters as
\begin{align}\label{eq:su2su2-defectH}
    \phi_1 \to \phi \, , \quad
    \phi_2 \to \phi + \epsilon_2/2 \, , \quad
    m_1 \to m_0 - x \, , \quad
    m_2 \to x \, , \quad
    m_3 \to -\epsilon_+ - \epsilon_2/2 \, .
\end{align}
By further shifting $ m_0 \to m_0-\epsilon_2 $ and $ x \to x + \epsilon_1/2 $, one finds that the partition function $ Z^{\mathrm{SU}(2)^2} $ reduces to the defect partition function $ \tilde{\Psi} $ in \eqref{eq:SU2-Psitilde} for the $ \theta=0 $ case up to an overall constant term which does not depend on the dynamical parameter $ \phi $. For instance, the $ k_1=0 $ sector of \eqref{eq:SU2SU2-Z} reduces to the q-hypergeometric function appearing in \eqref{eq:U1-vortex} as
\begin{align}
    1 + \sum_{k=1}^\infty \mathfrak{q}_2^k Z_{0,k}^{\mathrm{SU}(2)^2} \overset{\eqref{eq:su2su2-defectH}}{\longrightarrow} {}_0\phi_1(-; q_1Q^{-2}; q_1; q_1^2 Q^{-2} X) \, .
\end{align}
We have checked this identity up to 2-instanton order. Similarly, we reproduce $ \widetilde{\Psi}_k $ of the $ \mathrm{SU}(2)_0 $ theory studied in Section~\ref{subsubsec:su2S} up to 2-instanton order using the instanton partition functions of the $ \mathrm{SU}(2)_\pi \times \mathrm{SU}(2)_\pi $ theory.

\subsection{5d \texorpdfstring{$ \mathrm{SU}(N) + 1\mathbf{Adj} $}{SU(N)+1Adj}} \label{app:ADHM-Mstr}

Another interesting theory is the 5d $ \mathrm{SU}(N) $ gauge theory with an adjoint hypermultiplet, which is not  a genuine 5d SCFTs, but is instead UV-completed by 6d $ \mathcal{N}=(2,0) $  SCFT compactified on a circle. The 1-loop determinant for the adjoint hypermultiplet is
\begin{align}
    Z_{\mathrm{adj}} = \frac{\prod_{I=1}^k \prod_{i=1}^N 2\sinh(\frac{m \pm (\phi_I - a_i)}{2}) \cdot \prod_{I,J=1}^k 2\sinh(\frac{\phi_{IJ} \pm m - \epsilon_-}{2})}{\prod_{I,J=1}^k 2\sinh(\frac{\phi_{IJ} \pm m - \epsilon_+}{2})} \, ,
\end{align}
where $ m $ is the mass parameter of the hypermultiplet. The $ k $-instanton partition function is given by the residue integral
\begin{align}\label{eq:ADHM-adj}
    Z_k = \frac{1}{k!} \oint \left[ \prod_{I=1}^k \frac{d\phi_I}{2\pi i} \right] Z_{\mathrm{vec}} Z_{\mathrm{adj}} \, ,
\end{align}
where $ Z_{\mathrm{vec}} $ is given in \eqref{eq:SU-vector-adhm} and the integral contour is determined by the JK-residue prescription. This residue integral contains the following non-dynamical factor, which we factor out from the instanton partition function, 
\begin{align}\label{eq:msym-extra}
    \PE\left[ \frac{\sinh(\frac{m \pm \epsilon_-}{2})}{\sinh(\frac{\epsilon_1}{2}) \cdot \sinh(\frac{\epsilon_2}{2})} \cdot \frac{N}{1-\mathfrak{q}} \right] \, ,
\end{align}
where $ \mathfrak{q} $ is the instanton fugacity.

We now introduce a codimension-2 defect by coupling a 3d hypermultiplet that consists of two chiral multiplets each of which transforms in the fundamental and antifundamental representations of the $ \mathrm{SU}(2) $ gauge symmetry. The 1-loop determinant for the 3d fields is written as \cite{Gaiotto:2014ina, Bullimore:2014awa}
\begin{align}
    Z_{\mathrm{3d}} &= \prod_{I=1}^k \frac{2\sinh(\frac{-\phi_I - x - m/2 + (\epsilon_1+\epsilon_2)/4}{2})}{2\sinh(\frac{-\phi_I - x - m/2 + (\epsilon_1 - 3\epsilon_2)/4}{2})} \frac{2\sinh(\frac{\phi_I + x - m/2 + (\epsilon_1+5\epsilon_2)/4}{2})}{2\sinh(\frac{\phi_I + x - m/2 + (\epsilon_1+\epsilon_2)/4}{2})} \, ,
\end{align}
where $ x $ is the defect parameter. 

The instanton partition function in the presence of the codimension-2 defect is given by
\begin{align}
    \Psi_k = \frac{1}{k!} \oint \left[\prod_{I=1}^k \frac{d\phi_I}{2\pi i}\right] Z_{\mathrm{vec}} Z_{\mathrm{adj}} Z_{\mathrm{3d}} \, ,
\end{align}
where the integer contour is chosen by the JK-residue prescription. For instance, we choose the poles at
\begin{align}
    \phi_1 - a_i + \epsilon_+ = 0 \, , \quad
    \phi_1 + x - \frac{m}{2} + \frac{\epsilon_1 + \epsilon_2}{4} = 0\ ,
\end{align}
in the 1-instanton order. Furthermore, the following poles have non-zero residue contributions to the 2-instanton order: 
\begin{gather}
    \begin{gathered}\label{eq:inApp2inspole}
        \left\{\begin{array}{l}
                \phi_1 - a_i + \epsilon_+ = 0 \\
                \phi_2 + x - \frac{m}{2} + \frac{\epsilon_1 + \epsilon_2}{4} = 0
        \end{array}\right. \, , \qquad
        \left\{\begin{array}{l}
                \phi_1 - a_i + \epsilon_+ = 0 \\
                -\phi_1 + \phi_2 + \epsilon_{1,2} = 0
        \end{array}\right. , \\
        \left\{\begin{array}{l}
                \phi_1 + x - \frac{m}{2} + \frac{\epsilon_1 + \epsilon_2}{4} = 0 \\
                -\phi_1 + \phi_2 + \epsilon_{1,2} = 0
        \end{array}\right. , \quad
        \left\{\begin{array}{l}
                \phi_1 + x - \frac{m}{2} + \frac{\epsilon_1 + \epsilon_2}{4} = 0 \\
                -\phi_1 + \phi_2 \pm m - \epsilon_{+} = 0
        \end{array}\right. , \\
        \left\{\begin{array}{l}
                \phi_1 - a_i + \epsilon_+ = 0 \\
                \phi_2 - a_j + \epsilon_+ = 0
        \end{array}\right. , 
    \end{gathered}
\end{gather}
as well as those given by exchanging $ \phi_1 $ and $ \phi_2 $ except for the last poles in \eqref{eq:inApp2inspole}. We note that the defect partition function constructed in this way also has an extra factor similar to \eqref{eq:msym-extra}, 
\begin{align}
    \PE\left[ \left(N\frac{\sinh((m \pm \epsilon_-)/2)}{\sinh(\epsilon_1/2) \cdot \sinh(\epsilon_2/2)} + \frac{\sinh(\epsilon_+) \cdot \sinh((m+\epsilon_-)/2)}{\sinh(\epsilon_1/2) \cdot \sinh((m+\epsilon_+)/2)} \right) \cdot \frac{1}{1-\mathfrak{q}} \right] \, ,
\end{align}
where we used the defect partition function after factoring out this non-dynamical contribution in Section~\ref{sec:Mstr}.

\subsection{6d \texorpdfstring{$ \mathrm{SU}(N)+2N\mathbf{F} $}{SU(N)+2NF} theories}

We now consider 6d theories. Let us first consider the elliptic genus of 6d $ \mathrm{SU}(N) $ gauge theories coupled with $ 2N $ hypermultiplets in the fundamental representation, discussed in Section~\ref{sec:Mstr}. From the 2d gauged linear sigma model description for $ k $ D2-branes of the Type IIA brane configuration depicted in Figure~\ref{fig:Mstr-brane}(b), the $ k $-string elliptic genus of the theory can be expressed as the contour integral \cite{Haghighat:2013gba, Haghighat:2013tka}:
\begin{align}
    Z_k &= \frac{1}{k!} \oint \left[ \prod_{I=1}^k \frac{d\phi_I}{2\pi i} \frac{2\pi \eta(\tau)^2}{i} \frac{i\theta_1(2\epsilon_+)}{\eta(\tau)}\right] \left(\prod_{I \neq J}^k \frac{i\theta_1(\pm \phi_{IJ})}{\eta(\tau)^2} \frac{i\theta_1(2\epsilon_+ + \phi_{IJ})}{\eta(\tau)}\right) \\
    &\quad \times \left( \prod_{I,J=1}^k \frac{(i\eta(\tau))^2}{\theta_1(\epsilon_{1,2} + \phi_{IJ})} \right) \left( \prod_{I=1}^k \prod_{j=1}^N \frac{(i\eta(\tau))^2}{\theta_1(\epsilon_+ \pm (\phi_I - a_i))} \right) \left( \prod_{I=1}^k \prod_{l=1}^{2N} \frac{i\theta_1(\phi_I - m_l)}{\eta(\tau)} \right) \, , \nonumber
\end{align}
where $ \phi_{IJ} = \phi_I-\phi_J $, $ \sum a_i= \sum m_l=0 $, $ \eta(\tau) $ is the Dedekind eta function, and $ \theta_1(z) = \theta_1(\tau, z) $ is the Jacobi theta function. The integral contour is determined by the JK-residue prescription.

The Higgsing of the $N=2$ theory to the $N=1$ theory in the presence of the codimension-2 defect can be studied using this result. This is achieved by tuning the parameters as specified in \eqref{eq:Mstr-Higgs-parameter}, followed by the rescaling described subsequently in \eqref{eq:Mstr-Higgs-parameter}.\footnote{Here, $ a_1+a_2=0 $ in the contour integral and $ a_1 $ is identified with $ \phi_1 $ in \eqref{eq:Mstr-Higgs-parameter}.} After shifting the integration variable $ \phi_I \to \phi_I - (x-m-\epsilon_-)/2 - \epsilon_+ $, we obtain the contour integral expression of the codimension-2 defect partition functions of the M-string theory as
\begin{align}
        \Psi_k &= \frac{1}{k!} \oint \left[ \prod_{I=1}^k \frac{d\phi_I}{2\pi i} \frac{2\pi \eta(\tau)^2}{i} \frac{i\theta_1(2\epsilon_+)}{\eta(\tau)}\right] \left(\prod_{I \neq J}^k \frac{i\theta_1(\pm \phi_{IJ})}{\eta(\tau)^2} \frac{i\theta_1(2\epsilon_+ + \phi_{IJ})}{\eta(\tau)}\right) \\
        &\quad \times \left( \prod_{I,J=1}^k \frac{(i\eta(\tau))^2}{\theta_1(\epsilon_{1,2} + \phi_{IJ})} \right) \left( \prod_{I=1}^k \frac{\theta_1(\phi_I-m) \theta_1(\phi_I + m - x) \theta_1(\phi_I + m - \epsilon_2)}{-\theta_1(\pm \phi_I + \epsilon_+) \theta_1(\phi_I + m - x - \epsilon_2)}  \right) \, , \nonumber
\end{align}
where the contour is determined by the JK-residue prescription. For instance, one can choose the poles at
\begin{align}
    \phi_1 = -\epsilon_+ \, , \quad  x - m + \epsilon_2 \, ,
\end{align}
in the $ k=1 $ case, which yields \eqref{eq:Mstr-def-Psi1H}.

\subsection{6d \texorpdfstring{$ \mathrm{Sp}(N) $}{Sp(N)} theory and E-string theory} \label{app:ADHM-Sp}

Next, we consider 6d $ \mathrm{Sp}(N) + (2N+8)\mathbf{F} $ theory, which is related to the E-string theory by the Higgsing. The self-dual strings in the $ \mathrm{Sp}(N) $ gauge theories are realized by the M2-branes connecting the M5-brane and M9-plane in M-theory \cite{Kim:2014dza}. When there are $ k $ M2-branes, the corresponding worldvolume theory is described by the $ O(k) $ gauge theory, and the elliptic genus is given by
\begin{align}
    &Z_k^{\mathrm{Sp(N)}} = \sum_K \frac{1}{|W^{(K)}|} \frac{1}{(2\pi i)^r} \oint Z_{\text{1-loop}}^{\mathrm{Sp(N)}} \, , 
\end{align}
where $ K $ runs over the disconnected sectors of the $ \mathrm{O}(k) $ flat connections, $ W^{(K)} $ is the Weyl group, and $Z_{\text{1-loop}}^{\mathrm{Sp}(N)}$ is the 1-loop determinant of the supermultiplets in the 2d $ \mathrm{O}(k) $ gauge theory, given by, 
\begin{align}\label{eq:6dSpN-1loop}
    &Z_{\text{1-loop}}^{\mathrm{Sp}(N)} = \Bigg( \prod_{I=1}^{r} \frac{2\pi \eta^2 du_I}{i} \frac{i \theta_1(2\epsilon_+)}{\eta} \Bigg) \Bigg( \prod_{e \in \mathbf{root}} \frac{i \theta_1(e \cdot u)}{\eta} \frac{i \theta_1(2\epsilon_+ + e \cdot u)}{\eta} \Bigg) \\
    & \quad \times \Bigg( \prod_{\rho\in\mathbf{sym}} \frac{(i\eta)^2}{\theta_1(\epsilon_{1,2}+\rho(u))}  \Bigg) \Bigg(\prod_{\rho\in\mathbf{bifund}} \frac{i\eta}{\theta_1(\epsilon_+ + \rho(a,u))} \Bigg) \Bigg(\prod_{\rho\in\mathbf{fund}} \prod_{l=1}^{2N+8} \frac{i\theta_1(m_l+\rho(u))}{\eta} \Bigg) \, . \nonumber
\end{align}
Here, $ r $ is the number of the continuous complex moduli, and  $ \mathbf{root} $, $ \mathbf{sym} $, $ \mathbf{fund} $ are the root system, the symmetric, and fundamental representations of $ \mathrm{SO}(k) $, while $ \mathbf{bifund} $ corresponds to the bifundamental representation of $ \mathrm{SO}(k) \times \mathrm{Sp}(N) $. See \cite{Kim:2014dza} for details. For $ N=1 $ and at the 1-string order, one finds
\begin{align}
    Z_1^{\mathrm{Sp}(1)} = \sum_{I=1}^4 \frac{-\prod_{l=1}^{10} \theta_I(\tau,m_l)}{2\eta(\tau)^6 \theta_1(\tau,\epsilon_{1,2}) \theta_I(\tau, \pm \phi_1 + \epsilon_+)} \, ,
\end{align}
where $ \phi_1 $ is the holonomy for the $ \mathrm{Sp}(1) $ gauge symmetry. The residue integral at 2-string order yields
\begin{align}
    &Z_2^{\mathrm{Sp}(1)} = \frac{1}{4\eta(\tau)^{12} \theta_1(\tau,\epsilon_{1,2})} \left( \frac{\prod_{l=1}^{10} \theta_I(\tau,m_l \pm \frac{\epsilon_1}{2})}{\theta_1(\tau,2\epsilon_1) \theta_1(\tau,\epsilon_2-\epsilon_1) \prod_{s} \theta_I(\epsilon_+ \pm \frac{\epsilon_1}{2} + s \phi_1)} + (\epsilon_1 \leftrightarrow \epsilon_2) \right) \nonumber \\
    &\! + \frac{1}{2\eta(\tau)^{12} \theta_1(\tau,\epsilon_{1,2})} \left(\frac{\prod_{l=1}^{10} \theta_1(\tau, m_l \pm (\epsilon_+ + \phi_1))}{\prod_{s}\theta_1(\tau,2\phi_1 + (s+1)\epsilon_+) \theta_1(\tau,2\phi_1 + (s+1)\epsilon_+ + \epsilon_{1,2}) } \!+\! (\phi_1 \!\to\! -\phi_1) \! \right) \nonumber \\
    &\! + \sum_{I \leq J}^4 \frac{\theta_{\sigma(I,J)}(\tau,0) \theta_{\sigma(I,J)}(\tau,2\epsilon_+)}{4\eta(\tau)^{12} \theta_1(\tau,\epsilon_{1,2})^2} \frac{\prod_{l=1}^{10} \theta_I(\tau,m_l) \theta_J(\tau,m_l)}{\theta_{\sigma(I,J)}(\tau,\epsilon_{1,2}) \theta_I(\tau,\epsilon_+ \pm \phi_1) \theta_J(\tau,\epsilon_+ \pm \phi_1)} \, ,
\end{align}
where
\begin{alignat}{3}\label{eq:sigma}
    \begin{aligned}
        &\sigma(I, J) = \sigma(J, I) \, , \
        &&\sigma(I, I) = 0 \, , \
        &&\sigma(1, I) = I \, , \\
        &\sigma(2, 3) = 4 \, , \
        &&\sigma(2, 4) = 3 \, , \
        &&\sigma(3, 4) = 2 \, .
    \end{aligned}
\end{alignat}

When $ N=0 $, this theory corresponds to the E-string theory. The elliptic genus of the E-string theory at 1-string and 2-string orders is, respectively,  given by
\begin{align}
    Z_1 &= \sum_{I=1}^4 \frac{-\prod_{l=1}^8 \theta_I(\tau,m_l)}{2\eta(\tau)^6 \theta_1(\tau,\epsilon_{1,2})} \, , \label{eq:Estr-Z1} \\
    Z_{2} &= \frac{1}{4\eta(\tau)^{12} \theta_1(\tau,\epsilon_{1,2})} \sum_{I=1}^4 \left( \frac{\prod_{l=1}^8 \theta_I(\tau,m_l \pm \frac{\epsilon_1}{2})}{\theta_1(\tau,2\epsilon_1) \theta_1(\tau,\epsilon_2-\epsilon_1)} + \frac{\prod_{l=1}^8 \theta_I(m_l \pm \frac{\epsilon_2}{2})}{\theta_1(2\epsilon_2) \theta_1(\epsilon_1-\epsilon_2)} \right) \nonumber \\
    &\quad + \sum_{I\leq J}^4 \frac{\theta_{\sigma(I,J)}(\tau,0) \theta_{\sigma(I,J)}(\tau,2\epsilon_+) \prod_{l=1}^8 \theta_I(\tau,m_l) \theta_J(\tau,m_l)}{4\eta(\tau)^{12} \theta_1(\tau,\epsilon_{1,2})^2 \theta_{\sigma(I,J)}(\tau,\epsilon_{1,2})} \, . \label{eq:Estr-Z2}
\end{align}
This result can be also obtained from the Higgsing \eqref{eq:Estr-Higgs-usual} from the 6d $ \mathrm{Sp}(1) + 10\mathbf{F} $ theory.

Now we consider the defects. The codimension-4 defect partition function of the E-string theory is studied in \cite{Chen:2021ivd, Kim:2021gyj}. The elliptic genus in the presence of the codimension-4 defect can be represented as the residue integral
\begin{align}\label{eq:Estr-codim4}
    &Z_k^{(W)} = \sum_K \frac{1}{|W^{(K)}|} \frac{1}{(2\pi i)^r} \oint Z_{\text{1-loop}}^{\mathrm{Sp}(0)} \cdot \prod_{\rho \in \mathbf{fund}} \frac{\theta_1(\pm \epsilon_- + \rho(u) - z)}{\theta_1(\pm \epsilon_+ + \rho(u) - z)} \, ,
\end{align}
where the 1-loop determinant $ Z_{\text{1-loop}}^{\mathrm{Sp}(0)} $ is given in \eqref{eq:6dSpN-1loop}, $ \mathbf{fund} $ is the fundamental representation of $ \mathrm{SO}(k) $, $ z $ is the parameter for the defect, and the contour is chosen by the JK-residue prescription. At 1-string and 2-string orders, the integral gives
\begin{align}
    Z_1^{(W)} &= \sum_{I=1}^4 \frac{-\prod_{l=1}^8 \theta_I(\tau,m_l)}{2\eta(\tau)^6 \theta_1(\tau,\epsilon_{1,2})} \frac{\theta_I(\tau,z \pm \epsilon_-)}{\theta_I(\tau,z \pm \epsilon_+)} \, , \\
    Z_2^{(W)} &= \frac{1}{4\eta(\tau)^{12} \theta_1(\tau,\epsilon_{1,2})} \sum_{I=1}^4 \left( \frac{\prod_{l=1}^8 \theta_I(\tau,m_l \pm \frac{\epsilon_1}{2})}{\theta_1(\tau,2\epsilon_1) \theta_1(\tau,\epsilon_2-\epsilon_1)} \frac{\theta_I(\tau,z\pm(\epsilon_-+\frac{\epsilon_1}{2}))}{\theta_1(\tau,z \pm (\epsilon_+ + \frac{\epsilon_1}{2}))} + (\epsilon_1 \leftrightarrow \epsilon_2) \right) \nonumber \\
    &+ \sum_{I\leq J}^4 \frac{\theta_{\sigma(I,J)}(\tau,0) \theta_{\sigma(I,J)}(\tau,2\epsilon_+) \prod_{l=1}^8 \theta_I(\tau,m_l) \theta_J(\tau,m_l)}{4\eta(\tau)^{12} \theta_1(\tau,\epsilon_{1,2})^2 \theta_{\sigma(I,J)}(\tau, \epsilon_{1,2})} \frac{\theta_I(\tau,z \pm \epsilon_-) \theta_J(\tau,z \pm \epsilon_-)}{\theta_I(\tau,z \pm \epsilon_+) \theta_J(\tau,z \pm \epsilon_+)} \nonumber \\
    &+ \left( \frac{\prod_{l=1}^8 \theta_1(\tau,z + m_l + \epsilon_+)}{2\eta(\tau)^{12} \theta_1(\tau,2z) \theta_1(\tau,2z+2\epsilon_+) \theta_1(\tau,2z+2\epsilon_+ + \epsilon_{1,2})} + (\epsilon_{1,2} \to -\epsilon_{1,2}) \right) \, .
\end{align}
The expectation value $ \langle W \rangle $ of the codimension-4 defect is defined as
\begin{align}
    \langle W \rangle = e^{\phi} \left( 1 + \sum_{k=2}^\infty e^{-k\phi} W_k \right) = e^{\phi} \frac{1+\sum_{k=1}^\infty e^{-k\phi} Z_k^{(W)}}{1+\sum_{k=1}^\infty e^{-k\phi} Z_k} - V_0(z) \, ,
\end{align}
where $ W_1 $ is omitted from the summation since the $ k=1 $ order contribution does not depend on the dynamical parameter $ \phi $. This expectation value $ \langle W \rangle $ is independent of $ z $, where the subtraction of the overall constant is represented as $ V_0(z) = Z_1^{(W)} - Z_1 $.

Lastly, we present the explicit expression for the elliptic genus of the E-string theory in the presence of codimension-2 defect. Using the defect Higgsing \eqref{eq:Estr-defectHiggs}, one readily finds that the $ k $-string elliptic genus with codimension-2 defect takes the form,
\begin{align}
    \Psi_k &= \sum_K \frac{1}{|W^{(K)}|} \frac{1}{(2\pi i)^r} \oint Z_{\text{1-loop}}^{\mathrm{Sp}(0)} \cdot \prod_{\rho \in \mathbf{fund}} \frac{\theta_1(x - \frac{\epsilon_1}{2} + \rho(u) + \epsilon_2)}{\theta_1(x - \frac{\epsilon_1}{2} + \rho(u))} \, ,
\end{align}
where the contour is chosen by the JK-residue prescription. At 1-string order, the defect Higgsing gives
\begin{align}
    &\Psi_1 = \sum_{I=1}^4 \frac{\prod_{l=1}^8 \theta_I(\tau, m_l)}{2\eta(\tau)^6 \theta_1(\tau, \epsilon_1) \theta_1(\tau,\epsilon_2)} \frac{\theta_I(\tau, x- \epsilon_1/2- \epsilon_2)}{\theta_I(\tau,x-\epsilon_1/2)} \, .
\end{align}
Similarly, the result for the 2-string order is given by
\begin{align}
    &\Psi_2 = \frac{1}{4\eta(\tau)^{12} \theta_1(\tau,\epsilon_{1,2})} \sum_{I=1}^4 \Bigg( \frac{\prod_{l=1}^8 \theta_I(\tau,m_l \pm \frac{\epsilon_1}{2})}{\theta_1(\tau,2\epsilon_1) \theta_1(\tau,\epsilon_2-\epsilon_1)} \frac{\theta_I(\tau,x-2\epsilon_+) \theta_I(\tau,x-\epsilon_2)}{\theta_I(\tau,x-\epsilon_1) \theta_I(\tau,x)} \\
    & \qquad \qquad \qquad \qquad \qquad \qquad + \frac{\prod_{l=1}^8 \theta_I(m_l \pm \frac{\epsilon_2}{2})}{\theta_1(2\epsilon_2) \theta_1(\epsilon_1-\epsilon_2)} \frac{\theta_I(\tau,x - \epsilon_+ - \epsilon_2)}{\theta_I(\tau,x - \epsilon_-)}\Bigg) \nonumber \\
    &+ \sum_{I\leq J}^4 \frac{\theta_{\sigma(I,J)}(\tau,0) \theta_{\sigma(I,J)}(\tau,2\epsilon_+) \prod_{l=1}^8 \theta_I(\tau,m_l) \theta_J(\tau,m_l)}{4\eta(\tau)^{12} \theta_1(\tau,\epsilon_{1,2})^2 \theta_{\sigma(I,J)}(\tau, \epsilon_{1,2})} \frac{\theta_I(\tau,x \!-\! \frac{\epsilon_1}{2} \!-\! \epsilon_2) \theta_J(\tau,x \!-\! \frac{\epsilon_1}{2}\!-\!\epsilon_2)}{\theta_I(\tau,x\!-\!\frac{\epsilon_1}{2}) \theta_J(\tau,x\!-\!\frac{\epsilon_1}{2})} \nonumber \\
    &+ \frac{\theta_1(\tau,2\epsilon_+) \prod_{l=1}^8 \theta_1(\tau,m_l\pm(x-\frac{\epsilon_1}{2}))}{2\eta(\tau)^{12} \theta_1(\tau,\epsilon_1) \theta_1(\tau,2x-\epsilon_1) \theta_1(\tau,2x -2 \epsilon_1) \theta_1(\tau,2x) \theta_1(\tau,2x+2\epsilon_-)} \, . \nonumber
\end{align}

\section{Defect partition functions for \texorpdfstring{$\mathrm{SU}(2)_{\pi}+1\mathbf{Adj}$}{SU(2)pi+1Adj}} \label{app:piadj}

Here, we provide the explicit expressions for the partition functions of $ \mathrm{SU}(2)_\pi + 1\mathbf{Adj} $ theory with/without a defect, utilizing the map \eqref{eq:mapsthetapi6d} from the E-string theory. The 1-string elliptic genus \eqref{eq:Estr-Z1} of the E-string theory maps to
\begin{align}
    Z_1 = -\frac{\theta_4(2\tau,0)^4 \theta_1(2\tau, m \pm \epsilon_+) \theta_1(2\tau, m \pm \epsilon_-)}{\eta(\tau)^6 \theta_1(\tau,\epsilon_1) \theta_1(\tau,\epsilon_2)} \, ,
\end{align}
while the 2-string elliptic genus \eqref{eq:Estr-Z2} yields
\begin{align}
    \begin{aligned}
        Z_2 &= \frac{\theta_4(2\tau,0)^8}{2\eta(\tau)^{12} \theta_1(\tau,\epsilon_1) \theta_1(\tau,\epsilon_2)} \left( \prod_{s_{1,2}=\pm 1} \frac{\theta_1(2\tau, m + s_1 \epsilon_1 + s_2 \epsilon_\pm)}{\theta_1(\tau,2\epsilon_1) \theta_1(\tau,\epsilon_2-\epsilon_1)} + (\epsilon_1 \leftrightarrow \epsilon_2) \right) \\
        &\quad + \frac{\theta_2(\tau,0) \theta_4(2\tau,0)^8 \theta_2(\tau,2\epsilon_+)}{4\eta(\tau)^{12} \theta_1(\tau,\epsilon_{1,2})^2 \theta_2(\tau,\epsilon_{1,2})} \left( \theta_1(2\tau, m \pm \epsilon_+)^2 \theta_1(2\tau, m \pm \epsilon_-)^2  - (\theta_1 \to \theta_4) \right) \, .
    \end{aligned}
\end{align}
These correspond to the 1-string and 2-string elliptic genera of $ \mathbb{Z}_2 $ twist of the rank-2 M-string, which is UV dual to the 5d $ \mathrm{SU}(2)_\pi + 1\mathbf{Adj} $ theory. This result is consistent with the GV-invariant computation based on the blowup equation \cite{Kim:2020hhh}.

In a similar way, one writes down the expressions for the elliptic genus in the presence of codimension-4 defect. In the 5d picture, this defect corresponds to the Wilson loop operator in the fundamental representation of $ \mathrm{SU}(2) $. From the codimension-4 defect partition function \eqref{eq:Estr-codim4} of the E-string theory, one finds
\begin{align}
    Z_1^{(W)} &= \sum_{I=1}^4 \frac{(-1)^{1+\delta_{I4}} \theta_4(2\tau,0)^4}{2\eta(\tau)^6 \theta_1(\tau,\epsilon_{1,2})} \frac{\theta_{\mu(I)}(2\tau, m \pm \epsilon_+) \theta_{\mu(I)}(2\tau, m \pm \epsilon_-) \theta_I(\tau, z \pm \epsilon_-)}{\theta_I(\tau, z \pm \epsilon_+)} \, ,
\end{align}
and
\begin{align}
    Z_2^{(W)} &= \sum_{I=1}^4 \left( \frac{\theta_4(2\tau,0)^8 \theta_I(z \pm (\epsilon_1 - \frac{\epsilon_2}{2})) \prod_{s_{1,2}=\pm1} \theta_{\mu(I)}(2\tau, m + s_1 \epsilon_1 + s_2\epsilon_\pm)}{4\eta(\tau)^{12} \theta_1(\tau,\epsilon_1) \theta_1(\tau,\epsilon_2) \theta_I(\tau, z \pm (\epsilon_1 + \frac{\epsilon_2}{2}))} + (\epsilon_1 \leftrightarrow \epsilon_2) \right) \nonumber \\
    &+ \sum_{I \leq J}^4 \frac{(-1)^{\delta_{J4}} \theta_{\sigma(I,J)}(\tau,0) \theta_{\sigma(I,J)}(\tau,2\epsilon_+) \theta_4(2\tau,0)^8}{4\eta(\tau)^{12} \theta_1(\tau,\epsilon_{1,2})^2 \theta_{\sigma(I,J)}(\tau,\epsilon_{1,2})} \frac{\theta_I(\tau, z \pm \epsilon_-) \theta_J(\tau, z \pm \epsilon_-)}{\theta_I(\tau, z \pm \epsilon_+) \theta_J(\tau, z \pm \epsilon_+)} \nonumber \\
    &\quad \cdot \left( \prod_{s_1=\pm1} \theta_{\mu(I)}(2\tau, m + s_1 \epsilon_\pm) \theta_{\mu(J)}(2\tau, m + s_1 \epsilon_\pm) \right)  \\
    &+ \left( \frac{\theta_4(2\tau,0)^8 \prod_{s_{1,2}=\pm1} \theta_1(2\tau, 2z+2\epsilon_+ + s_1m + s_2\epsilon_\pm)}{2\eta(\tau)^{12} \theta_1(\tau,2z) \theta_1(\tau,2z+2\epsilon_+) \theta_1(\tau,2z+2\epsilon_++\epsilon_{1,2})} + (\epsilon_{1,2} \to -\epsilon_{1,2})  \right) , \nonumber
\end{align}
where the indices $ s_1 $ and $ s_2 $ run over $ \pm 1 $, and $\mu(I)$ take the following values:
\begin{align}
    \mu(1) = \mu(2) = 1 \, , \quad
    \mu(3) = \mu(4) = 4 \, .
\end{align}
The expectation value of the Wilson loop operator is defined as
\begin{align}
    \langle W \rangle = e^{\phi} \left( 1 + \sum_{k=2}^\infty e^{-k\phi} W_k \right) = e^\phi \frac{1+\sum_{k=1}^\infty Z_k^{(W)}}{1+\sum_{k=1}^\infty e^{-k\phi} Z_k} - V_0(z) \, ,
\end{align}
where we omit $ W_1 $ in the summation because the $ k=1 $ order contribution does not depend on the dynamical parameter $ \phi $. Then $ \langle W \rangle $ is independent of $ z $, and the subtraction of the constant for $ \langle W \rangle $ is expressed as $ V_0(z) = Z_1^{(W)} - Z_1 $.

The codimension-2 defect partition function can be treated in a similar manner using \eqref{eq:mapsthetapi6d} and \eqref{eq:xmappi}. From the E-string defect partition function, one obtains the 1-string elliptic genus in the presence of codimension-2 defect is expressed as 
\begin{align}
    \Psi_1 &= \sum_{I=1}^4 (-1)^{1+\delta_{I4}} \frac{\theta_4(2\tau,0)^4 \theta_{\mu(I)}(2\tau, m \pm \epsilon_+) \theta_{\mu(I)}(2\tau, m \pm \epsilon_-) \theta_I(x - \epsilon_+)}{2\eta(\tau)^6 \theta_1(\tau,\epsilon_{1}) \theta_1(\tau, \epsilon_2) \theta_I(x - \epsilon_-)} \, ,
\end{align}
while the 2-string elliptic genus in the presence of codimension-2 defect is expressed as
\begin{align}
    \Psi_2 &= \frac{\theta_4(2\tau,0)^8}{4\eta(\tau)^{12} \theta_1(\tau,\epsilon_{1,2})} \Bigg( \sum_{I=1}^4 \frac{\prod_{s_{1,2}=\pm1} \theta_{\mu(I)}(2\tau, m + s_1 \epsilon_1 + s_2 \epsilon_\pm)}{\theta_1(\tau,2\epsilon_1) \theta_1(\tau,\epsilon_2-\epsilon_1)} \frac{\theta_I(\tau,x - \epsilon_+ \pm \frac{\epsilon_1}{2})}{\theta_I(\tau,x - \epsilon_- \pm \frac{\epsilon_1}{2})} \nonumber \\
    & \qquad \qquad \qquad \qquad \qquad + \frac{\prod_{s_{1,2}=\pm1} \theta_{\mu(I)}(2\tau, m + s_2 \epsilon_1 + s_2 \epsilon_\pm)}{\theta_1(\tau,2\epsilon_1) \theta_1(\tau,\epsilon_2-\epsilon_1)} \frac{\theta_I(\tau,x - \epsilon_+ - \frac{\epsilon_2}{2})}{\theta_I(\tau,x - \epsilon_- + \frac{\epsilon_2}{2})} \Bigg) \nonumber \\
    & + \sum_{I\leq J}^4 \frac{(-1)^{\delta_{J4}} \theta_{\sigma(I,J)}(\tau,0) \theta_{\sigma(I,J)}(\tau,2\epsilon_+) \theta_4(2\tau,0)^8 }{4\eta(\tau)^{12} \theta_1(\tau,\epsilon_{1,2})^2} \nonumber \\
    & \qquad \cdot \frac{\prod_{s_1=\pm1} \theta_{\mu(I)}(2\tau, m + s_1 \epsilon_\pm) \theta_{\mu(J)}(2\tau, m + s_1 \epsilon_\pm)}{ \theta_{\sigma(I,J)}(\epsilon_{1,2})} \frac{\theta_I(\tau,x - \epsilon_+) \theta_J(\tau,x - \epsilon_+)}{\theta_I(\tau,x-\epsilon_-) \theta_J(\tau,x-\epsilon_-)} \nonumber \\
    &+ \frac{\theta_1(\tau,2\epsilon_+) \theta_4(2\tau,0)^8 \prod_{s_{1,2}=\pm1} \theta_1(2\tau, m + s_1(2x-\epsilon_-) + s_2 \epsilon_\pm)}{2\eta(\tau)^{12} \theta_1(\tau,\epsilon_1) \theta_1(\tau,2x-2\epsilon_-) \theta_1(\tau,2x - 2\epsilon_- \pm \epsilon_1) \theta_1(\tau,2x - 2\epsilon_- + \epsilon_2) } \, .
\end{align}

\section{Hanany-Witten transition in the \texorpdfstring{$\mathrm{SU}(2)$}{SU(2)} defect partition function} \label{app:hw}

In Section~\ref{sec:qcurve_HW}, we demonstrated that the eigenfunction for the quantum curve of the $\mathrm{SU}(2)_{\theta+2\pi}$ theory is related to that for the curve of the $\mathrm{SU}(2)_{\theta}$ theory via \eqref{eq:psitheta}. In this appendix, we utilize this relation to derive the defect partition function of the $\mathrm{SU}(2)_{\theta}$ theory with any theta angles, starting from the defect partition functions of the $\mathrm{SU}(2)_{0,\pi}$ theories.

To begin with, we write the codimension-2 defect partition function for the $ \mathrm{SU}(2)_\theta $ theory as
\begin{align}\label{eq:Psitheta}
    \Psi^\theta(x) = e^{\frac{x^2}{2\epsilon_1}-\frac{x}{2}}\cdot (X/\mu_1;q_1)_{\infty}^{-1}(X/\mu_2;q_1)_{\infty}^{-1}\cdot\Psi^{\theta}_{\mathrm{inst}}(x) \, ,
\end{align}
where $ \Psi^\theta_{\mathrm{inst}} $ is the instanton partition function in the presence of the defect, and the prefactors are the effective prepotential $({\frac{x^2}{2\epsilon_1}-\frac{x}{2}})$ and the 1-loop contributions with $ \mu_j = e^{-a_j} $. Unlike \eqref{eq:SU2-E1}, we here omit $ \phi^2/2\epsilon_1  - \pi i x/\epsilon_1 $ in the effective prepotential where $- \pi i x/\epsilon_1 $ affects only the sign of $ Y $-dependent terms in the quantum curve, while the $\phi^2/2\epsilon_1$ term does not affect to the curve. As we are interested in the quantum curve, we shall only keep the prefactors that are relevant to the quantum curve, dropping the irrelevant factors from here on. For instance, for $ \theta=0 $, \eqref{eq:Psitheta} is an eigenfunction of the following curve,
\begin{align}\label{eq:theta0curve}
    -X-X^{-1}+Y+\mathfrak{q}\,Y^{-1} + E = 0 \, .
\end{align}
Consider now a slight modified version of \eqref{eq:psitheta}:
\begin{align}\label{eq:psitheta2}
    \widetilde{\Psi}(x) \equiv \Psi^{\theta+2\pi}(x)=e^{-\frac{x^2}{2\epsilon_1}}\int d{s_1}d{s_2} \: e^{\frac{s_2x}{\epsilon_1}}e^{-\frac{s_1s_2}{\epsilon_1}}e^{-\frac{s_1^2}{2\epsilon_1}}(\sqrt{q_1}\sigma_2^{-1};q_1)_{\infty}^{-1}\Psi^{\theta}(s_1) \, ,
\end{align}
where $ \sigma_j = e^{-s_j} $ and $ q_j = e^{-\epsilon_j} $. It follows from this that for $\theta=0$, $ \widetilde{\Psi} $ is an eigenfunction of the curve, 
\begin{align}
    -X - X^{-1} + Y + \mathfrak{q} \left( X Y^{-1} X - X \right) + E = 0 \, .
\end{align}

In the following, we study the integral \eqref{eq:psitheta2} order by order in instanton number. Let us consider the integral at zero-instanton order, $ \mathfrak{q} \to 0 $:
\begin{align}
    \widetilde{\Psi}_0(x) = e^{-\frac{x^2}{2\epsilon_1}}\int d{s_1}d{s_2} \, \frac{e^{\frac{s_2x}{\epsilon_1}}e^{-\frac{s_1s_2}{\epsilon_1}}e^{-\frac{s_1}{2}}}{(\sqrt{q_1}\sigma_2^{-1};q_1)_{\infty}(\sigma_1/\mu_1;q_1)_{\infty} (\sigma_1/\mu_2;q_1)_{\infty}} \, .
\end{align}
We first perform the integral over $ s_2 $ by choosing the poles at $ \sigma_2 = \sqrt{q_1} \cdot q_1^n $ for $ n \in \mathbb{Z}_{\geq 0} $. Using \eqref{eq:poch-residue} and \eqref{eq:poch-id}, we obtain
\begin{align}\label{eq:123}
    \widetilde{\Psi}_0(x) = \frac{e^{-\frac{x^2}{2\epsilon_1} + \frac{x}{2}}}{(q_1;q_1)_{\infty}}\int ds_1 \: \frac{-e^{-s_1}(q_1\sigma_1/X;q_1)_{\infty}}{{(\sigma_1/\mu_1;q_1)_{\infty} (\sigma_1/\mu_2;q_1)_{\infty}}} \, .
\end{align}
Next, we perform a Fourier transformation on both sides by multiplying $ e^{a_2y/\epsilon_1} $ and integrating over $ a_2 $. The integral over $ a_2 $ is evaluated by choosing the poles at $ \mu_2 = \sigma_1 q_1^n $ for $ n \in \mathbb{Z}_{\geq 0} $ as
\begin{align}
    \int da_2 \: e^{\frac{a_2 y}{\epsilon_1}} \widetilde{\Psi}_0(x)
    &= \frac{e^{-\frac{x^2}{2\epsilon_1}+\frac{x}{2}}(q_1e^{y};q_1)_{\infty}}{(q_1;q_1)_{\infty}^2} \int d{s_1}e^{-s_1+s_1 y/\epsilon_1}\frac{(q_1\sigma_1/X;q_1)_{\infty}}{(\sigma_1/\mu_1;q_1)_{\infty}}  \\
    &= \frac{e^{-\frac{x^2}{2\epsilon_1}+\frac{x}{2}}(q_1e^{y};q_1)_{\infty}}{(q_1;q_1)_{\infty}^3} \frac{e^{-a_1(1-y/\epsilon_1)} (q_1 \mu_1/X; q_1)_\infty (e^{-y}; q_1)_\infty}{(e^{-y} \mu_1/X; q_1)_\infty} \, . \nonumber
\end{align}
In the second line, the integral over $ s_1 $ was evaluated using the poles at $ \sigma_1 = \mu_1 q_1^{-n} $ for $ n\in \mathbb{Z}_{\geq 0} $. Then we perform the inverse Fourier transformation as
\begin{align}\label{eq:Psi0-fourier}
    \widetilde{\Psi}_0(x)
    &= \frac{e^{-\frac{x^2}{2\epsilon_1}+\frac{x}{2}-a_1} (q_1 \mu_1/X; q_1)_\infty }{(q_1;q_1)_{\infty}^3} \int dy \: \frac{(q_1 e^{y};q_1)_\infty (e^{-y}; q_1)_\infty}{(e^{-y} \mu_1/X; q_1)_\infty} e^{(a_1-a_2)y/\epsilon_1} \\
    &= \frac{e^{-\frac{x^2}{2\epsilon_1}+\frac{x}{2}-a_1} (q_1 \mu_1/X; q_1)_\infty }{(q_1;q_1)_{\infty}^4} \frac{e^{(x-a_1)(a_1-a_2)/\epsilon_1} (q_1 \mu_1/X;q_1)_\infty (X/\mu_1; q_1)_\infty}{(X/\mu_2; q_1)_\infty} \, . \nonumber
\end{align}
From the analytic continuation formula\footnote{This formula is derived from the relation
\begin{align*}
    (X; q_1)_\infty = \exp\left(\sum_{n=0}^\infty (-1)^{n-1} \frac{B_n}{n!} \operatorname{Li}_{2-n}(X) \epsilon_1^{n-1}\right) \, ,
\end{align*}
where $ \operatorname{Li}_n(z) = \sum z^k/k^n $ is the polylogarithm function and $ B_n $ is the $ n $-th Bernoulli number. To convert $ X $ into $ q_1/X $, we use the inversion formula of the polylogarithm, given by
\begin{align*}
    \operatorname{Li}_n(X) + (-1)^n \operatorname{Li}_n(1/X) = -\frac{(2\pi i)^n}{n!} B_n\left(\frac{1}{2} \pm \frac{\log(-X^{\pm 1})}{2\pi i} \right) \, ,
\end{align*}
where $ B_n(z) $ is the Bernoulli polynomial with $ B_n(z)=0 $ for the negative integers $ n $. The plus and minus signs correspond to $ X \notin (0,1) $ and $ X \notin (1,\infty) $, respectively. This sign change only affects \eqref{eq:AC} by altering the signs of $ -i\pi x/\epsilon_1 $ and $ i\pi/2 $ terms, which is irrelevant for our purpose.}
\begin{align}\label{eq:AC}
    (X;q_1)_{\infty}=\frac{e^{\frac{x^2}{2\epsilon_1} - \frac{x}{2}-\frac{i\pi x}{\epsilon_1} -\frac{\pi^2}{3\epsilon_1} + \frac{i\pi}{2}+\frac{\epsilon_1}{12}}}{(q_1/X;q_1)_{\infty}} \, ,
\end{align}
the defect partition function \eqref{eq:Psi0-fourier} can be written as
\begin{align}\label{eq:Psi0-transform}
    \widetilde{\Psi}_0(x)
    &= \frac{e^{S}}{(q_1;q_1)_{\infty}^4} \frac{e^{\frac{x^2}{2\epsilon_1} - \frac{x}{2}}}{(X/\mu_1; q_1)_\infty (X/\mu_2; q_1)_\infty} \, .
\end{align}
Here, $ S =  -x (a_1+a_2+2\pi i)/\epsilon_1  +\cdots$, where $\cdots$ is $x$-independent part. After identifying $ a_1+a_2=0 $, we find that the $ x $-dependent term in $ S $ does not contribute to the quantum curve. Thus, we obtain the perturbative part of the defect partition function for the $\mathrm{SU}(2)_{\theta+2\pi} $ theory using the transformation \eqref{eq:psitheta2}, up to an overall factor that does not affect to the quantum curve.

The higher-instanton computation closely follows the structure of the zero-instanton case. We express the instanton partition function of the $ \mathrm{SU}(2)_\theta $ theory as $ \Psi_{\mathrm{inst}}^\theta = 1 + \sum_{k \geq 1} \mathfrak{q}^k \Psi_k^\theta $, where the $ k $-instanton partition function $ \Psi_k^\theta $ exhibits the pole structures given by
\begin{align}\label{eq:Psik-pole}
    \Psi_{k}^\theta(s_1) = \frac{\sum_{j,l,m}{c_{jlm}} \sigma_1^j \mu_1^l \mu_2^m}{f(\mu_1,\mu_2;q_1,q_2)\prod_{n=1}^k(1-q_1^{-n} \sigma_1/\mu_1)(1-q_1^{-n}\sigma_1/\mu_2)} \, .
\end{align}
Here, the numerator is a finite summation,  and $ f(\mu_1,\mu_2;q_1,q_2) $ is a function that introduces additional poles involving $ q_2 $. The $ k $-instanton contribution to the integral \eqref{eq:psitheta2} is given by
\begin{align}\label{eq:12}
    \widetilde{\Psi}_0(x) \widetilde{\Psi}_k(x) &= e^{-\frac{x^2}{2\epsilon_1}}\int d{s_1}d{s_2} \: \frac{e^{\frac{s_2x}{\epsilon_1}} e^{-\frac{s_1s_2}{\epsilon_1}} e^{\frac{s_1}{2}}}{(\sqrt{q_1} \sigma_2;q_1)_{\infty} (\sigma_1/\mu_1;q_1)_{\infty} (\sigma_1/\mu_2;q_1)_{\infty}} \Psi_{k}^\theta(s_1) \\
    &=\frac{e^{-\frac{x^2}{2\epsilon_1}+\frac{x}{2}}}{(q_1;q_1)_{\infty}}\int d{s_1} \: \frac{ -e^{-s_1}(q_1\sigma_1/X;q_1)_{\infty}}{(q_1^{-k} \sigma_1/\mu_1;q_1)_{\infty} (q_1^{-k} \sigma_1/\mu_2;q_1)_{\infty}} \frac{\sum_{j,l,m} c_{jlm} \sigma_1^j \mu_1^l \mu_2^m}{f(\mu_1,\mu_2; q_1,q_2)} \, , \nonumber
\end{align}
where $ \widetilde{\Psi}_0(x) $ is the perturbative partition function given in \eqref{eq:Psi0-transform} and the integral over $ s_2 $ was evaluated using the poles at $ \sigma_2 = q_1^{(n+1/2)} $ for $ n \in \mathbb{Z}_{\geq 0} $. Next, we multiply $ f(\mu_1,\mu_2,q_1,q_2) $ on both sides and perform the Fourier transformation on $ a_2 $. This involves summing up residues over the poles at $ \mu_2 = \sigma_1 q_1^{n-k} $ for $ n \in \mathbb{Z}_{\geq 0} $:
\begin{align}
    &\int d a_2 \: e^{\frac{a_2 y}{\epsilon_1}} f(\mu_1,\mu_2;q_1,q_2) \widetilde{\Psi}_0(x) \widetilde{\Psi}_k(x) \\
    &= \sum_{j,l,m} c_{jlm} \frac{e^{-\frac{x^2}{2\epsilon_1} + \frac{x}{2}} (q_1^{m+1}e^y; q_1)_\infty }{(q_1; q_1)_\infty^2} \int ds_1 \: e^{-s_1+s_1y/\epsilon_1} \frac{(q_1 \sigma_1/X; q_1)_\infty}{(q_1^{-k} \sigma_1/\mu_1; q_1)_\infty} \sigma_1^{j+m} \mu_1^l q_1^{-km} e^{-ky} \nonumber \\
    &= \sum_{j,l,m} c_{jlm} \frac{e^{-\frac{x^2}{2\epsilon_1} + \frac{x}{2}} (q_1^{m+1}e^y; q_1)_\infty }{(q_1; q_1)_\infty^3} \frac{(q_1^{k+1} \mu_1/X; q_1)_\infty (q_1^{-j-m} e^{-y}; q_1)_\infty}{(q_1^{k-j-m} e^{-y} \mu_1/X; q_1)_\infty} e^{\frac{y a_1}{\epsilon_1}} q_1^{k(1+j)} \mu_1^{1+j+l+m}. \nonumber
\end{align}
The last line is obtained by evaluating the integral over $ s_1 $ by selecting the poles at $ \sigma_1 = \mu_1 q_1^{-n+k} $, where $ n\in \mathbb{Z}_{\geq 0} $. The inverse Fourier transformation, performed by multiplying $ e^{-a_2y/\epsilon_1} $ and integrating over $ y $, yields
\begin{align}
    &f(\mu_1,\mu_2;q_1,q_2) \widetilde{\Psi}_0(x) \widetilde{\Psi}_k(x) \\
    &= \sum_{j,l,m} \frac{c_{jlm} e^{-\frac{x^2}{2\epsilon_1} + \frac{x}{2}} (q_1^{k+1} \mu_1/X; q_1)_\infty}{(q_1; q_1)_\infty^3 \cdot e^{k(1+j)\epsilon_1 + (1+j+l+m)a_1}} \int dy\: \frac{(q_1^{m+1}e^y; q_1)_\infty (q_1^{-j-m} e^{-y}; q_1)_\infty}{(q_1^{k-j-m} e^{-y} \mu_1/X; q_1)_\infty} e^{\frac{y(a_1-a_2)}{\epsilon_1}} \nonumber \\
    &= \sum_{j,l,m} \sum_{n=0}^\infty \frac{c_{jlm} e^{-\frac{x^2}{2\epsilon_1} + \frac{x}{2} + \frac{(x-a_1+(j-k+m)\epsilon_1)(a_1-a_2)}{\epsilon_1}}}{(q_1; q_1)_\infty^4 \cdot e^{k(1+j)\epsilon_1 + (1+j+l+m)a_1}} \frac{(q_1^{k+1}\mu_1/X;q_1)_\infty^2 (q_1^{-k} X/\mu_1; q_1)_\infty}{(q_1;q_1)_n (q_1^{k+n+1}\mu_1/X;q_1)_{-j}} \left(\frac{X}{q_1^k \mu_2}\right)^n . \nonumber
\end{align}
By applying the analytic continuation formula \eqref{eq:AC}, the obtained expression can be further simplified to yield
\begin{align}\label{eq:inst-transform}
    \widetilde{\Psi}_k(x) = \sum_{j,l,m,h} \frac{c_{jlm} \mu_1^l \mu_2^{j+m} q_1^{jk-k^2} X^{2k} }{(\mu_1\mu_2)^{k}} \frac{b_{jh} (X/\mu_1; q_1)_{-k} (X/\mu_2; q_1)_{h-k}}{f(\mu_1,\mu_2; q_1,q_2)} \left(\frac{q_1^k \mu_1}{X} \right)^h \, ,
\end{align}
where the summations over $ j,l,m,h $ are all finite, and $ b_{jh} $ is defined by
\begin{align}
    \frac{1}{(q_1 z; q_1)_{-j}} = \sum_{h=0}^j b_{jh} z^h \, .
\end{align}

We now compare the defect partition functions obtained from \eqref{eq:inst-transform} with the results in Section~\ref{subsubsec:su2-blowup}. Suppose that we start from the codimension-2 defect partition function $ \Psi_{\mathrm{inst}}^{\theta=0} $, computed using either the blowup equation or the ADHM computation in Appendix~\ref{app:ADHM}. Up to 2-instanton order, we have verified that the transformation \eqref{eq:inst-transform} yields
\begin{align}
     \widetilde{\Psi}_k^{\theta=2\pi} = q_2^{-k} \Psi_k^{\theta=\mathrm{2\pi}} \, ,
\end{align}
where $ \Psi_k^{\theta=2\pi} $ is the $ k $-instanton defect partition function of the $ \mathrm{SU}(2)_{\theta=2\pi} $ theory, computed from the blowup equation, as shown at 1-instanton order, given in \eqref{eq:su2-2pi-psi1}. Suppose that we instead start from $ \Psi_{\mathrm{inst}}^{\theta=2\pi} $. By applying the transformation \eqref{eq:inst-transform}, we obtain
\begin{align}
    1 + \sum_{k=1}^\infty \mathfrak{q}^k \widetilde{\Psi}_k^{\theta=4\pi} =\PE\left[ -\frac{q_1 \mathfrak{q}}{1-q_1} \right] \cdot \left( 1 + \sum_{k=1}^\infty (q_2^{-1} \mathfrak{q})^k \Psi_k^{\theta=4\pi} \right) \, ,
\end{align}
where $ \Psi_k^{\theta=4\pi} $ is the $ k $-instanton defect partition function of the $ \mathrm{SU}(2)_{\theta=4\pi} $ theory, computed from the blowup equation, as shown to 1-instanton order in \eqref{eq:su2-4pi-psi1}. We checked this relation up to 2-instanton order. Similarly, starting from the $ \Psi_{\mathrm{inst}}^{\theta=\pi} $ and applying the transformation \eqref{eq:inst-transform}, we find
\begin{align}
     \widetilde{\Psi}_k^{\theta=3\pi} = q_2^{-k} \Psi_k^{\theta=\mathrm{3\pi}} \, ,
\end{align}
where $ \Psi_k^{\theta=\mathrm{3\pi}} $ is the $ \mathrm{SU}(2)_{\theta=3\pi} $ defect partition function computed from the blowup equation as \eqref{eq:su2-3-defect}, and we checked the equivalence up to 2-instanton order. Thus, the transformation \eqref{eq:psitheta2} reproduces the defect partition functions of higher theta angles, up to an overall factor that does not affect to the quantum curves.

For larger theta angles $ \theta>4\pi $, the result of transformation \eqref{eq:inst-transform} includes unphysical terms that are independent of $ x $. For instance, applying the transformation \eqref{eq:inst-transform} to $ \Psi_1^{\theta=4\pi} $ yields: 
\begin{align}
    \widetilde{\Psi}_1^{\theta=6\pi} = q_2^{-1} \Psi_1^{\theta=6\pi} - \frac{q_1^2 q_2 (Q^2+Q^{-2})}{1-q_1} - \frac{q_1(1+q_1q_2)}{1-q_1} \, .
\end{align}
Here, $ Q=e^{-\phi} $ and $ \Psi_1^{\theta=6\pi} $ is the conjectured codimension-2 defect partition function for the $ \mathrm{SU}(2)_{\theta=6\pi} $ theory, whose zeroth-order in the $ e^{-x} $-expansion corresponds to the partition function without a defect, as shown in \eqref{eq:su2-m2-defect}-\eqref{eq:su2-4pi-psi1}. Since the transformation \eqref{eq:psitheta2} is motivated by the transition of quantum curves, we expect that the result for larger theta angles, such as $ \Psi_1^{\theta=6\pi} $, at least captures the NS-limit of the defect partition function.

\bibliographystyle{JHEP}
\bibliography{refs}

\providecommand{\href}[2]{#2}\begingroup\raggedright\begin{thebibliography}{100}

\bibitem{Seiberg:1996bd}
N.~Seiberg, \emph{{Five-dimensional SUSY field theories, nontrivial fixed points and string dynamics}}, \href{https://doi.org/10.1016/S0370-2693(96)01215-4}{\emph{Phys. Lett. B} {\bfseries 388} (1996) 753} [\href{https://arxiv.org/abs/hep-th/9608111}{{\ttfamily hep-th/9608111}}].

\bibitem{Douglas:1996xp}
M.R.~Douglas, S.H.~Katz and C.~Vafa, \emph{{Small instantons, Del Pezzo surfaces and type I-prime theory}}, \href{https://doi.org/10.1016/S0550-3213(97)00281-2}{\emph{Nucl. Phys. B} {\bfseries 497} (1997) 155} [\href{https://arxiv.org/abs/hep-th/9609071}{{\ttfamily hep-th/9609071}}].

\bibitem{Intriligator:1997pq}
K.A.~Intriligator, D.R.~Morrison and N.~Seiberg, \emph{{Five-dimensional supersymmetric gauge theories and degenerations of Calabi-Yau spaces}}, \href{https://doi.org/10.1016/S0550-3213(97)00279-4}{\emph{Nucl. Phys. B} {\bfseries 497} (1997) 56} [\href{https://arxiv.org/abs/hep-th/9702198}{{\ttfamily hep-th/9702198}}].

\bibitem{Heckman:2013pva}
J.J.~Heckman, D.R.~Morrison and C.~Vafa, \emph{{On the Classification of 6D SCFTs and Generalized ADE Orbifolds}}, \href{https://doi.org/10.1007/JHEP05(2014)028}{\emph{JHEP} {\bfseries 05} (2014) 028} [\href{https://arxiv.org/abs/1312.5746}{{\ttfamily 1312.5746}}].

\bibitem{Heckman:2015bfa}
J.J.~Heckman, D.R.~Morrison, T.~Rudelius and C.~Vafa, \emph{{Atomic Classification of 6D SCFTs}}, \href{https://doi.org/10.1002/prop.201500024}{\emph{Fortsch. Phys.} {\bfseries 63} (2015) 468} [\href{https://arxiv.org/abs/1502.05405}{{\ttfamily 1502.05405}}].

\bibitem{Bhardwaj:2015oru}
L.~Bhardwaj, M.~Del~Zotto, J.J.~Heckman, D.R.~Morrison, T.~Rudelius and C.~Vafa, \emph{{F-theory and the Classification of Little Strings}}, \href{https://doi.org/10.1103/PhysRevD.93.086002}{\emph{Phys. Rev. D} {\bfseries 93} (2016) 086002} [\href{https://arxiv.org/abs/1511.05565}{{\ttfamily 1511.05565}}].

\bibitem{Bhardwaj:2015xxa}
L.~Bhardwaj, \emph{{Classification of 6d $ \mathcal{N}=\left(1,0\right) $ gauge theories}}, \href{https://doi.org/10.1007/JHEP11(2015)002}{\emph{JHEP} {\bfseries 11} (2015) 002} [\href{https://arxiv.org/abs/1502.06594}{{\ttfamily 1502.06594}}].

\bibitem{DelZotto:2017pti}
M.~Del~Zotto, J.J.~Heckman and D.R.~Morrison, \emph{{6D SCFTs and Phases of 5D Theories}}, \href{https://doi.org/10.1007/JHEP09(2017)147}{\emph{JHEP} {\bfseries 09} (2017) 147} [\href{https://arxiv.org/abs/1703.02981}{{\ttfamily 1703.02981}}].

\bibitem{Xie:2017pfl}
D.~Xie and S.-T.~Yau, \emph{{Three dimensional canonical singularity and five dimensional $ \mathcal{N} $ = 1 SCFT}}, \href{https://doi.org/10.1007/JHEP06(2017)134}{\emph{JHEP} {\bfseries 06} (2017) 134} [\href{https://arxiv.org/abs/1704.00799}{{\ttfamily 1704.00799}}].

\bibitem{Jefferson:2017ahm}
P.~Jefferson, H.-C.~Kim, C.~Vafa and G.~Zafrir, \emph{{Towards classification of 5d SCFTs: Single gauge node}}, \href{https://doi.org/10.21468/SciPostPhys.14.5.122}{\emph{SciPost Phys.} {\bfseries 14} (2023) 122} [\href{https://arxiv.org/abs/1705.05836}{{\ttfamily 1705.05836}}].

\bibitem{Jefferson:2018irk}
P.~Jefferson, S.~Katz, H.-C.~Kim and C.~Vafa, \emph{{On Geometric Classification of 5d SCFTs}}, \href{https://doi.org/10.1007/JHEP04(2018)103}{\emph{JHEP} {\bfseries 04} (2018) 103} [\href{https://arxiv.org/abs/1801.04036}{{\ttfamily 1801.04036}}].

\bibitem{Bhardwaj:2019fzv}
L.~Bhardwaj, P.~Jefferson, H.-C.~Kim, H.-C.~Tarazi and C.~Vafa, \emph{{Twisted Circle Compactifications of 6d SCFTs}}, \href{https://doi.org/10.1007/JHEP12(2020)151}{\emph{JHEP} {\bfseries 12} (2020) 151} [\href{https://arxiv.org/abs/1909.11666}{{\ttfamily 1909.11666}}].

\bibitem{Apruzzi:2019vpe}
F.~Apruzzi, C.~Lawrie, L.~Lin, S.~Sch\"afer-Nameki and Y.-N.~Wang, \emph{{5d Superconformal Field Theories and Graphs}}, \href{https://doi.org/10.1016/j.physletb.2019.135077}{\emph{Phys. Lett. B} {\bfseries 800} (2020) 135077} [\href{https://arxiv.org/abs/1906.11820}{{\ttfamily 1906.11820}}].

\bibitem{Apruzzi:2019opn}
F.~Apruzzi, C.~Lawrie, L.~Lin, S.~Sch\"afer-Nameki and Y.-N.~Wang, \emph{{Fibers add Flavor, Part I: Classification of 5d SCFTs, Flavor Symmetries and BPS States}}, \href{https://doi.org/10.1007/JHEP11(2019)068}{\emph{JHEP} {\bfseries 11} (2019) 068} [\href{https://arxiv.org/abs/1907.05404}{{\ttfamily 1907.05404}}].

\bibitem{Apruzzi:2019enx}
F.~Apruzzi, C.~Lawrie, L.~Lin, S.~Sch\"afer-Nameki and Y.-N.~Wang, \emph{{Fibers add Flavor, Part II: 5d SCFTs, Gauge Theories, and Dualities}}, \href{https://doi.org/10.1007/JHEP03(2020)052}{\emph{JHEP} {\bfseries 03} (2020) 052} [\href{https://arxiv.org/abs/1909.09128}{{\ttfamily 1909.09128}}].

\bibitem{Apruzzi:2019kgb}
F.~Apruzzi, S.~Schafer-Nameki and Y.-N.~Wang, \emph{{5d SCFTs from Decoupling and Gluing}}, \href{https://doi.org/10.1007/JHEP08(2020)153}{\emph{JHEP} {\bfseries 08} (2020) 153} [\href{https://arxiv.org/abs/1912.04264}{{\ttfamily 1912.04264}}].

\bibitem{Bhardwaj:2020gyu}
L.~Bhardwaj and G.~Zafrir, \emph{{Classification of 5d $ \mathcal{N} $ = 1 gauge theories}}, \href{https://doi.org/10.1007/JHEP12(2020)099}{\emph{JHEP} {\bfseries 12} (2020) 099} [\href{https://arxiv.org/abs/2003.04333}{{\ttfamily 2003.04333}}].

\bibitem{Aharony:1997ju}
O.~Aharony and A.~Hanany, \emph{{Branes, superpotentials and superconformal fixed points}}, \href{https://doi.org/10.1016/S0550-3213(97)00472-0}{\emph{Nucl. Phys. B} {\bfseries 504} (1997) 239} [\href{https://arxiv.org/abs/hep-th/9704170}{{\ttfamily hep-th/9704170}}].

\bibitem{Aharony:1997bh}
O.~Aharony, A.~Hanany and B.~Kol, \emph{{Webs of (p,q) five-branes, five-dimensional field theories and grid diagrams}}, \href{https://doi.org/10.1088/1126-6708/1998/01/002}{\emph{JHEP} {\bfseries 01} (1998) 002} [\href{https://arxiv.org/abs/hep-th/9710116}{{\ttfamily hep-th/9710116}}].

\bibitem{Bergman:2013aca}
O.~Bergman, D.~Rodr\'\i{}guez-G\'omez and G.~Zafrir, \emph{{5-Brane Webs, Symmetry Enhancement, and Duality in 5d Supersymmetric Gauge Theory}}, \href{https://doi.org/10.1007/JHEP03(2014)112}{\emph{JHEP} {\bfseries 03} (2014) 112} [\href{https://arxiv.org/abs/1311.4199}{{\ttfamily 1311.4199}}].

\bibitem{Bergman:2014kza}
O.~Bergman and G.~Zafrir, \emph{{Lifting 4d dualities to 5d}}, \href{https://doi.org/10.1007/JHEP04(2015)141}{\emph{JHEP} {\bfseries 04} (2015) 141} [\href{https://arxiv.org/abs/1410.2806}{{\ttfamily 1410.2806}}].

\bibitem{Hayashi:2015fsa}
H.~Hayashi, S.-S.~Kim, K.~Lee, M.~Taki and F.~Yagi, \emph{{A new 5d description of 6d D-type minimal conformal matter}}, \href{https://doi.org/10.1007/JHEP08(2015)097}{\emph{JHEP} {\bfseries 08} (2015) 097} [\href{https://arxiv.org/abs/1505.04439}{{\ttfamily 1505.04439}}].

\bibitem{Zafrir:2015rga}
G.~Zafrir, \emph{{Brane webs, $5d$ gauge theories and $6d$ $\mathcal{N}=(1,0)$ SCFT's}}, \href{https://doi.org/10.1007/JHEP12(2015)157}{\emph{JHEP} {\bfseries 12} (2015) 157} [\href{https://arxiv.org/abs/1509.02016}{{\ttfamily 1509.02016}}].

\bibitem{Hayashi:2018lyv}
H.~Hayashi, S.-S.~Kim, K.~Lee and F.~Yagi, \emph{{Dualities and 5-brane webs for 5d rank 2 SCFTs}}, \href{https://doi.org/10.1007/JHEP12(2018)016}{\emph{JHEP} {\bfseries 12} (2018) 016} [\href{https://arxiv.org/abs/1806.10569}{{\ttfamily 1806.10569}}].

\bibitem{Nekrasov:2002qd}
N.A.~Nekrasov, \emph{{Seiberg-Witten prepotential from instanton counting}}, \href{https://doi.org/10.4310/ATMP.2003.v7.n5.a4}{\emph{Adv. Theor. Math. Phys.} {\bfseries 7} (2003) 831} [\href{https://arxiv.org/abs/hep-th/0206161}{{\ttfamily hep-th/0206161}}].

\bibitem{Nekrasov:2003rj}
N.~Nekrasov and A.~Okounkov, \emph{{Seiberg-Witten theory and random partitions}}, \href{https://doi.org/10.1007/0-8176-4467-9_15}{\emph{Prog. Math.} {\bfseries 244} (2006) 525} [\href{https://arxiv.org/abs/hep-th/0306238}{{\ttfamily hep-th/0306238}}].

\bibitem{Mitev:2014jza}
V.~Mitev, E.~Pomoni, M.~Taki and F.~Yagi, \emph{{Fiber-Base Duality and Global Symmetry Enhancement}}, \href{https://doi.org/10.1007/JHEP04(2015)052}{\emph{JHEP} {\bfseries 04} (2015) 052} [\href{https://arxiv.org/abs/1411.2450}{{\ttfamily 1411.2450}}].

\bibitem{Hayashi:2014wfa}
H.~Hayashi and G.~Zoccarato, \emph{{Exact partition functions of Higgsed 5d $T_N$ theories}}, \href{https://doi.org/10.1007/JHEP01(2015)093}{\emph{JHEP} {\bfseries 01} (2015) 093} [\href{https://arxiv.org/abs/1409.0571}{{\ttfamily 1409.0571}}].

\bibitem{Hayashi:2016abm}
H.~Hayashi, S.-S.~Kim, K.~Lee and F.~Yagi, \emph{{Equivalence of several descriptions for 6d SCFT}}, \href{https://doi.org/10.1007/JHEP01(2017)093}{\emph{JHEP} {\bfseries 01} (2017) 093} [\href{https://arxiv.org/abs/1607.07786}{{\ttfamily 1607.07786}}].

\bibitem{Kim:2020hhh}
H.-C.~Kim, M.~Kim, S.-S.~Kim and K.-H.~Lee, \emph{{Bootstrapping BPS spectra of 5d/6d field theories}}, \href{https://doi.org/10.1007/JHEP04(2021)161}{\emph{JHEP} {\bfseries 04} (2021) 161} [\href{https://arxiv.org/abs/2101.00023}{{\ttfamily 2101.00023}}].

\bibitem{Bhattacharya:2008zy}
J.~Bhattacharya, S.~Bhattacharyya, S.~Minwalla and S.~Raju, \emph{{Indices for Superconformal Field Theories in 3,5 and 6 Dimensions}}, \href{https://doi.org/10.1088/1126-6708/2008/02/064}{\emph{JHEP} {\bfseries 02} (2008) 064} [\href{https://arxiv.org/abs/0801.1435}{{\ttfamily 0801.1435}}].

\bibitem{Kim:2012gu}
H.-C.~Kim, S.-S.~Kim and K.~Lee, \emph{{5-dim Superconformal Index with Enhanced En Global Symmetry}}, \href{https://doi.org/10.1007/JHEP10(2012)142}{\emph{JHEP} {\bfseries 10} (2012) 142} [\href{https://arxiv.org/abs/1206.6781}{{\ttfamily 1206.6781}}].

\bibitem{Seiberg:1994rs}
N.~Seiberg and E.~Witten, \emph{{Electric - magnetic duality, monopole condensation, and confinement in N=2 supersymmetric Yang-Mills theory}}, \href{https://doi.org/10.1016/0550-3213(94)90124-4}{\emph{Nucl. Phys. B} {\bfseries 426} (1994) 19} [\href{https://arxiv.org/abs/hep-th/9407087}{{\ttfamily hep-th/9407087}}].

\bibitem{Seiberg:1994aj}
N.~Seiberg and E.~Witten, \emph{{Monopoles, duality and chiral symmetry breaking in N=2 supersymmetric QCD}}, \href{https://doi.org/10.1016/0550-3213(94)90214-3}{\emph{Nucl. Phys. B} {\bfseries 431} (1994) 484} [\href{https://arxiv.org/abs/hep-th/9408099}{{\ttfamily hep-th/9408099}}].

\bibitem{Katz:1996fh}
S.H.~Katz, A.~Klemm and C.~Vafa, \emph{{Geometric engineering of quantum field theories}}, \href{https://doi.org/10.1016/S0550-3213(97)00282-4}{\emph{Nucl. Phys. B} {\bfseries 497} (1997) 173} [\href{https://arxiv.org/abs/hep-th/9609239}{{\ttfamily hep-th/9609239}}].

\bibitem{Chiang:1999tz}
T.M.~Chiang, A.~Klemm, S.-T.~Yau and E.~Zaslow, \emph{{Local mirror symmetry: Calculations and interpretations}}, \href{https://doi.org/10.4310/ATMP.1999.v3.n3.a3}{\emph{Adv. Theor. Math. Phys.} {\bfseries 3} (1999) 495} [\href{https://arxiv.org/abs/hep-th/9903053}{{\ttfamily hep-th/9903053}}].

\bibitem{Gorsky:1995zq}
A.~Gorsky, I.~Krichever, A.~Marshakov, A.~Mironov and A.~Morozov, \emph{{Integrability and Seiberg-Witten exact solution}}, \href{https://doi.org/10.1016/0370-2693(95)00723-X}{\emph{Phys. Lett. B} {\bfseries 355} (1995) 466} [\href{https://arxiv.org/abs/hep-th/9505035}{{\ttfamily hep-th/9505035}}].

\bibitem{Martinec:1995by}
E.J.~Martinec and N.P.~Warner, \emph{{Integrable systems and supersymmetric gauge theory}}, \href{https://doi.org/10.1016/0550-3213(95)00588-9}{\emph{Nucl. Phys. B} {\bfseries 459} (1996) 97} [\href{https://arxiv.org/abs/hep-th/9509161}{{\ttfamily hep-th/9509161}}].

\bibitem{Donagi:1995cf}
R.~Donagi and E.~Witten, \emph{{Supersymmetric Yang-Mills theory and integrable systems}}, \href{https://doi.org/10.1016/0550-3213(95)00609-5}{\emph{Nucl. Phys. B} {\bfseries 460} (1996) 299} [\href{https://arxiv.org/abs/hep-th/9510101}{{\ttfamily hep-th/9510101}}].

\bibitem{Aganagic:2003qj}
M.~Aganagic, R.~Dijkgraaf, A.~Klemm, M.~Marino and C.~Vafa, \emph{{Topological strings and integrable hierarchies}}, \href{https://doi.org/10.1007/s00220-005-1448-9}{\emph{Commun. Math. Phys.} {\bfseries 261} (2006) 451} [\href{https://arxiv.org/abs/hep-th/0312085}{{\ttfamily hep-th/0312085}}].

\bibitem{Bershadsky:1993cx}
M.~Bershadsky, S.~Cecotti, H.~Ooguri and C.~Vafa, \emph{{Kodaira-Spencer theory of gravity and exact results for quantum string amplitudes}}, \href{https://doi.org/10.1007/BF02099774}{\emph{Commun. Math. Phys.} {\bfseries 165} (1994) 311} [\href{https://arxiv.org/abs/hep-th/9309140}{{\ttfamily hep-th/9309140}}].

\bibitem{Nekrasov:1996cz}
N.~Nekrasov, \emph{{Five dimensional gauge theories and relativistic integrable systems}}, \href{https://doi.org/10.1016/S0550-3213(98)00436-2}{\emph{Nucl. Phys. B} {\bfseries 531} (1998) 323} [\href{https://arxiv.org/abs/hep-th/9609219}{{\ttfamily hep-th/9609219}}].

\bibitem{Nekrasov:2009rc}
N.A.~Nekrasov and S.L.~Shatashvili, \emph{{Quantization of Integrable Systems and Four Dimensional Gauge Theories}},  in \emph{{16th International Congress on Mathematical Physics}}, pp.~265--289, 2010, \href{https://doi.org/10.1142/9789814304634_0015}{DOI} [\href{https://arxiv.org/abs/0908.4052}{{\ttfamily 0908.4052}}].

\bibitem{Nakajima:2003pg}
H.~Nakajima and K.~Yoshioka, \emph{{Instanton counting on blowup. 1.}}, \href{https://doi.org/10.1007/s00222-005-0444-1}{\emph{Invent. Math.} {\bfseries 162} (2005) 313} [\href{https://arxiv.org/abs/math/0306198}{{\ttfamily math/0306198}}].

\bibitem{Nakajima:2005fg}
H.~Nakajima and K.~Yoshioka, \emph{{Instanton counting on blowup. II. K-theoretic partition function}},  \href{https://arxiv.org/abs/math/0505553}{{\ttfamily math/0505553}}.

\bibitem{Gottsche:2006bm}
L.~Gottsche, H.~Nakajima and K.~Yoshioka, \emph{{K-theoretic Donaldson invariants via instanton counting}}, \href{https://doi.org/10.4310/PAMQ.2009.v5.n3.a5}{\emph{Pure Appl. Math. Quart.} {\bfseries 5} (2009) 1029} [\href{https://arxiv.org/abs/math/0611945}{{\ttfamily math/0611945}}].

\bibitem{Huang:2017mis}
M.-x.~Huang, K.~Sun and X.~Wang, \emph{{Blowup Equations for Refined Topological Strings}}, \href{https://doi.org/10.1007/JHEP10(2018)196}{\emph{JHEP} {\bfseries 10} (2018) 196} [\href{https://arxiv.org/abs/1711.09884}{{\ttfamily 1711.09884}}].

\bibitem{Kim:2019uqw}
J.~Kim, S.-S.~Kim, K.-H.~Lee, K.~Lee and J.~Song, \emph{{Instantons from Blow-up}}, \href{https://doi.org/10.1007/JHEP11(2019)092}{\emph{JHEP} {\bfseries 11} (2019) 092} [\href{https://arxiv.org/abs/1908.11276}{{\ttfamily 1908.11276}}].

\bibitem{Gu:2018gmy}
J.~Gu, B.~Haghighat, K.~Sun and X.~Wang, \emph{{Blowup Equations for 6d SCFTs. I}}, \href{https://doi.org/10.1007/JHEP03(2019)002}{\emph{JHEP} {\bfseries 03} (2019) 002} [\href{https://arxiv.org/abs/1811.02577}{{\ttfamily 1811.02577}}].

\bibitem{Gu:2019dan}
J.~Gu, A.~Klemm, K.~Sun and X.~Wang, \emph{{Elliptic blowup equations for 6d SCFTs. Part II. Exceptional cases}}, \href{https://doi.org/10.1007/JHEP12(2019)039}{\emph{JHEP} {\bfseries 12} (2019) 039} [\href{https://arxiv.org/abs/1905.00864}{{\ttfamily 1905.00864}}].

\bibitem{Gu:2019pqj}
J.~Gu, B.~Haghighat, A.~Klemm, K.~Sun and X.~Wang, \emph{{Elliptic blowup equations for 6d SCFTs. Part III. E-strings, M-strings and chains}}, \href{https://doi.org/10.1007/JHEP07(2020)135}{\emph{JHEP} {\bfseries 07} (2020) 135} [\href{https://arxiv.org/abs/1911.11724}{{\ttfamily 1911.11724}}].

\bibitem{Gu:2020fem}
J.~Gu, B.~Haghighat, A.~Klemm, K.~Sun and X.~Wang, \emph{{Elliptic blowup equations for 6d SCFTs. Part IV. Matters}}, \href{https://doi.org/10.1007/JHEP11(2021)090}{\emph{JHEP} {\bfseries 11} (2021) 090} [\href{https://arxiv.org/abs/2006.03030}{{\ttfamily 2006.03030}}].

\bibitem{Kim:2023glm}
H.-C.~Kim, M.~Kim and Y.~Sugimoto, \emph{{Blowup equations for little strings}}, \href{https://doi.org/10.1007/JHEP05(2023)029}{\emph{JHEP} {\bfseries 05} (2023) 029} [\href{https://arxiv.org/abs/2301.04151}{{\ttfamily 2301.04151}}].

\bibitem{Kim:2021gyj}
H.-C.~Kim, M.~Kim and S.-S.~Kim, \emph{{5d/6d Wilson loops from blowups}}, \href{https://doi.org/10.1007/JHEP08(2021)131}{\emph{JHEP} {\bfseries 08} (2021) 131} [\href{https://arxiv.org/abs/2106.04731}{{\ttfamily 2106.04731}}].

\bibitem{Jeong:2020uxz}
S.~Jeong and N.~Nekrasov, \emph{{Riemann-Hilbert correspondence and blown up surface defects}}, \href{https://doi.org/10.1007/JHEP12(2020)006}{\emph{JHEP} {\bfseries 12} (2020) 006} [\href{https://arxiv.org/abs/2007.03660}{{\ttfamily 2007.03660}}].

\bibitem{Nekrasov:2020qcq}
N.~Nekrasov, \emph{{Blowups in BPS/CFT Correspondence, and Painlev\'e VI}}, \href{https://doi.org/10.1007/s00023-023-01301-5}{\emph{Annales Henri Poincare} {\bfseries 25} (2024) 1123} [\href{https://arxiv.org/abs/2007.03646}{{\ttfamily 2007.03646}}].

\bibitem{Hwang:2014uwa}
C.~Hwang, J.~Kim, S.~Kim and J.~Park, \emph{{General instanton counting and 5d SCFT}}, \href{https://doi.org/10.1007/JHEP07(2015)063}{\emph{JHEP} {\bfseries 07} (2015) 063} [\href{https://arxiv.org/abs/1406.6793}{{\ttfamily 1406.6793}}].

\bibitem{Gaiotto:2014ina}
D.~Gaiotto and H.-C.~Kim, \emph{{Surface defects and instanton partition functions}}, \href{https://doi.org/10.1007/JHEP10(2016)012}{\emph{JHEP} {\bfseries 10} (2016) 012} [\href{https://arxiv.org/abs/1412.2781}{{\ttfamily 1412.2781}}].

\bibitem{Gaiotto:2012xa}
D.~Gaiotto, L.~Rastelli and S.S.~Razamat, \emph{{Bootstrapping the superconformal index with surface defects}}, \href{https://doi.org/10.1007/JHEP01(2013)022}{\emph{JHEP} {\bfseries 01} (2013) 022} [\href{https://arxiv.org/abs/1207.3577}{{\ttfamily 1207.3577}}].

\bibitem{Hayashi:2023boy}
H.~Hayashi, S.-S.~Kim, K.~Lee and F.~Yagi, \emph{{Seiberg-Witten curves with O7$^{\pm}$-planes}}, \href{https://doi.org/10.1007/JHEP11(2023)178}{\emph{JHEP} {\bfseries 11} (2023) 178} [\href{https://arxiv.org/abs/2306.11631}{{\ttfamily 2306.11631}}].

\bibitem{Kim:2023qwh}
H.-C.~Kim, M.~Kim, S.-S.~Kim and G.~Zafrir, \emph{{Superconformal indices for non-Lagrangian theories in five dimensions}}, \href{https://doi.org/10.1007/JHEP03(2024)164}{\emph{JHEP} {\bfseries 03} (2024) 164} [\href{https://arxiv.org/abs/2307.03231}{{\ttfamily 2307.03231}}].

\bibitem{Kim:2024vci}
S.-S.~Kim, X.~Li, S.~Nawata and F.~Yagi, \emph{{Freezing and BPS jumping}}, \href{https://doi.org/10.1007/JHEP05(2024)340}{\emph{JHEP} {\bfseries 05} (2024) 340} [\href{https://arxiv.org/abs/2403.12525}{{\ttfamily 2403.12525}}].

\bibitem{Hanany:1996ie}
A.~Hanany and E.~Witten, \emph{{Type IIB superstrings, BPS monopoles, and three-dimensional gauge dynamics}}, \href{https://doi.org/10.1016/S0550-3213(97)00157-0}{\emph{Nucl. Phys. B} {\bfseries 492} (1997) 152} [\href{https://arxiv.org/abs/hep-th/9611230}{{\ttfamily hep-th/9611230}}].

\bibitem{Morrison:1996xf}
D.R.~Morrison and N.~Seiberg, \emph{{Extremal transitions and five-dimensional supersymmetric field theories}}, \href{https://doi.org/10.1016/S0550-3213(96)00592-5}{\emph{Nucl. Phys. B} {\bfseries 483} (1997) 229} [\href{https://arxiv.org/abs/hep-th/9609070}{{\ttfamily hep-th/9609070}}].

\bibitem{Bergman:2013ala}
O.~Bergman, D.~Rodr\'\i{}guez-G\'omez and G.~Zafrir, \emph{{Discrete $\theta$ and the 5d superconformal index}}, \href{https://doi.org/10.1007/JHEP01(2014)079}{\emph{JHEP} {\bfseries 01} (2014) 079} [\href{https://arxiv.org/abs/1310.2150}{{\ttfamily 1310.2150}}].

\bibitem{Bergman:2015dpa}
O.~Bergman and G.~Zafrir, \emph{{5d fixed points from brane webs and O7-planes}}, \href{https://doi.org/10.1007/JHEP12(2015)163}{\emph{JHEP} {\bfseries 12} (2015) 163} [\href{https://arxiv.org/abs/1507.03860}{{\ttfamily 1507.03860}}].

\bibitem{Hayashi:2019yxj}
H.~Hayashi, S.-S.~Kim, K.~Lee and F.~Yagi, \emph{{Rank-3 antisymmetric matter on 5-brane webs}}, \href{https://doi.org/10.1007/JHEP05(2019)133}{\emph{JHEP} {\bfseries 05} (2019) 133} [\href{https://arxiv.org/abs/1902.04754}{{\ttfamily 1902.04754}}].

\bibitem{Benini:2009gi}
F.~Benini, S.~Benvenuti and Y.~Tachikawa, \emph{{Webs of five-branes and N=2 superconformal field theories}}, \href{https://doi.org/10.1088/1126-6708/2009/09/052}{\emph{JHEP} {\bfseries 09} (2009) 052} [\href{https://arxiv.org/abs/0906.0359}{{\ttfamily 0906.0359}}].

\bibitem{Kim:2014nqa}
S.-S.~Kim and F.~Yagi, \emph{{5d E$_{n}$ Seiberg-Witten curve via toric-like diagram}}, \href{https://doi.org/10.1007/JHEP06(2015)082}{\emph{JHEP} {\bfseries 06} (2015) 082} [\href{https://arxiv.org/abs/1411.7903}{{\ttfamily 1411.7903}}].

\bibitem{Arias-Tamargo:2024fjt}
G.~Arias-Tamargo, S.~Franco and D.~Rodr{\'\i}guez-G{\'o}mez, \emph{{The geometry of GTPs and 5d SCFTs}}, \href{https://doi.org/10.1007/JHEP07(2024)159}{\emph{JHEP} {\bfseries 07} (2024) 159} [\href{https://arxiv.org/abs/2403.09776}{{\ttfamily 2403.09776}}].

\bibitem{Ghim:2024asj}
D.~Ghim, M.~Kho and R.-K.~Seong, \emph{{Combinatorial and algebraic mutations of toric Fano 3-folds and mass deformations of 2d(0,2) quiver gauge theories}}, \href{https://doi.org/10.1103/PhysRevD.110.086001}{\emph{Phys. Rev. D} {\bfseries 110} (2024) 086001} [\href{https://arxiv.org/abs/2407.19924}{{\ttfamily 2407.19924}}].

\bibitem{Witten:2003ya}
E.~Witten, \emph{{SL(2,Z) action on three-dimensional conformal field theories with Abelian symmetry}},  in \emph{{From Fields to Strings: Circumnavigating Theoretical Physics: A Conference in Tribute to Ian Kogan}}, pp.~1173--1200, 7, 2003 [\href{https://arxiv.org/abs/hep-th/0307041}{{\ttfamily hep-th/0307041}}].

\bibitem{Jeong:2024onv}
S.~Jeong and N.~Lee, \emph{{Bispectral duality and separation of variables from surface defect transition}}, \href{https://doi.org/10.1007/JHEP12(2024)142}{\emph{JHEP} {\bfseries 12} (2024) 142} [\href{https://arxiv.org/abs/2402.13889}{{\ttfamily 2402.13889}}].

\bibitem{Gopakumar:1998ii}
R.~Gopakumar and C.~Vafa, \emph{{M theory and topological strings. 1.}},  \href{https://arxiv.org/abs/hep-th/9809187}{{\ttfamily hep-th/9809187}}.

\bibitem{Gopakumar:1998jq}
R.~Gopakumar and C.~Vafa, \emph{{M theory and topological strings. 2.}},  \href{https://arxiv.org/abs/hep-th/9812127}{{\ttfamily hep-th/9812127}}.

\bibitem{Bonetti:2013ela}
F.~Bonetti, T.W.~Grimm and S.~Hohenegger, \emph{{One-loop Chern-Simons terms in five dimensions}}, \href{https://doi.org/10.1007/JHEP07(2013)043}{\emph{JHEP} {\bfseries 07} (2013) 043} [\href{https://arxiv.org/abs/1302.2918}{{\ttfamily 1302.2918}}].

\bibitem{Grimm:2015zea}
T.W.~Grimm and A.~Kapfer, \emph{{Anomaly Cancelation in Field Theory and F-theory on a Circle}}, \href{https://doi.org/10.1007/JHEP05(2016)102}{\emph{JHEP} {\bfseries 05} (2016) 102} [\href{https://arxiv.org/abs/1502.05398}{{\ttfamily 1502.05398}}].

\bibitem{BenettiGenolini:2019zth}
P.~Benetti~Genolini, M.~Honda, H.-C.~Kim, D.~Tong and C.~Vafa, \emph{{Evidence for a Non-Supersymmetric 5d CFT from Deformations of 5d $SU(2)$ SYM}}, \href{https://doi.org/10.1007/JHEP05(2020)058}{\emph{JHEP} {\bfseries 05} (2020) 058} [\href{https://arxiv.org/abs/2001.00023}{{\ttfamily 2001.00023}}].

\bibitem{Young:2011aa}
D.~Young, \emph{{Wilson Loops in Five-Dimensional Super-Yang-Mills}}, \href{https://doi.org/10.1007/JHEP02(2012)052}{\emph{JHEP} {\bfseries 02} (2012) 052} [\href{https://arxiv.org/abs/1112.3309}{{\ttfamily 1112.3309}}].

\bibitem{Assel:2012nf}
B.~Assel, J.~Estes and M.~Yamazaki, \emph{{Wilson Loops in 5d N=1 SCFTs and AdS/CFT}}, \href{https://doi.org/10.1007/s00023-013-0249-5}{\emph{Annales Henri Poincare} {\bfseries 15} (2014) 589} [\href{https://arxiv.org/abs/1212.1202}{{\ttfamily 1212.1202}}].

\bibitem{Ooguri:1999bv}
H.~Ooguri and C.~Vafa, \emph{{Knot invariants and topological strings}}, \href{https://doi.org/10.1016/S0550-3213(00)00118-8}{\emph{Nucl. Phys. B} {\bfseries 577} (2000) 419} [\href{https://arxiv.org/abs/hep-th/9912123}{{\ttfamily hep-th/9912123}}].

\bibitem{Gukov:2004hz}
S.~Gukov, A.S.~Schwarz and C.~Vafa, \emph{{Khovanov-Rozansky homology and topological strings}}, \href{https://doi.org/10.1007/s11005-005-0008-8}{\emph{Lett. Math. Phys.} {\bfseries 74} (2005) 53} [\href{https://arxiv.org/abs/hep-th/0412243}{{\ttfamily hep-th/0412243}}].

\bibitem{Aganagic:2011sg}
M.~Aganagic and S.~Shakirov, \emph{{Knot Homology and Refined Chern-Simons Index}}, \href{https://doi.org/10.1007/s00220-014-2197-4}{\emph{Commun. Math. Phys.} {\bfseries 333} (2015) 187} [\href{https://arxiv.org/abs/1105.5117}{{\ttfamily 1105.5117}}].

\bibitem{Aganagic:2012ne}
M.~Aganagic and S.~Shakirov, \emph{{Refined Chern-Simons Theory and Knot Homology}}, \href{https://doi.org/10.1090/pspum/085/1372}{\emph{Proc. Symp. Pure Math.} {\bfseries 85} (2012) 3} [\href{https://arxiv.org/abs/1202.2489}{{\ttfamily 1202.2489}}].

\bibitem{Aganagic:2012hs}
M.~Aganagic and S.~Shakirov, \emph{{Refined Chern-Simons Theory and Topological String}},  \href{https://arxiv.org/abs/1210.2733}{{\ttfamily 1210.2733}}.

\bibitem{Cheng:2021nex}
S.~Cheng and P.~Su\l{}kowski, \emph{{Refined open topological strings revisited}}, \href{https://doi.org/10.1103/PhysRevD.104.106012}{\emph{Phys. Rev. D} {\bfseries 104} (2021) 106012} [\href{https://arxiv.org/abs/2104.00713}{{\ttfamily 2104.00713}}].

\bibitem{Kashani-Poor:2016edc}
A.-K.~Kashani-Poor, \emph{{Quantization condition from exact WKB for difference equations}}, \href{https://doi.org/10.1007/JHEP06(2016)180}{\emph{JHEP} {\bfseries 06} (2016) 180} [\href{https://arxiv.org/abs/1604.01690}{{\ttfamily 1604.01690}}].

\bibitem{Aganagic:2001nx}
M.~Aganagic, A.~Klemm and C.~Vafa, \emph{{Disk instantons, mirror symmetry and the duality web}}, \href{https://doi.org/10.1515/zna-2002-1-201}{\emph{Z. Naturforsch. A} {\bfseries 57} (2002) 1} [\href{https://arxiv.org/abs/hep-th/0105045}{{\ttfamily hep-th/0105045}}].

\bibitem{Bullimore:2014awa}
M.~Bullimore, H.-C.~Kim and P.~Koroteev, \emph{{Defects and Quantum Seiberg-Witten Geometry}}, \href{https://doi.org/10.1007/JHEP05(2015)095}{\emph{JHEP} {\bfseries 05} (2015) 095} [\href{https://arxiv.org/abs/1412.6081}{{\ttfamily 1412.6081}}].

\bibitem{Lerche:2001cw}
W.~Lerche and P.~Mayr, \emph{{On N=1 mirror symmetry for open type 2 strings}},  \href{https://arxiv.org/abs/hep-th/0111113}{{\ttfamily hep-th/0111113}}.

\bibitem{Chen:2020jla}
J.~Chen, B.~Haghighat, H.-C.~Kim and M.~Sperling, \emph{{Elliptic quantum curves of class $ {\mathcal{S}}_k $}}, \href{https://doi.org/10.1007/JHEP03(2021)028}{\emph{JHEP} {\bfseries 03} (2021) 028} [\href{https://arxiv.org/abs/2008.05155}{{\ttfamily 2008.05155}}].

\bibitem{Kim:2014dza}
J.~Kim, S.~Kim, K.~Lee, J.~Park and C.~Vafa, \emph{{Elliptic Genus of E-strings}}, \href{https://doi.org/10.1007/JHEP09(2017)098}{\emph{JHEP} {\bfseries 09} (2017) 098} [\href{https://arxiv.org/abs/1411.2324}{{\ttfamily 1411.2324}}].

\bibitem{Kim:2020npz}
S.-S.~Kim, Y.~Sugimoto and F.~Yagi, \emph{{Surface defects on E-string from 5-brane webs}}, \href{https://doi.org/10.1007/JHEP12(2020)183}{\emph{JHEP} {\bfseries 12} (2020) 183} [\href{https://arxiv.org/abs/2008.06428}{{\ttfamily 2008.06428}}].

\bibitem{Chen:2021ivd}
J.~Chen, B.~Haghighat, H.-C.~Kim, M.~Sperling and X.~Wang, \emph{{E-string quantum curve}}, \href{https://doi.org/10.1016/j.nuclphysb.2021.115602}{\emph{Nucl. Phys. B} {\bfseries 973} (2021) 115602} [\href{https://arxiv.org/abs/2103.16996}{{\ttfamily 2103.16996}}].

\bibitem{Diejen:1994}
J.F.~van Diejen, \emph{{Integrability of difference Calogero–Moser systems}}, \href{https://doi.org/10.1063/1.530498}{\emph{J. Math. Phys.} {\bfseries 35} (1994) 2983}.

\bibitem{Nazzal:2018brc}
B.~Nazzal and S.S.~Razamat, \emph{{Surface Defects in E-String Compactifications and the van Diejen Model}}, \href{https://doi.org/10.3842/SIGMA.2018.036}{\emph{SIGMA} {\bfseries 14} (2018) 036} [\href{https://arxiv.org/abs/1801.00960}{{\ttfamily 1801.00960}}].

\bibitem{Bhardwaj:2019jtr}
L.~Bhardwaj, \emph{{On the classification of 5d SCFTs}}, \href{https://doi.org/10.1007/JHEP09(2020)007}{\emph{JHEP} {\bfseries 09} (2020) 007} [\href{https://arxiv.org/abs/1909.09635}{{\ttfamily 1909.09635}}].

\bibitem{Bonelli:2024wha}
G.~Bonelli, P.~Gavrylenko, I.~Majtara and A.~Tanzini, \emph{{Surface observables in gauge theories, modular Painlev\'e tau functions and non-perturbative topological strings}},  \href{https://arxiv.org/abs/2410.17868}{{\ttfamily 2410.17868}}.

\bibitem{Bonelli:2025owb}
G.~Bonelli, A.~Shchechkin and A.~Tanzini, \emph{{Refined Painlev\'e/gauge theory correspondence and quantum tau functions}},  \href{https://arxiv.org/abs/2502.01499}{{\ttfamily 2502.01499}}.

\bibitem{Bershtein:2016aef}
M.A.~Bershtein and A.I.~Shchechkin, \emph{{q-deformed Painlev\'e $\tau$ function and q-deformed conformal blocks}}, \href{https://doi.org/10.1088/1751-8121/aa5572}{\emph{J. Phys. A} {\bfseries 50} (2017) 085202} [\href{https://arxiv.org/abs/1608.02566}{{\ttfamily 1608.02566}}].

\bibitem{Bershtein:2017swf}
M.~Bershtein, P.~Gavrylenko and A.~Marshakov, \emph{{Cluster integrable systems, $q$-Painlev\'e equations and their quantization}}, \href{https://doi.org/10.1007/JHEP02(2018)077}{\emph{JHEP} {\bfseries 02} (2018) 077} [\href{https://arxiv.org/abs/1711.02063}{{\ttfamily 1711.02063}}].

\bibitem{Bonelli:2017gdk}
G.~Bonelli, A.~Grassi and A.~Tanzini, \emph{{Quantum curves and $q$-deformed Painlev\'e equations}}, \href{https://doi.org/10.1007/s11005-019-01174-y}{\emph{Lett. Math. Phys.} {\bfseries 109} (2019) 1961} [\href{https://arxiv.org/abs/1710.11603}{{\ttfamily 1710.11603}}].

\bibitem{Gamayun:2012ma}
O.~Gamayun, N.~Iorgov and O.~Lisovyy, \emph{{Conformal field theory of Painlev\'e VI}}, \href{https://doi.org/10.1007/JHEP10(2012)038}{\emph{JHEP} {\bfseries 10} (2012) 038} [\href{https://arxiv.org/abs/1207.0787}{{\ttfamily 1207.0787}}].

\bibitem{Gavrylenko:2016zlf}
P.~Gavrylenko and O.~Lisovyy, \emph{{Fredholm Determinant and Nekrasov Sum Representations of Isomonodromic Tau Functions}}, \href{https://doi.org/10.1007/s00220-018-3224-7}{\emph{Commun. Math. Phys.} {\bfseries 363} (2018) 1} [\href{https://arxiv.org/abs/1608.00958}{{\ttfamily 1608.00958}}].

\bibitem{Aganagic:2006wq}
M.~Aganagic, V.~Bouchard and A.~Klemm, \emph{{Topological Strings and (Almost) Modular Forms}}, \href{https://doi.org/10.1007/s00220-007-0383-3}{\emph{Commun. Math. Phys.} {\bfseries 277} (2008) 771} [\href{https://arxiv.org/abs/hep-th/0607100}{{\ttfamily hep-th/0607100}}].

\bibitem{Wang:2023zcb}
X.~Wang, \emph{{Wilson loops, holomorphic anomaly equations and blowup equations}},  \href{https://arxiv.org/abs/2305.09171}{{\ttfamily 2305.09171}}.

\bibitem{Fang:2018ett}
B.~Fang, Y.~Ruan, Y.~Zhang and J.~Zhou, \emph{{Open Gromov\textendash{}Witten Theory of $K_{{\mathbb {P}}^2}, K_{{{\mathbb {P}}^1}\times {{\mathbb {P}}^1}}, K_{W{\mathbb {P}}\left[ 1,1,2\right] }, K_{{{\mathbb {F}}}_1}$ and Jacobi Forms}}, \href{https://doi.org/10.1007/s00220-019-03440-5}{\emph{Commun. Math. Phys.} {\bfseries 369} (2019) 675} [\href{https://arxiv.org/abs/1805.04894}{{\ttfamily 1805.04894}}].

\bibitem{Ruan:2019aug}
Y.~Ruan, Y.~Zhang and J.~Zhou, \emph{{Genus Two Siegel Quasi-Modular Forms and Gromov\textendash{}Witten Theory of Toric Calabi\textendash{}Yau Threefolds}}, \href{https://doi.org/10.1007/s00220-022-04534-3}{\emph{Commun. Math. Phys.} {\bfseries 398} (2023) 757} [\href{https://arxiv.org/abs/1911.07204}{{\ttfamily 1911.07204}}].

\bibitem{Bouchard:2007ys}
V.~Bouchard, A.~Klemm, M.~Marino and S.~Pasquetti, \emph{{Remodeling the B-model}}, \href{https://doi.org/10.1007/s00220-008-0620-4}{\emph{Commun. Math. Phys.} {\bfseries 287} (2009) 117} [\href{https://arxiv.org/abs/0709.1453}{{\ttfamily 0709.1453}}].

\bibitem{Eynard:2007kz}
B.~Eynard and N.~Orantin, \emph{{Invariants of algebraic curves and topological expansion}}, \href{https://doi.org/10.4310/CNTP.2007.v1.n2.a4}{\emph{Commun. Num. Theor. Phys.} {\bfseries 1} (2007) 347} [\href{https://arxiv.org/abs/math-ph/0702045}{{\ttfamily math-ph/0702045}}].

\bibitem{Marino:2016rsq}
M.~Marino and S.~Zakany, \emph{{Exact eigenfunctions and the open topological string}}, \href{https://doi.org/10.1088/1751-8121/aa791e}{\emph{J. Phys. A} {\bfseries 50} (2017) 325401} [\href{https://arxiv.org/abs/1606.05297}{{\ttfamily 1606.05297}}].

\bibitem{Francois:2025wwd}
M.~Fran\c{c}ois and A.~Grassi, \emph{{On the open TS/ST correspondence}},  \href{https://arxiv.org/abs/2503.21762}{{\ttfamily 2503.21762}}.

\bibitem{Marino:2017gyg}
M.~Marino and S.~Zakany, \emph{{Wavefunctions, integrability, and open strings}}, \href{https://doi.org/10.1007/JHEP05(2019)014}{\emph{JHEP} {\bfseries 05} (2019) 014} [\href{https://arxiv.org/abs/1706.07402}{{\ttfamily 1706.07402}}].

\bibitem{Shadchin:2004yx}
S.~Shadchin, \emph{{Saddle point equations in Seiberg-Witten theory}}, \href{https://doi.org/10.1088/1126-6708/2004/10/033}{\emph{JHEP} {\bfseries 10} (2004) 033} [\href{https://arxiv.org/abs/hep-th/0408066}{{\ttfamily hep-th/0408066}}].

\bibitem{Benini:2013nda}
F.~Benini, R.~Eager, K.~Hori and Y.~Tachikawa, \emph{{Elliptic genera of two-dimensional N=2 gauge theories with rank-one gauge groups}}, \href{https://doi.org/10.1007/s11005-013-0673-y}{\emph{Lett. Math. Phys.} {\bfseries 104} (2014) 465} [\href{https://arxiv.org/abs/1305.0533}{{\ttfamily 1305.0533}}].

\bibitem{Benini:2013xpa}
F.~Benini, R.~Eager, K.~Hori and Y.~Tachikawa, \emph{{Elliptic Genera of 2d ${\mathcal{N}}$ = 2 Gauge Theories}}, \href{https://doi.org/10.1007/s00220-014-2210-y}{\emph{Commun. Math. Phys.} {\bfseries 333} (2015) 1241} [\href{https://arxiv.org/abs/1308.4896}{{\ttfamily 1308.4896}}].

\bibitem{Haghighat:2013gba}
B.~Haghighat, A.~Iqbal, C.~Koz\c{c}az, G.~Lockhart and C.~Vafa, \emph{{M-Strings}}, \href{https://doi.org/10.1007/s00220-014-2139-1}{\emph{Commun. Math. Phys.} {\bfseries 334} (2015) 779} [\href{https://arxiv.org/abs/1305.6322}{{\ttfamily 1305.6322}}].

\bibitem{Haghighat:2013tka}
B.~Haghighat, C.~Kozcaz, G.~Lockhart and C.~Vafa, \emph{{Orbifolds of M-strings}}, \href{https://doi.org/10.1103/PhysRevD.89.046003}{\emph{Phys. Rev. D} {\bfseries 89} (2014) 046003} [\href{https://arxiv.org/abs/1310.1185}{{\ttfamily 1310.1185}}].

\end{thebibliography}\endgroup

\end{document}